\pdfoutput=1
\documentclass[prd,showpacs,letterpaper,10pt,aps,notitlepage,nofootinbib,superscriptaddress]{revtex4-2}

\usepackage{amsfonts,amsmath,graphicx}
\usepackage{slashed}
\usepackage{hyperref}
\usepackage{placeins}
\usepackage{mathrsfs}
\usepackage{rotating}
\usepackage{color}

\hypersetup{colorlinks, linkcolor = [rgb]{0,0.0,0.75}, citecolor = [rgb]{0,0.0,0.75}, urlcolor = [rgb]{0,0.0,0.75}}

\newcommand{\nb}{\phantom{0}}
\newcommand{\wm}{\phantom{-}}

\begin{document}

\title{$\Lambda_b \to \Lambda \, \ell^+ \ell^-$ form factors, differential branching fraction, and angular observables \\ from lattice QCD with relativistic $b$ quarks}

\author{William Detmold}
\affiliation{Center for Theoretical Physics, Massachusetts Institute of Technology, Cambridge, MA 02139, USA}

\author{Stefan Meinel\:}
\email{smeinel@email.arizona.edu}
\affiliation{Department of Physics, University of Arizona, Tucson, AZ 85721, USA}
\affiliation{RIKEN BNL Research Center, Brookhaven National Laboratory, Upton, NY 11973, USA}


\begin{abstract}
Using $(2+1)$-flavor lattice QCD, we compute the 10 form factors describing the $\Lambda_b\to\Lambda$ matrix elements of the $b\to s$ vector, axial vector, and tensor currents.
The calculation is based on gauge field ensembles generated by the RBC and UKQCD Collaborations with a domain-wall action for the $u$, $d$, and $s$ quarks and the Iwasaki gauge action. The
$b$ quark is implemented using an anisotropic clover action, tuned nonperturbatively to the physical point, and the currents are renormalized with a mostly nonperturbative method. We perform simultaneous
chiral, continuum, and kinematic extrapolations of the form factors through modified $z$ expansions. Using our form factor results, we obtain precise predictions for the
$\Lambda_b \to \Lambda(\to p^+ \pi^-) \, \mu^+ \mu^-$ differential branching fraction and angular observables in the Standard Model.
\end{abstract}

\maketitle

\FloatBarrier
\section{Introduction}
\FloatBarrier

Decays of bottom hadrons involving the flavor-changing neutral-current transition $b \to s\, \ell^+ \ell^-$ are sensitive probes of physics beyond the Standard Model.
The properties of multiple exclusive $b \to s\, \ell^+ \ell^-$ decay channels have been measured with unprecedented precision at the Large Hadron Collider \cite{Blake:2015tda}, and further substantial increases in statistics
are expected in the near future. Some tensions between the experimental data and calculations in the Standard Model have been found, including in the $B \to K^* (\to K\,\pi) \mu^+\mu^-$
angular distribution \cite{Descotes-Genon:2013wba, Aaij:2013qta, Aaij:2015oid}, the $B \to K^{(*)} \mu^+\mu^-$ and $B_s \to \phi \mu^+\mu^-$ differential decay rates \cite{Bouchard:2013mia, Horgan:2013pva, Straub:2015ica, Du:2015tda, Aaij:2013iag, Aaij:2013aln, Aaij:2015esa},
and the ratio $R_K$ of $B\to K\mu^+\mu^-$ and $B\to K e^+e^-$ decay rates \cite{Aaij:2014ora}. Global fits of all relevant $B$ and $B_s$ meson decay modes
show that a better agreement with the data can be obtained if the Wilson coefficient $C_9$ in the $b \to s\, \mu^+ \mu^-$ effective Hamiltonian is shifted by approximately $-25\%$ from
its Standard-Model value (other scenarios involving deviations in multiple Wilson coefficients have also been considered) \cite{Altmannshofer:2014rta, Descotes-Genon:2015uva}. However, with the exception of $R_K$, the decay observables are strongly
affected by hadronic physics, and it is important to carefully examine all sources of the theory uncertainties. The hadronic contributions include local matrix elements
of the $b\to s$ tensor, vector, and axial vector currents from the operators $O_{7,9,10}$ in the effective
Hamiltonian, as well as nonlocal matrix elements of products of the operators $O_{1\text{-}6;8}$ with the quark electromagnetic current. The local hadronic
matrix elements are expressed in terms of form factors, and can be calculated directly using lattice QCD \cite{Detmold:2012vy, Bouchard:2013pna, Horgan:2013hoa, Bailey:2015dka}.
The treatment of the nonlocal hadronic matrix elements is significantly more challenging, and is usually based on an operator product expansion at high $q^2$ \cite{Grinstein:2004vb, Beylich:2011aq},
and QCD factorization \cite{Beneke:2001at} combined with light-cone sum rules \cite{Khodjamirian:2010vf, Khodjamirian:2012rm} at low $q^2$. A particular problem is that the contributions from $O_1$ and $O_2$
are enhanced by a multitude of charmonium resonances, with unexpectedly large deviations from naive factorization \cite{Lyon:2014hpa}, and can mimic a shift in $C_9$. To distinguish
an apparent shift in $C_9$ caused by uncontrolled charm contributions from a shift due to new physics, one can study its $q^2$-dependence \cite{Altmannshofer:2015sma} and compare the effects seen in multiple different decay modes.

The baryonic decay mode $\Lambda_b \to \Lambda(\to p\,\pi) \ell^+ \ell^-$ \cite{Aaltonen:2011qs, Aaij:2013mna, Aaij:2015xza} can shed new light on these puzzles. Similarly to $B \to K^* (\to K\,\pi) \ell^+\ell^-$, this decay
provides a wealth of angular observables that can be used to disentangle the contributions from individual operators in the $b \to s\, \ell^+ \ell^-$ effective Hamiltonian \cite{Gutsche:2013pp, Boer:2014kda}
(see also Refs.~\cite{Gremm:1995nx, Mannel:1997xy, Chua:1998dx, Huang:1998ek, Hiller:2001zj, Chen:2001ki, Chen:2002rg, Aliev:2005np, Wang:2008sm, Aslam:2008hp, Mannel:2011xg, Azizi:2015hoa, Mott:2015zma, Kumar:2015tnz}).
The theoretical description of $\Lambda_b \to \Lambda(\to p\,\pi) \ell^+ \ell^-$ is cleaner than that of $B \to K^* (\to K\,\pi) \ell^+\ell^-$ because the $\Lambda$ is stable under the strong interactions.
The $\Lambda_b \to \Lambda$ form factors are thus ``gold-plated'' quantities for lattice QCD that can be calculated to high precision using standard methods. In contrast,
a rigorous analysis of $B \to K^* (\to K\,\pi) \ell^+\ell^-$ would require the computation of $B \to K \pi$ matrix elements; lattice methods for $1\to 2$ transition form factors
have recently been developed \cite{Lellouch:2000pv, Briceno:2014uqa, Briceno:2015csa}, but numerical results for $B \to K \pi$ are not yet available. Besides these simplifications
in terms of the hadronic physics, the angular distribution of the decay $\Lambda_b \to \Lambda(\to p\,\pi) \ell^+ \ell^-$ is sensitive to new combinations of Wilson coefficients that do not appear
in $B \to K^* (\to K\,\pi) \ell^+\ell^-$; this is a consequence of parity violation in the secondary weak decay $\Lambda\to p\,\pi$ \cite{Boer:2014kda}. A further difference
compared to the $B$ decays is the nonzero spin of the initial hadron. While the $\Lambda_b$ polarization at LHCb was found to be small and consistent
with zero \cite{Aaij:2013oxa}, in principle, decays of polarized $\Lambda_b$ baryons give access to even more observables (taking into account the $\Lambda_b$ polarization direction,
the decay distribution of $\Lambda_b \to \Lambda(\to p\,\pi) \ell^+ \ell^-$ depends on five angles \cite{Korner:2014bca}). Finally, just like the mesonic $b \to s\, \ell^+ \ell^-$ decays,
the baryonic mode $\Lambda_b \to \Lambda(\to p\,\pi) \ell^+ \ell^-$ is affected by charmonium resonances, which may contribute here with different phases, providing a new
handle on this difficult issue.

Given all of these motivations, there is clearly a need for precise determinations of the $\Lambda_b \to \Lambda$ form factors. These form factors have been studied
using continuum-based methods in Refs.~\cite{Mannel:1990vg, Hussain:1990uu, Hussain:1992rb, Cheng:1994kp, Cheng:1995fe, Huang:1998ek, Mohanta:1999id, He:2006ud, Wang:2008sm, Wang:2009hra, Aliev:2010uy, Mott:2011cx,
Feldmann:2011xf, Mannel:2011xg, Wang:2011uv, Gutsche:2013pp, Liu:2015qfa, Mott:2015zma, Wang:2015ndk}. In Ref.~\cite{Detmold:2012vy}, we published a first lattice QCD
calculation of the $\Lambda_b \to \Lambda$ form factors, where the $b$ quark was treated at leading order in heavy-quark effective theory to simplify the analysis. In the following,
we present a new lattice QCD calculation in which we do not make this approximation (early progress was shown in Ref.~\cite{Meinel:2014wua}). Using a relativistic heavy-quark action \cite{ElKhadra:1996mp, Aoki:2001ra, Aoki:2003dg, Lin:2006ur, Aoki:2012xaa},
we now work directly at the physical $b$ quark mass and compute all 10 QCD form factors describing the $\Lambda_b \to \Lambda$ matrix elements of the
$b\to s$ vector, axial vector, and tensor currents. The methods closely follow Ref.~\cite{Detmold:2015aaa}, where we computed the $\Lambda_b \to p$ and $\Lambda_b
\to \Lambda_c$ form factors that were used in Ref.~\cite{Aaij:2015bfa} to determine the ratio $|V_{ub}/V_{cb}|$ from the decays $\Lambda_b \to p\, \mu^- \bar{\nu}_\mu$
and $\Lambda_b \to \Lambda_c\, \mu^- \bar{\nu}_\mu$ at the LHC. Our lattice calculations utilize a domain-wall action \cite{Kaplan:1992bt, Furman:1994ky, Shamir:1993zy} for the $u$, $d$, and $s$ quarks, and are based
on gauge field ensembles generated by the RBC and UKQCD Collaborations \cite{Aoki:2010dy}.

We proceed by summarizing the relevant definitions of the $\Lambda_b \to \Lambda$ form factors in Sec.~\ref{sec:FFdefinitions}, before presenting our computation
of the form factors on the lattice in Sec.~\ref{sec:lattice}. We discuss the fits of the form factors using the modified $z$ expansion
and the estimates of systematic uncertainties in Sec.~\ref{sec:ccextrap}. Our predictions for the $\Lambda_b \to \Lambda(\to p\,\pi) \mu^+ \mu^-$
differential branching fraction and angular observables are given in Sec.~\ref{sec:observables}.

\FloatBarrier
\section{\label{sec:FFdefinitions}Definitions of the form factors}
\FloatBarrier

In this work, we mainly use the helicity-based definition of the form factors from Ref.~\cite{Feldmann:2011xf}, which is given by
\begin{eqnarray}
 \nonumber \langle \Lambda(p^\prime,s^\prime) | \overline{s} \,\gamma^\mu\, b | \Lambda_b(p,s) \rangle &=&
 \overline{u}_\Lambda(p^\prime,s^\prime) \bigg[ f_0(q^2)\: (m_{\Lambda_b}-m_\Lambda)\frac{q^\mu}{q^2} \\
 \nonumber && \phantom{\overline{u}_\Lambda \bigg[}+ f_+(q^2) \frac{m_{\Lambda_b}+m_\Lambda}{s_+}\left( p^\mu + p^{\prime \mu} - (m_{\Lambda_b}^2-m_\Lambda^2)\frac{q^\mu}{q^2}  \right) \\
 && \phantom{\overline{u}_\Lambda \bigg[}+ f_\perp(q^2) \left(\gamma^\mu - \frac{2m_\Lambda}{s_+} p^\mu - \frac{2 m_{\Lambda_b}}{s_+} p^{\prime \mu} \right) \bigg] u_{\Lambda_b}(p,s), \\
 \nonumber \langle \Lambda(p^\prime,s^\prime) | \overline{s} \,\gamma^\mu\gamma_5\, b | \Lambda_b(p,s) \rangle &=&
 -\overline{u}_\Lambda(p^\prime,s^\prime) \:\gamma_5 \bigg[ g_0(q^2)\: (m_{\Lambda_b}+m_\Lambda)\frac{q^\mu}{q^2} \\
 \nonumber && \phantom{\overline{u}_\Lambda \bigg[}+ g_+(q^2)\frac{m_{\Lambda_b}-m_\Lambda}{s_-}\left( p^\mu + p^{\prime \mu} - (m_{\Lambda_b}^2-m_\Lambda^2)\frac{q^\mu}{q^2}  \right) \\
 && \phantom{\overline{u}_\Lambda \bigg[}+ g_\perp(q^2) \left(\gamma^\mu + \frac{2m_\Lambda}{s_-} p^\mu - \frac{2 m_{\Lambda_b}}{s_-} p^{\prime \mu} \right) \bigg]  u_{\Lambda_b}(p,s), \\
 \nonumber \langle \Lambda(p^\prime,s^\prime) | \overline{s} \,i\sigma^{\mu\nu} q_\nu \, b | \Lambda_b(p,s) \rangle &=&
 - \overline{u}_\Lambda(p^\prime,s^\prime) \bigg[  h_+(q^2) \frac{q^2}{s_+} \left( p^\mu + p^{\prime \mu} - (m_{\Lambda_b}^2-m_{\Lambda}^2)\frac{q^\mu}{q^2} \right) \\
 && \phantom{\overline{u}_\Lambda \bigg[} + h_\perp(q^2)\, (m_{\Lambda_b}+m_\Lambda) \left( \gamma^\mu -  \frac{2  m_\Lambda}{s_+} \, p^\mu - \frac{2m_{\Lambda_b}}{s_+} \, p^{\prime \mu}   \right) \bigg] u_{\Lambda_b}(p,s), \\
 \nonumber \langle \Lambda(p^\prime,s^\prime)| \overline{s} \, i\sigma^{\mu\nu}q_\nu \gamma_5  \, b|\Lambda_b(p,s)\rangle &=&
 -\overline{u}_{\Lambda}(p^\prime,s^\prime) \, \gamma_5 \bigg[   \widetilde{h}_+(q^2) \, \frac{q^2}{s_-} \left( p^\mu + p^{\prime \mu} -  (m_{\Lambda_b}^2-m_{\Lambda}^2) \frac{q^\mu}{q^2} \right) \\
 && \phantom{\overline{u}_\Lambda \bigg[}  + \widetilde{h}_\perp(q^2)\,  (m_{\Lambda_b}-m_\Lambda) \left( \gamma^\mu +  \frac{2 m_\Lambda}{s_-} \, p^\mu - \frac{2 m_{\Lambda_b}}{s_-} \, p^{\prime \mu}  \right) \bigg]  u_{\Lambda_b}(p,s),
\end{eqnarray}
with $q=p-p^\prime$, $\sigma^{\mu\nu}=\frac{i}{2}(\gamma^\mu\gamma^\nu-\gamma^\nu\gamma^\mu)$ and $s_\pm =(m_{\Lambda_b} \pm m_\Lambda)^2-q^2$.
The helicity form factors describe the contractions of the matrix elements with virtual polarization vectors that are given explicitly in
Ref.~\cite{Boer:2014kda} (note that Ref.~\cite{Boer:2014kda} uses different labels for the form factors, which are related to the notation of
Ref.~\cite{Feldmann:2011xf} adopted here as follows: $f_t^V=f_0$, $f_0^V=f_+$, $f_\perp^V=f_\perp$, $f_t^A=g_0$, $f_0^A=g_+$, $f_\perp^A=g_\perp$,
$f_0^T=h_+$, $f_\perp^T=h_\perp$, $f_0^{T5}=\widetilde{h}_+$, $f_\perp^{T5}=\widetilde{h}_\perp$). The helicity form factors satisfy the endpoint
relations
\begin{eqnarray}
 f_0(0) &=& f_+(0), \label{eq:FFC1} \\
 g_0(0) &=& g_+(0), \label{eq:FFC2} \\
 g_\perp(q^2_{\rm max}) &=& g_+(q^2_{\rm max}), \label{eq:FFC3} \\
 \widetilde{h}_\perp(q^2_{\rm max}) &=& \widetilde{h}_+(q^2_{\rm max}), \label{eq:FFC4}
\end{eqnarray}
where $q^2_{\rm max}=(m_{\Lambda_b}-m_{\Lambda})^2$. As in Ref.~\cite{Detmold:2015aaa}, in some
parts of our data analysis we simultaneously work with an alternative basis that decomposes the matrix elements into form factors of the first
and second class according to Weinberg's classification \cite{Weinberg:1958ut}, and is given by \cite{Gutsche:2013pp}
\begin{eqnarray}
 \langle \Lambda(p^\prime,s^\prime) | \overline{s} \,\gamma^\mu\, b | \Lambda_b(p,s) \rangle &=& \overline{u}_\Lambda(p^\prime,s^\prime) \left[ f_1^V(q^2)\: \gamma^\mu - \frac{f_2^V(q^2)}{m_{\Lambda_b}} i\sigma^{\mu\nu}q_\nu + \frac{f_3^V(q^2)}{m_{\Lambda_b}} q^\mu \right] u_{\Lambda_b}(p,s),  \label{eq:WeinbergFF1} \\
 \langle \Lambda(p^\prime,s^\prime) | \overline{s} \,\gamma^\mu\gamma_5\, b | \Lambda_b(p,s) \rangle &=& \overline{u}_\Lambda(p^\prime,s^\prime) \left[ f_1^A(q^2)\: \gamma^\mu - \frac{f_2^A(q^2)}{m_{\Lambda_b}} i\sigma^{\mu\nu}q_\nu + \frac{f_3^A(q^2)}{m_{\Lambda_b}} q^\mu \right]\gamma_5\: u_{\Lambda_b}(p,s),  \label{eq:WeinbergFF2} \\
 \langle \Lambda(p^\prime,s^\prime) | \overline{s} \,i\sigma^{\mu\nu}q_\nu\, b | \Lambda_b(p,s) \rangle &=& \overline{u}_\Lambda(p^\prime,s^\prime) \left[  \frac{f_1^{TV}(q^2)}{m_{\Lambda_b}} \left(\gamma^\mu q^2 - q^\mu \slashed{q} \right) - f_2^{TV}(q^2) i\sigma^{\mu\nu}q_\nu  \right] u_{\Lambda_b}(p,s), \label{eq:WeinbergFF3} \\
 \langle \Lambda(p^\prime,s^\prime) | \overline{s} \,i\sigma^{\mu\nu}q_\nu\,\gamma_5\, b | \Lambda_b(p,s) \rangle &=& \overline{u}_\Lambda(p^\prime,s^\prime) \left[  \frac{f_1^{TA}(q^2)}{m_{\Lambda_b}} \left(\gamma^\mu q^2 - q^\mu \slashed{q} \right) - f_2^{TA}(q^2) i\sigma^{\mu\nu}q_\nu  \right]\gamma_5\: u_{\Lambda_b}(p,s). \label{eq:WeinbergFF4}
\end{eqnarray}
These ``Weinberg form factors'' are related to the helicity form factors introduced above as follows:
\begin{eqnarray}
 f_+(q^2)     &=& f_1^V(q^2) + \frac{q^2}{m_{\Lambda_b}(m_{\Lambda_b}+m_\Lambda)} f_2^V(q^2), \label{eq:FFR1} \\
 f_\perp(q^2) &=& f_1^V(q^2) + \frac{m_{\Lambda_b}+m_\Lambda}{m_{\Lambda_b}} f_2^V(q^2),  \\
 f_0(q^2)     &=& f_1^V(q^2) + \frac{q^2}{m_{\Lambda_b}(m_{\Lambda_b}-m_\Lambda)} f_3^V(q^2), \\
 g_+(q^2)     &=& f_1^A(q^2) - \frac{q^2}{m_{\Lambda_b}(m_{\Lambda_b}-m_\Lambda)} f_2^A(q^2), \\
 g_\perp(q^2) &=& f_1^A(q^2) - \frac{m_{\Lambda_b}-m_\Lambda}{m_{\Lambda_b}} f_2^A(q^2), \\
 g_0(q^2)     &=& f_1^A(q^2) - \frac{q^2}{m_{\Lambda_b}(m_{\Lambda_b}+m_\Lambda)} f_3^A(q^2), \\
 h_+(q^2)     &=& -f_2^{TV}(q^2) - \frac{m_{\Lambda_b}+m_\Lambda}{m_{\Lambda_b}} f_1^{TV}(q^2),  \\
 h_\perp(q^2) &=& -f_2^{TV}(q^2) - \frac{q^2}{m_{\Lambda_b}(m_{\Lambda_b}+m_\Lambda)} f_1^{TV}(q^2),  \\
 \widetilde{h}_+(q^2)     &=& -f_2^{TA}(q^2) + \frac{m_{\Lambda_b}-m_\Lambda}{m_{\Lambda_b}} f_1^{TA}(q^2),  \\
 \widetilde{h}_\perp(q^2) &=& -f_2^{TA}(q^2) + \frac{q^2}{m_{\Lambda_b}(m_{\Lambda_b}-m_\Lambda)} f_1^{TA}(q^2). \label{eq:FFR10}
\end{eqnarray}

\FloatBarrier
\section{\label{sec:lattice}Lattice calculation}
\FloatBarrier

The lattice calculation was performed using the same actions, parameters, and analysis methods as in Ref.~\cite{Detmold:2015aaa}, with a few
modifications to accommodate the $\Lambda$ final state and the tensor currents as explained in the following. The strange quark
was implemented using the same domain-wall action as the up and down quarks, with masses given
in Table \ref{tab:params}.
We used the interpolating field
\begin{equation}
\Lambda_{\alpha} = \epsilon^{abc}\:(C\gamma_5)_{\beta\gamma}\:\widetilde{d}^a_\beta\:\widetilde{u}^b_\gamma\: \widetilde{s}^c_\alpha
\end{equation}
for the $\Lambda$ baryon, with smearing parameters $(\sigma, n_S)=(4.35, 30)$ for all three quark fields \cite{Detmold:2015aaa, Detmold:2012ge}.
The renormalized, $\mathcal{O}(a)$-improved $b\to s$ vector and axial vector currents are defined as in Eqs.~(18)-(21) of Ref.~\cite{Detmold:2015aaa},
with matching coefficients equal to those for $b \to u$ (see Table III of Ref.~\cite{Detmold:2015aaa}). For the tensor current, we also
use the mostly nonperturbative renormalization method introduced in Refs.~\cite{Hashimoto:1999yp, ElKhadra:2001rv}, but we set the residual
matching factors and $\mathcal{O}(a)$-improvement coefficients to their mean-field-improved tree-level values (because one-loop results were not
available). That is, we write the tensor current as
\begin{equation}
 T_{\mu\nu}=\sqrt{Z_V^{(ss)} Z_V^{(bb)}} \bigg[ \bar{s} \sigma_{\mu\nu} b + a\, d_1\, \sum_{j=1}^3\bar{s} \sigma_{\mu\nu} \gamma_j \overrightarrow{\nabla}_j  b \bigg],
\end{equation}
with $Z_V^{(bb)}$ and $Z_V^{(ss)}=Z_V^{(uu)}$ as given in Table IV of Ref.~\cite{Detmold:2015aaa}, and with $d_1=0.0740$ for the coarse lattice
spacing and $d_1=0.0718$ for the fine lattice spacing (for $d_1$, we use the averages of the values computed with $u_0$ from either the
Landau-gauge mean link or the plaquette). As discussed in Secs.~\ref{sec:ccextrap} and \ref{sec:observables}, this approximation introduces a systematic uncertainty of approximately
5\% in the tensor form factors, which however has negligible impact on the $\Lambda_b \to \Lambda(\to p\,\pi) \mu^+ \mu^-$ observables.

\begin{table}
\begin{tabular}{cccccccccccccccccccccc}
\hline\hline
Set & \hspace{1ex} & $\beta$ & \hspace{1ex} & $N_s^3\times N_t$ & \hspace{1ex} & $am_{s}^{(\mathrm{sea})}$
& \hspace{1ex} & $am_{u,d}^{(\mathrm{sea})}$   & \hspace{1ex} & $a$ [fm] & \hspace{1ex} & $am_{u,d}^{(\mathrm{val})}$ 
& \hspace{1ex} & $m_\pi^{(\mathrm{val})}$ [MeV] & \hspace{1ex} & $am_{s}^{(\mathrm{val})}$ 
& \hspace{1ex} & $m_{\eta_s}^{(\mathrm{val})}$ [MeV] & \hspace{1ex} & $N_{\rm meas}$ \\
\hline
\texttt{C14} && $2.13$ && $24^3\times64$ && $0.04$ && $0.005$ && $0.1119(17)$ && $0.001$ && 245(4) && $0.04$ && 761(12) && 2672 \\
\texttt{C24} && $2.13$ && $24^3\times64$ && $0.04$ && $0.005$ && $0.1119(17)$ && $0.002$ && 270(4) && $0.04$ && 761(12) && 2676 \\
\texttt{C54} && $2.13$ && $24^3\times64$ && $0.04$ && $0.005$ && $0.1119(17)$ && $0.005$ && 336(5) && $0.04$ && 761(12) && 2782 \\
\texttt{C53} && $2.13$ && $24^3\times64$ && $0.04$ && $0.005$ && $0.1119(17)$ && $0.005$ && 336(5) && $0.03$ && 665(10) && 1205 \\
\texttt{F23} && $2.25$ && $32^3\times64$ && $0.03$ && $0.004$ && $0.0849(12)$ && $0.002$ && 227(3) && $0.03$ && 747(10) && 1907 \\
\texttt{F43} && $2.25$ && $32^3\times64$ && $0.03$ && $0.004$ && $0.0849(12)$ && $0.004$ && 295(4) && $0.03$ && 747(10) && 1917 \\
\texttt{F63} && $2.25$ && $32^3\times64$ && $0.03$ && $0.006$ && $0.0848(17)$ && $0.006$ && 352(7) && $0.03$ && 749(14) && 2782 \\
\hline\hline
\end{tabular}
\caption{\label{tab:params} Parameters of the seven data sets used in this work.
The quark masses $am_{u,d}^{(\mathrm{sea})}$ and $am_{s}^{(\mathrm{sea})}$ were used in the generation of the ensembles \cite{Aoki:2010dy},
while the quark masses $am_{u,d}^{(\mathrm{val})}$ and $am_{s}^{(\mathrm{val})}$ were used in the computation of the propagators.
The resulting pion and $\eta_s$ masses are denoted as $m_\pi^{(\mathrm{val})}$ and $m_{\eta_s}^{(\mathrm{val})}$. The $\eta_s$ is an
artificial pseudoscalar $s\bar{s}$ meson that is obtained by treating the $s$ and $\bar{s}$ as different, but mass-degenerate flavors. We use
this state as an intermediate quantity to tune the strange-quark mass \cite{Davies:2009tsa}; the $\eta_s$ mass at the physical point has
been computed precisely by the HPQCD collaboration and is $m_{\eta_s}^{(\mathrm{phys})}=689.3(1.2)\:\:{\rm MeV}$ \cite{Dowdall:2011wh}. The
values of the lattice spacing, $a$, were taken from Ref.~\cite{Meinel:2010pv}. The parameters of the anisotropic clover
action used for the bottom quark can be found in Ref.~\cite{Aoki:2012xaa}.}
\end{table}

\begin{figure}
 \includegraphics[width=0.95\linewidth]{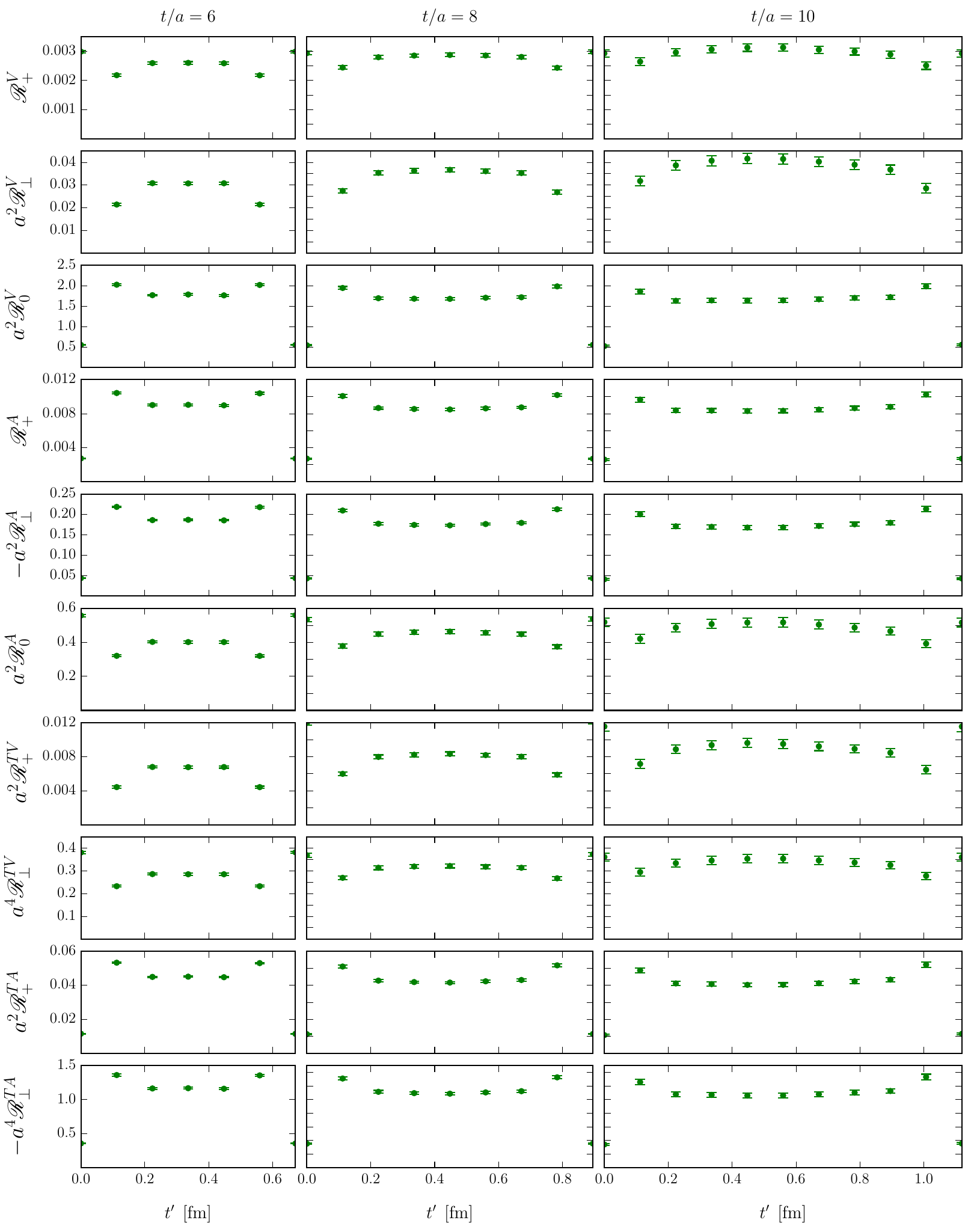}
 \caption{\label{fig:ratios}Examples of numerical results for the ratios $\mathscr{R}_{+,\,\perp,\,0}^{V,\,A}(|\mathbf{p}^\prime|,t,t^\prime)$, $\mathscr{R}_{+,\,\perp}^{TV,\,TA}(|\mathbf{p}^\prime|,t,t^\prime)$
 for three different source-sink separations, plotted as a function of the current insertion time, $t^\prime$. The data shown here are from the \texttt{C24} data set at $|\mathbf{p}^\prime|^2=3(2\pi/L)^2$.}
\end{figure}

\begin{figure}
\begin{center}
 \hfill \includegraphics[width=0.46\linewidth]{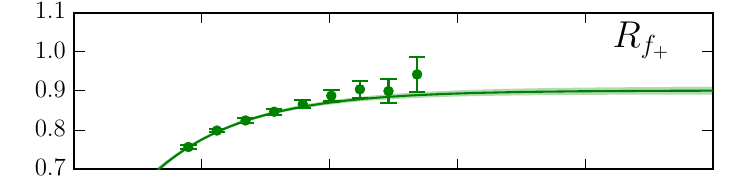} \hfill \includegraphics[width=0.46\linewidth]{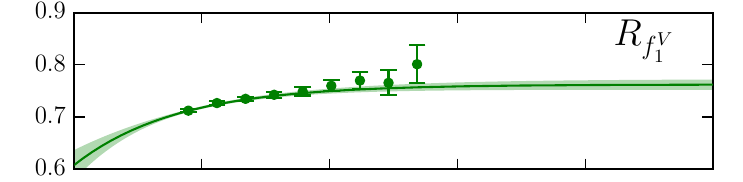} \hfill \null \\
 \hfill \includegraphics[width=0.46\linewidth]{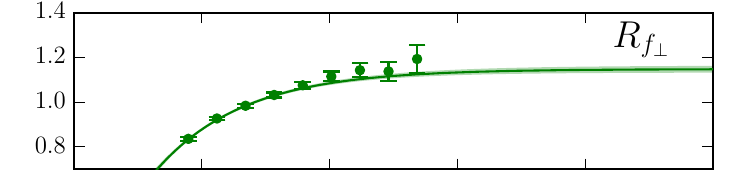} \hfill \includegraphics[width=0.46\linewidth]{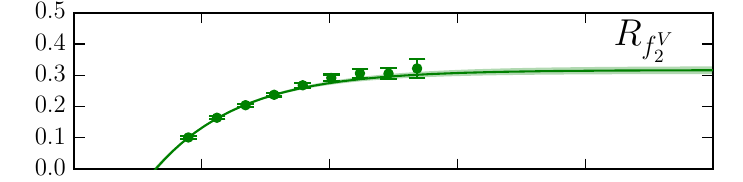} \hfill \null \\
 \hfill \includegraphics[width=0.46\linewidth]{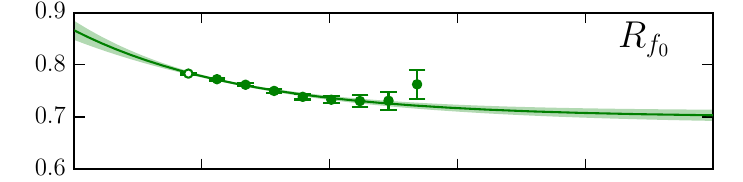}    \hfill \includegraphics[width=0.46\linewidth]{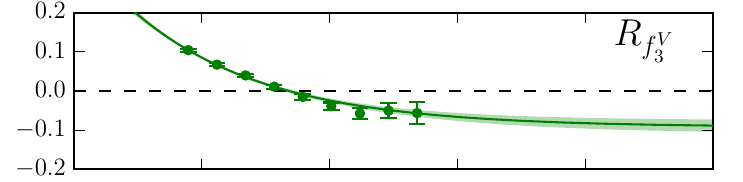} \hfill \null \\
 \hfill \includegraphics[width=0.46\linewidth]{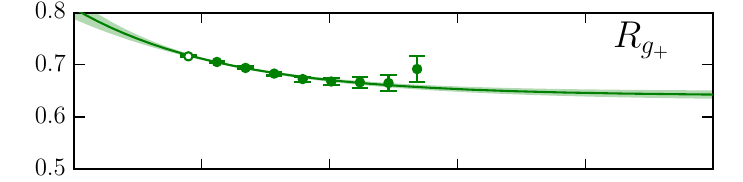} \hfill \includegraphics[width=0.46\linewidth]{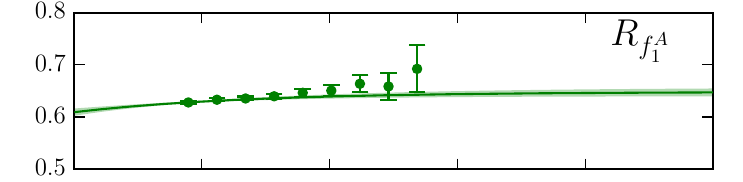} \hfill \null \\
 \hfill \includegraphics[width=0.46\linewidth]{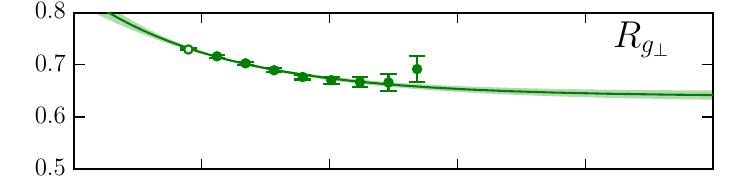} \hfill \includegraphics[width=0.46\linewidth]{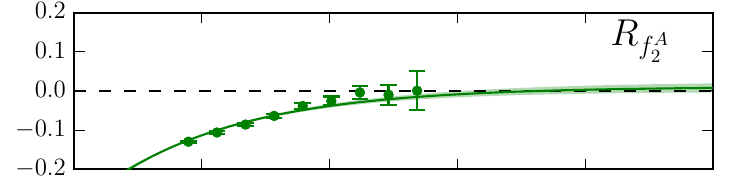} \hfill \null \\
 \hfill \includegraphics[width=0.46\linewidth]{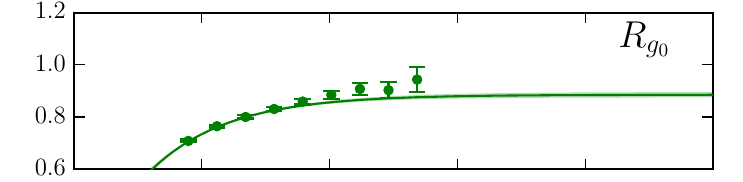}    \hfill \includegraphics[width=0.46\linewidth]{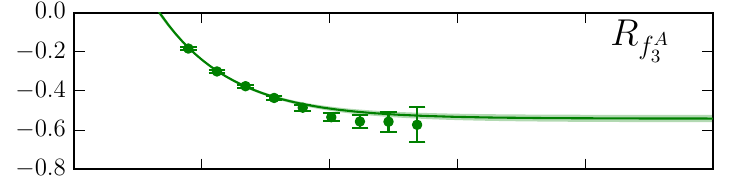} \hfill \null \\
 \hfill \includegraphics[width=0.46\linewidth]{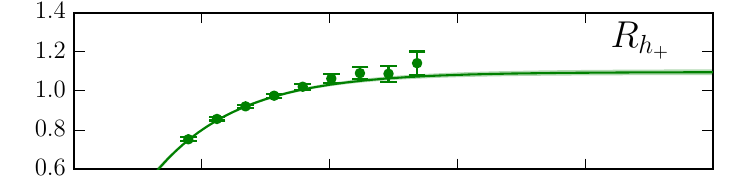} \hfill \includegraphics[width=0.46\linewidth]{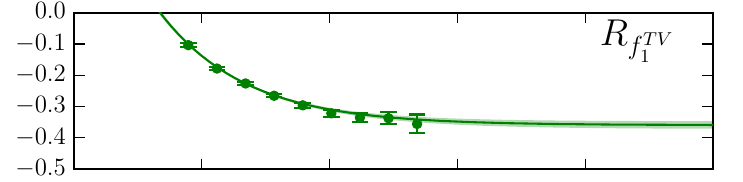} \hfill \null \\
 \hfill \includegraphics[width=0.46\linewidth]{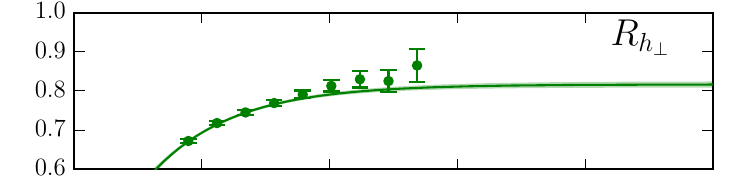} \hfill \includegraphics[width=0.46\linewidth]{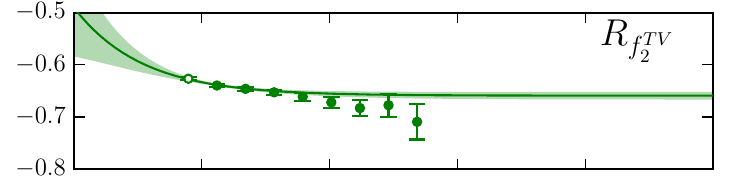} \hfill \null \\
 \hfill \includegraphics[width=0.46\linewidth]{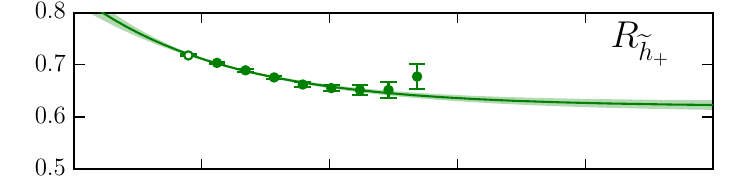}    \hfill \includegraphics[width=0.46\linewidth]{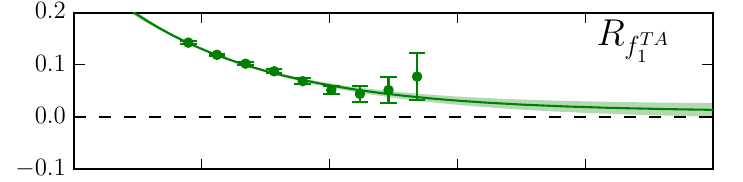} \hfill \null \\
 \hfill \includegraphics[width=0.46\linewidth]{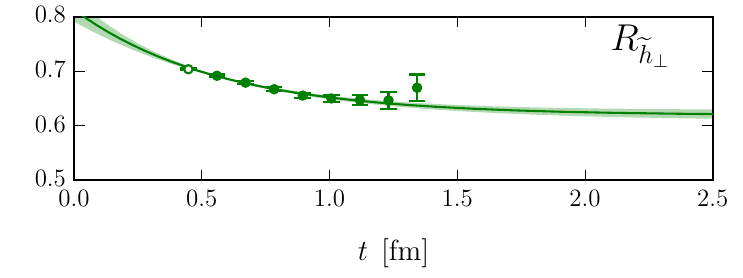}    \hfill \includegraphics[width=0.46\linewidth]{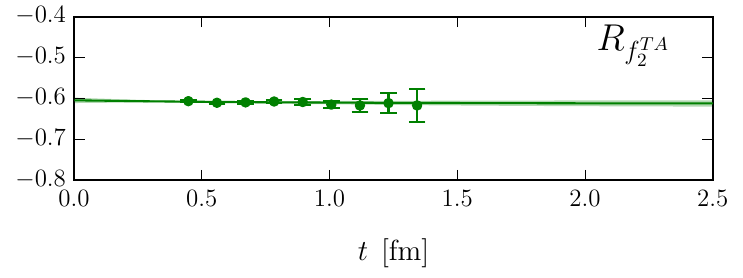} \hfill \null \\
 \caption{\label{fig:tsepextrap} Examples of fits of the quantities $R_f(|\mathbf{p}^\prime|,t)$ as a function of the source-sink separation, $t$.
 These quantities approach the ground-state form factors, $f$, for large $t$.  The data shown here are from the \texttt{C24} data set at $|\mathbf{p}^\prime|^2=3(2\pi/L)^2$; points plotted with open symbols are excluded from the fit.}
\end{center}
\end{figure}

The extraction of the form factors from ratios of three-point and two-point functions is performed as in Ref.~\cite{Detmold:2015aaa}.
In addition to the ratios $\mathscr{R}_{+,\,\perp,\,0}^{V,A}(\mathbf{p}^\prime,t,t^\prime)$ for the vector and axial vector currents, given in Eqs.~(46)-(48) of Ref.~\cite{Detmold:2015aaa}, we now
define
\begin{eqnarray}
\mathscr{R}_{+}^{TV}(\mathbf{p}^\prime,t,t^\prime) &=& \frac{ r_\mu[(1,\mathbf{0})] \: r_\nu[(1,\mathbf{0})] \:
\mathrm{Tr}\Big[   C^{(3,{\rm fw})}(\mathbf{p'},\:i\sigma^{\mu\rho}q_\rho, t, t') \:    C^{(3,{\rm bw})}(\mathbf{p'},\:i\sigma^{\nu\lambda}q_\lambda, t, t-t')  \Big] }
{\mathrm{Tr}\Big[C^{(2,\Lambda,{\rm av})}(\mathbf{p'}, t)\Big] \mathrm{Tr}\Big[C^{(2,\Lambda_b,{\rm av})}(t)\Big] }, \\
\mathscr{R}_{\perp}^{TV}(\mathbf{p}^\prime,t,t^\prime) &=& \frac{ r_\mu[(0,\mathbf{e}_j\times \mathbf{p}')] \:   r_\nu[(0,\mathbf{e}_k\times \mathbf{p}')] \:
\nonumber \mathrm{Tr}\Big[  C^{(3,{\rm fw})}(\mathbf{p'},\:i\sigma^{\mu\rho}q_\rho, t, t') \gamma_5 \gamma^j \:    C^{(3,{\rm bw})}(\mathbf{p'},\:i\sigma^{\nu\lambda}q_\lambda, t, t-t') \gamma_5 \gamma^k  \Big] }
{\mathrm{Tr}\Big[C^{(2,\Lambda,{\rm av})}(\mathbf{p'}, t)\Big] \mathrm{Tr}\Big[C^{(2,\Lambda_b,{\rm av})}(t)\Big] }, \\
\end{eqnarray}
where the current $i T^{\mu\nu} q_\nu$ is used in the three-point functions, as well as the ratios $\mathscr{R}_{+,\,\perp}^{TA}(\mathbf{p}^\prime,t,t^\prime)$ with the replacement
$\sigma^{\mu\nu} \mapsto \sigma^{\mu\nu}\gamma_5$ in the current. Here, $\mathbf{p}^\prime$ is the spatial momentum of the $\Lambda$ baryon, $t$ is the source-sink separation,
and $t^\prime$ is the time at which the current is inserted in the three-point function \cite{Detmold:2015aaa}. We average the data at fixed $|\mathbf{p}^\prime|$ over the directions of
$\mathbf{p}^\prime$, and denote the direction-averaged ratios by $\mathscr{R}_{+,\,\perp,\,0}^{V,\,A}(|\mathbf{p}^\prime|,t,t^\prime)$, $\mathscr{R}_{+,\,\perp}^{TV,\,TA}(|\mathbf{p}^\prime|,t,t^\prime)$.
As in Ref.~\cite{Detmold:2015aaa}, we generated data for all source-sink separations in the range $t/a=4...15$ (\texttt{C14}, \texttt{C24}, \texttt{C54}, \texttt{C53} data sets),
$t/a=5...15$ (\texttt{F23}, \texttt{F43} data sets), and $t/a=5...17$ (\texttt{F63} data set). Examples of numerical results for $\mathscr{R}_{+,\,\perp,\,0}^{V,\,A}(|\mathbf{p}^\prime|,t,t^\prime)$, $\mathscr{R}_{+,\,\perp}^{TV,\,TA}(|\mathbf{p}^\prime|,t,t^\prime)$
from the \texttt{C24} data set are shown in Fig.~\ref{fig:ratios}. We then evaluate these ratios at the midpoint $t^\prime=t/2$ (or, in the case of odd $t/a$, average over $t^\prime=(t-a)/2$ and $t^\prime=(t+a)/2$),
and compute the quantities $R_{f_+}(|\mathbf{p}^\prime|, t)$, $R_{f_\perp}(|\mathbf{p}^\prime|, t)$, $R_{f_0}(|\mathbf{p}^\prime|, t)$, $R_{g_+}(|\mathbf{p}^\prime|, t)$, $R_{g_\perp}(|\mathbf{p}^\prime|, t)$, $R_{g_0}(|\mathbf{p}^\prime|, t)$,
which are defined as in Eqs.~(52), (53), (54), (58), (59), (60) of Ref.~\cite{Detmold:2015aaa}, and
\begin{eqnarray}
R_{h_+}(|\mathbf{p}^\prime|, t)     &=& \frac{2}{E_\Lambda-m_\Lambda} \sqrt{ \frac{ E_\Lambda}{ (E_\Lambda+m_\Lambda)} \mathscr{R}_{+}^{TV}(|\mathbf{p}^\prime|, t, t/2)} , \\
R_{h_\perp}(|\mathbf{p}^\prime|, t) &=&  \frac{1}{(E_\Lambda-m_\Lambda) (m_{\Lambda_b}+m_\Lambda)} \sqrt{\frac{ E_\Lambda}{E_\Lambda+m_\Lambda} \mathscr{R}_{\perp}^{TV}(|\mathbf{p}^\prime|, t, t/2)}, \\
R_{\widetilde{h}_+}(|\mathbf{p}^\prime|, t)    &=& \frac{2}{E_\Lambda+m_\Lambda} \sqrt{\frac{E_\Lambda}{E_\Lambda-m_\Lambda} \mathscr{R}_{+}^{TA}(|\mathbf{p}^\prime|, t, t/2)} , \\
R_{\widetilde{h}_\perp}(|\mathbf{p}^\prime|, t) &=& \frac{1}{(E_\Lambda+m_\Lambda) (m_{\Lambda_b}-m_\Lambda)}\sqrt{-\frac{E_\Lambda}{E_\Lambda-m_\Lambda} \mathscr{R}_{\perp}^{TA}(|\mathbf{p}^\prime|, t, t/2)} .
\end{eqnarray}
These quantities are equal to the desired helicity form factors at the given momentum and lattice parameters, up to excited-state contamination that decays exponentially with the source-sink separation, $t$.
Here, we use bootstrap samples for the lattice baryon masses, $a m_{\Lambda_b}$ and $a m_\Lambda$, from fits to the two-point functions of the individual data sets
(see Table \ref{tab:hadronmasses}), and compute the energies $a E_\Lambda$ at nonzero momentum using the relativistic continuum dispersion relation.
Following Ref.~\cite{Detmold:2015aaa}, we also constructed the linear combinations of the above quantities that yield the Weinberg form factors by
inverting Eqs.~(\ref{eq:FFR1})-(\ref{eq:FFR10}), for example
\begin{equation}
 R_{f_2^{TV}}(|\mathbf{p}^\prime|, t)  = \frac{q^2 R_{h_+}(|\mathbf{p}^\prime|, t) -(m_{\Lambda_b}+m_\Lambda)^2 R_{h_\perp}(|\mathbf{p}^\prime|, t) }{s_+}.
\end{equation}
Denoting the data by $R_{f,i,n}(t)$, where $f$ labels the helicity and Weinberg form factors, $i=\texttt{C14},\:\texttt{C24},\, ...$ labels the data set, and $n$ labels the $\Lambda$-momentum via $|\mathbf{p^\prime}|^2=n\, (2\pi)^2/L^2$, we then performed fits using the functions
\begin{equation}
 R_{f,i,n}(t) = f_{i,n} + A_{f,i,n} \: e^{-\delta_{f,i,n}\:t},\hspace{2ex}\delta_{f,i,n}=\delta_{\rm min} + e^{\,l_{f,i,n}}\:\:{\rm GeV},
\end{equation}
with parameters $f_{i,n}$, $A_{f,i,n}$, and $l_{f,i,n}$. Here $f_{i,n}$ are the ground-state form factors we aim to extract, and the term with the
exponential $t$-dependence describes the leading excited-state contamination. Writing the energy gaps $\delta_{f,i,n}$ in the above form
imposes the constraint $\delta_{f,i,n}>\delta_{\rm min}$, where we set $\delta_{\rm min}=170\:\:{\rm MeV}$ \cite{Detmold:2015aaa}.
This constraint has negligible effect in most cases, but prevents numerical instabilities for some form factors at certain momenta where the data
show no discernible $t$-dependence. At each momentum $n$, we performed one coupled fit to the data for
all the vector form factors ($f_{+,\,\perp,\,0}$, $f_{1,\,2,\,3}^V$), one coupled fit to the data for all the axial vector form factors
($g_{+,\,\perp,\,0}$, $f_{1,\,2,\,3}^A$), one coupled fit to the data for all the ``tensor-vector'' form factors ($h_{+,\,\perp}$, $f_{1,\,2}^{TV}$), and
one coupled fit for all the ``tensor-axial-vector'' form factors ($\widetilde{h}_{+,\,\perp}$, $f_{1,\,2}^{TA}$). As discussed in detail in Ref.~\cite{Detmold:2015aaa},
in these coupled fits we impose the constraint that the form factor parameters $f_{i,n}$ satisfy the relations (\ref{eq:FFR1})-(\ref{eq:FFR10}), and we include
Gaussian priors that limit the variation of the energy gap parameters between the different data sets to reasonable ranges (we generalize Eq.~(71) of
Ref.~\cite{Detmold:2015aaa} to also include strange-quark mass dependence by writing $[\sigma_m^{i,j}]^2 = w_m^2 [ (m_\pi^i)^2-(m_\pi^j)^2 ]^2 + w_m^2 [ (m_{\eta_s}^i)^2-(m_{\eta_s}^j)^2 ]^2$,
with $w_m = 4\:{\rm GeV}^{-2}$ as before). Examples of these fits are shown in Fig.~\ref{fig:tsepextrap}. Following Ref.~\cite{Detmold:2015aaa}, we estimated the systematic uncertainties resulting from
neglected higher excited states by computing the shifts in $f_{i,n}$ when removing the points with the smallest values of $t$ from the fits, and
added these uncertainties in quadrature to the statistical uncertainties. Tables of the extracted lattice form factors are given in Appendix \ref{sec:tables}.

\begin{table}
\begin{tabular}{ccccccccccccc}
\hline\hline
Set  & \hspace{1ex} & $a m_{\Lambda_b}$   & \hspace{1ex} & $a m_{\Lambda}$  & \hspace{1ex} & $am_{B_s}$    & \hspace{1ex} & $am_{B}$ \\
\hline
\texttt{C14}        && $3.305(11)\nb$   && $0.7064(38)$ && $3.1135(11)$ && $3.0649(27)$   \\
\texttt{C24}        && $3.299(10)\nb$   && $0.7159(37)$ && $3.1117(11)$ && $3.0628(29)$   \\
\texttt{C54}        && $3.3161(71)$     && $0.7348(30)$ && $3.1115(11)$ && $3.0638(33)$   \\ 
\texttt{C53}        && $3.3257(92)$     && $0.7096(47)$ && $3.0994(14)$ && $3.0682(43)$   \\ 
\texttt{F23}        && $2.469(16)\nb$   && $0.5190(42)$ && $2.3546(16)$ && $2.3198(32)$   \\ 
\texttt{F43}        && $2.492(11)\nb$   && $0.5354(29)$ && $2.3542(16)$ && $2.3230(26)$   \\ 
\texttt{F63}        && $2.5089(70)$     && $0.5514(23)$ && $2.3554(11)$ && $2.3221(22)$   \\ 
\hline\hline
\end{tabular}
\caption{\label{tab:hadronmasses}Hadron masses in lattice units.}
\end{table}

\FloatBarrier
\section{\label{sec:ccextrap}Chiral/continuum/kinematic extrapolation of the form factors}
\FloatBarrier

To obtain parametrizations of the $\Lambda_b \to \Lambda$ helicity form factors in the physical limit $a=0$, $m_\pi=m_{\pi,{\rm phys}}$, $m_{\eta_s}=m_{\eta_s,{\rm phys}}$ (see the caption of Table \ref{tab:params}),
we performed global fits of the lattice data using the simplified $z$ expansion \cite{Bourrely:2008za}, augmented with
additional terms that allow for quark-mass and lattice-spacing dependence. In this approach, the $q^2$-dependence
is described by a Taylor expansion in the variable
\begin{equation}
z(q^2) = \frac{\sqrt{t_+-q^2}-\sqrt{t_+-t_0}}{\sqrt{t_+-q^2}+\sqrt{t_+-t_0}}.
\end{equation}
Here, $t_0$ determines the value of $q^2$ that is mapped to $z=0$; we set
\begin{equation}
t_0 = q^2_{\rm max} = (m_{\Lambda_b} - m_{\Lambda})^2.
\end{equation}
Values of $q^2$ greater than $t_+$ will be mapped onto the unit circle in the complex $z$ plane. We set $t_+$ equal
to the onset location of the branch cut associated with the $B\, K$ threshold,
\begin{equation}
t_+ = (m_B + m_K)^2. \label{eq:tplus}
\end{equation}
Note that $B_s\, \pi$ intermediate states are forbidden by isospin symmetry, which is exact in our calculation. The three-particle $B_s\, \pi\, \pi$ threshold
also lies slightly below the $B\, K$ threshold and is not forbidden by isospin symmetry, but its contributions to dispersive bounds on
the $z$-expansion coefficients are expected to be highly suppressed \cite{Boyd:1995tg}.
Before expanding the form factors in a series in $z$, we factor out the poles associated with the lowest relevant $B_s$ states;
their masses are given in Table \ref{tab:polemasses}. We find that the lattice data are well described by a fit to first order in $z$ using the functions
\begin{eqnarray}
\nonumber f(q^2) &=& \frac{1}{1-q^2/(m_{\rm pole}^f)^2} \bigg[ a_0^f\bigg(1+c_0^f \frac{m_\pi^2-m_{\pi,{\rm phys}}^2}{\Lambda_\chi^2}+c_{s,0}^f \frac{m_{\eta_s}^2-m_{\eta_s,{\rm phys}}^2}{\Lambda_\chi^2}\bigg) + a_1^f\:z(q^2)  \bigg] \\
 & & \times \bigg[1  + b^f\, a^2 |\mathbf{p^\prime}|^2 + d^f\, a^2 \Lambda_{\rm QCD}^2 \bigg],  \label{eq:FFccfit}
\end{eqnarray}
with parameters $a_0^f$, $a_1^f$, $c_0^f$, $c_s^f$, $b^f$, and $d^f$. Terms $\sim\mathcal{O}(a)$ are absent because of the form of the actions and currents that are used.
Above, we introduced the scales $\Lambda_\chi = 4\pi f_\pi$  with $f_\pi = 132\:\,{\rm MeV}$
and $\Lambda_{\rm QCD}=300\:\,{\rm MeV}$ to make all parameters dimensionless. We evaluate $a^2 q^2$ and $z$ using the lattice QCD results for
the hadron masses from each individual data set, taking into account their uncertainties and correlations. We evaluate the pole factor in Eq.~(\ref{eq:FFccfit}) as
\begin{equation}
 \frac{1}{1-(a^2 q^2)/(a m_{B_s} + a \Delta^f)^2}, \label{eq:polefactorlattice}
\end{equation}
where $a m_{B_s}$ are the individual lattice QCD results for the pseudoscalar $B_s$ mass, and $\Delta^f=m_{\rm pole}^f-m_{B_s,{\rm phys}}$ with
$m_{\rm pole}^f$ given in Table \ref{tab:polemasses} and $m_{B_s,{\rm phys}}=5.367\:{\rm GeV}$. In this way, the explicit values of the lattice spacing are needed
only for the small term $a \Delta^f$, minimizing the resulting uncertainty. We implement the constraints (\ref{eq:FFC3}) and (\ref{eq:FFC4}) at $q^2=q^2_{\rm max}$ (corresponding
to $z=0$) by using shared parameters $a_0^{g_\perp,g_+}$, $c_0^{g_\perp,g_+}$, $c_{0,s}^{g_\perp,g_+}$ and
$a_0^{\widetilde{h}_\perp,\widetilde{h}_+}$, $c_0^{\widetilde{h}_\perp,\widetilde{h}_+}$, $c_{0,s}^{\widetilde{h}_\perp,\widetilde{h}_+}$
for the form factors $g_\perp$, $g_+$ and $\widetilde{h}_\perp$, $\widetilde{h}_+$, respectively. The constraints (\ref{eq:FFC1}) and (\ref{eq:FFC2})
at $q^2=0$ are included using Gaussian priors with widths equal to $z(0)^2$ to allow for the missing higher-order terms in the $z$ expansion.

\begin{table}
\begin{tabular}{ccccc}
\hline\hline
 $f$     & \hspace{1ex} & $J^P$ & \hspace{1ex}  & $m_{\rm pole}^f$ [GeV]  \\
\hline
$f_+$, $f_\perp$, $h_+$, $h_\perp$                         && $1^-$   && $5.416$  \\
$f_0$                                                      && $0^+$   && $5.711$  \\
$g_+$, $g_\perp$, $\widetilde{h}_+$, $\widetilde{h}_\perp$ && $1^+$   && $5.750$  \\
$g_0$                                                      && $0^-$   && $5.367$  \\
\hline\hline
\end{tabular}
\caption{\label{tab:polemasses} Values of the $B_s$ meson pole masses, $m_{\rm pole}^f$. The $0^-$ and $1^-$ masses are from the Particle Data Group \protect\cite{Agashe:2014kda},
while the $0^+$ and $1^+$ masses were taken from the lattice QCD calculation of Ref.~\protect\cite{Lang:2015hza}. To evaluate $t_+$ [defined in Eq.~(\protect\ref{eq:tplus})], the values $m_B=5.279\:{\rm GeV}$ and $m_K=494\:{\rm MeV}$ should be used.}
\end{table}

\begin{figure}
 \includegraphics[width=\linewidth]{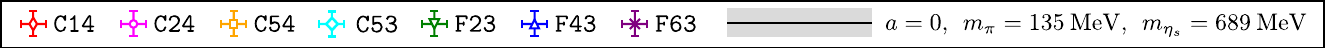}
 
 \vspace{2ex}

 \includegraphics[width=0.95\linewidth]{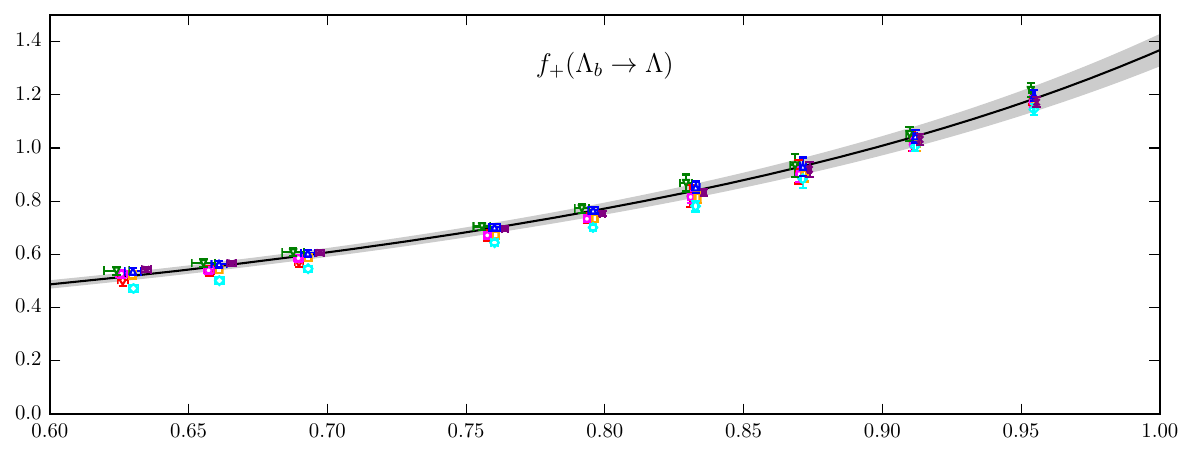}
 
 \includegraphics[width=0.95\linewidth]{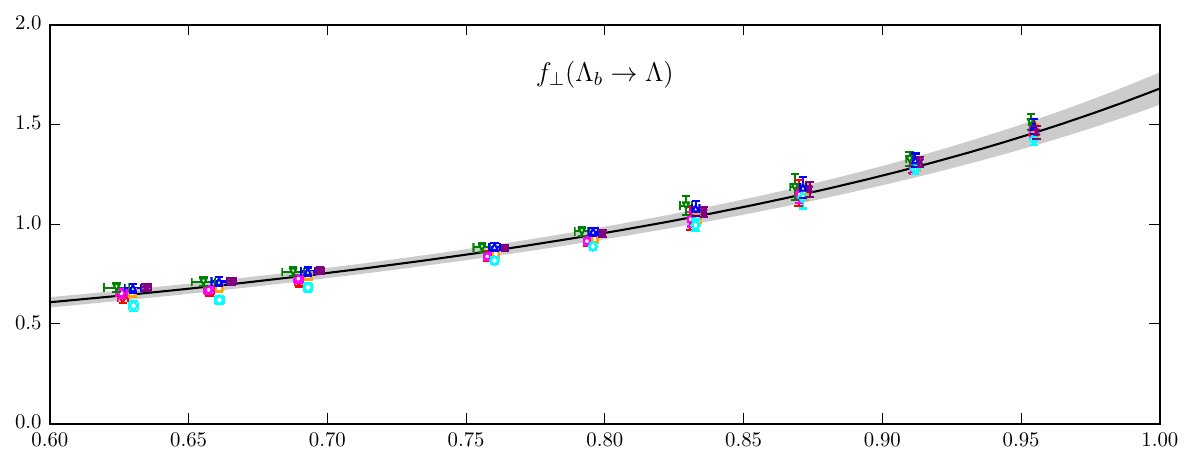}

 \includegraphics[width=0.95\linewidth]{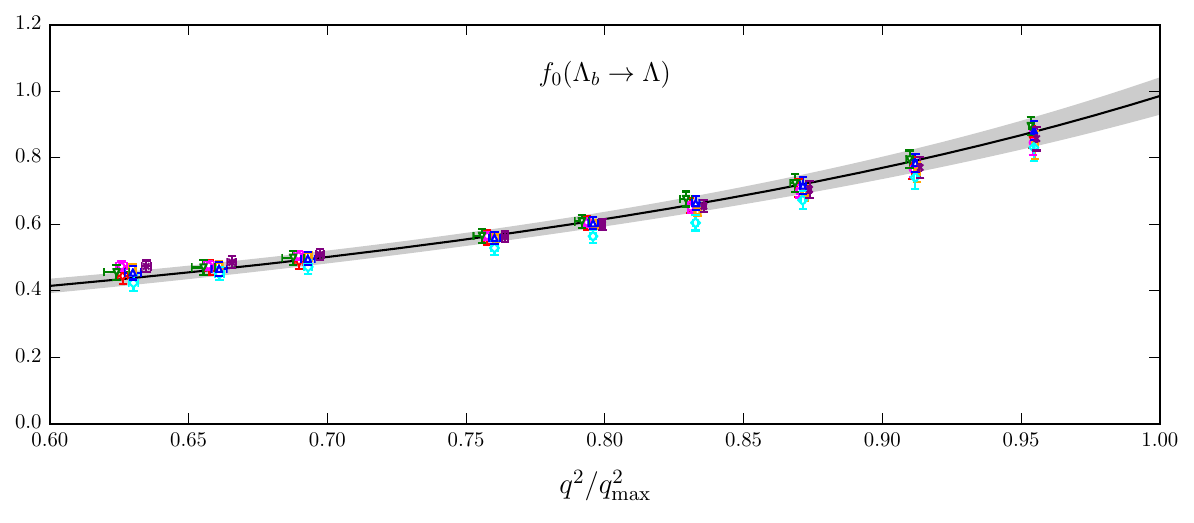}
 
 \caption{\label{fig:vectorFFnominal}Vector form factors in the high-$q^2$ region: lattice results and nominal fit function in the physical limit.
 The bands indicate the statistical uncertainty.}
\end{figure}

\begin{figure}
 \includegraphics[width=\linewidth]{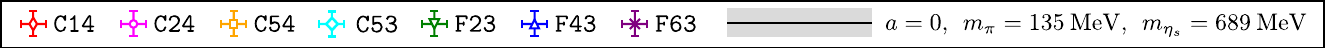}
 
 \vspace{2ex}

 \includegraphics[width=0.95\linewidth]{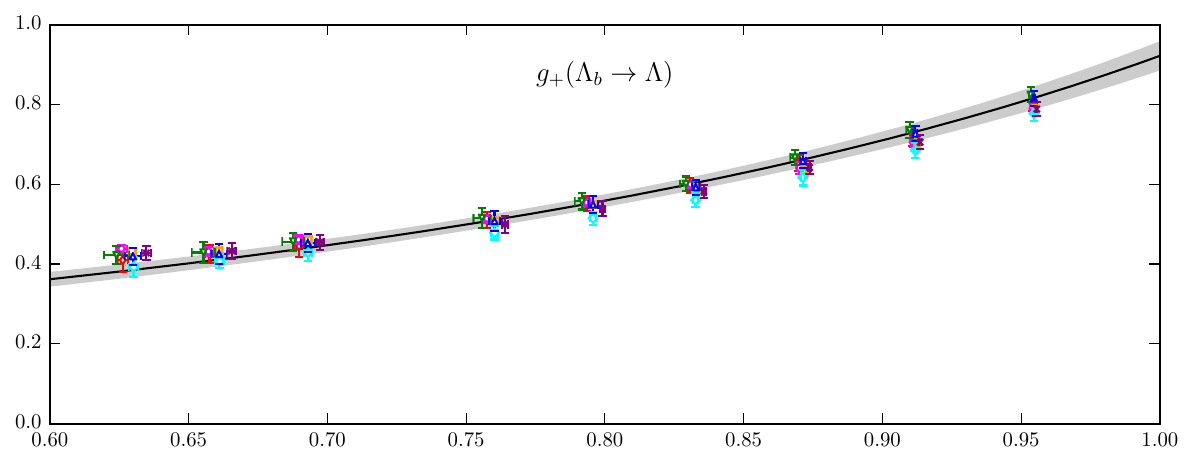}
 
 \includegraphics[width=0.95\linewidth]{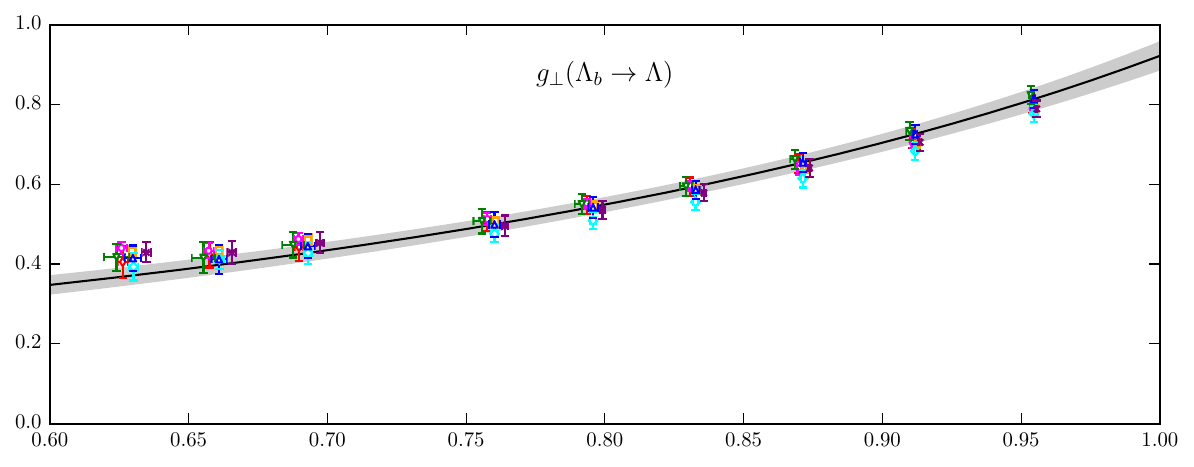}

 \includegraphics[width=0.95\linewidth]{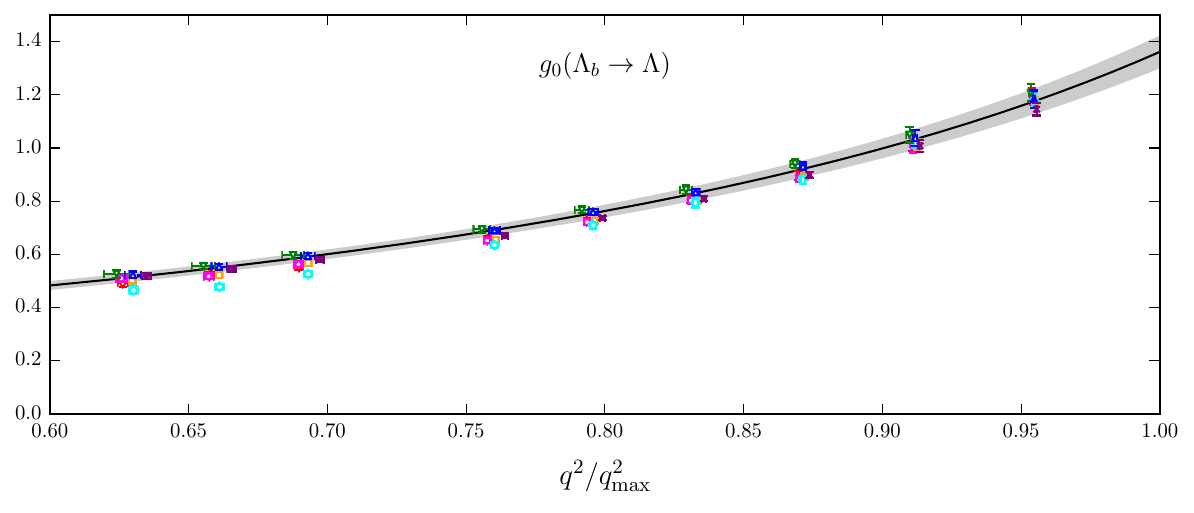}

 \caption{\label{fig:axialvectorFFnominal}Axial vector form factors in the high-$q^2$ region: lattice results and nominal fit function in the physical limit.
 The bands indicate the statistical uncertainty.}
\end{figure}

\begin{figure}
 \includegraphics[width=\linewidth]{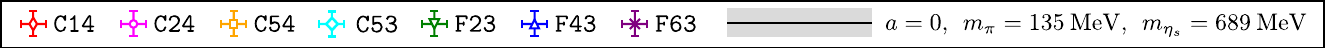}
 
 \vspace{2ex}

 \includegraphics[width=0.95\linewidth]{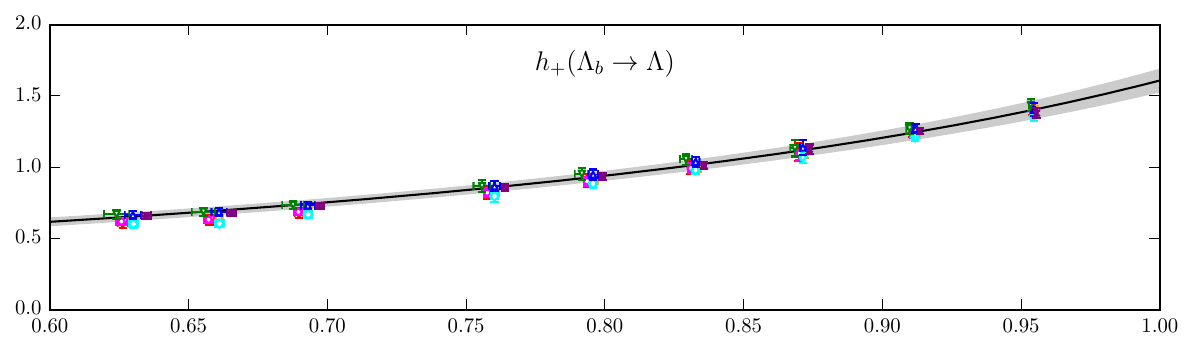}
 
 \includegraphics[width=0.95\linewidth]{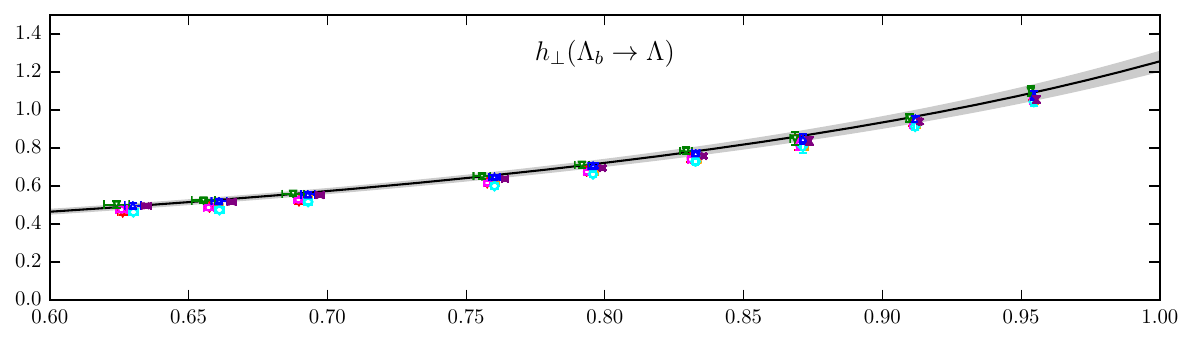}

 \includegraphics[width=0.95\linewidth]{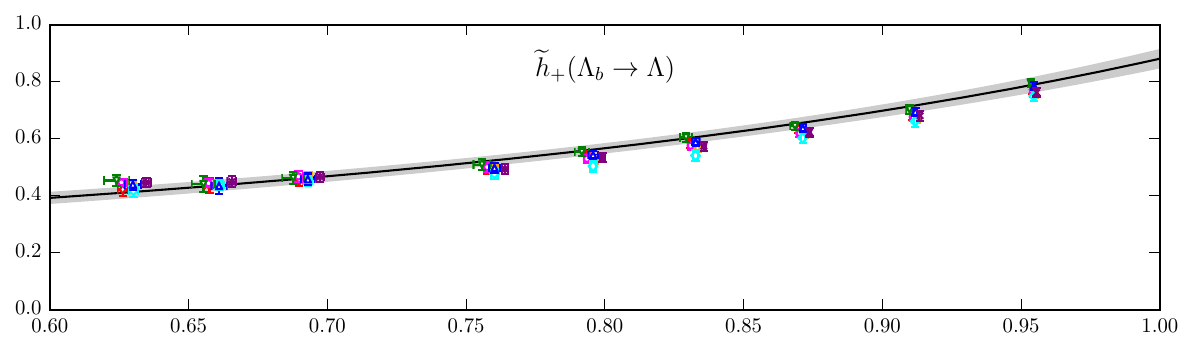}
 
 \includegraphics[width=0.95\linewidth]{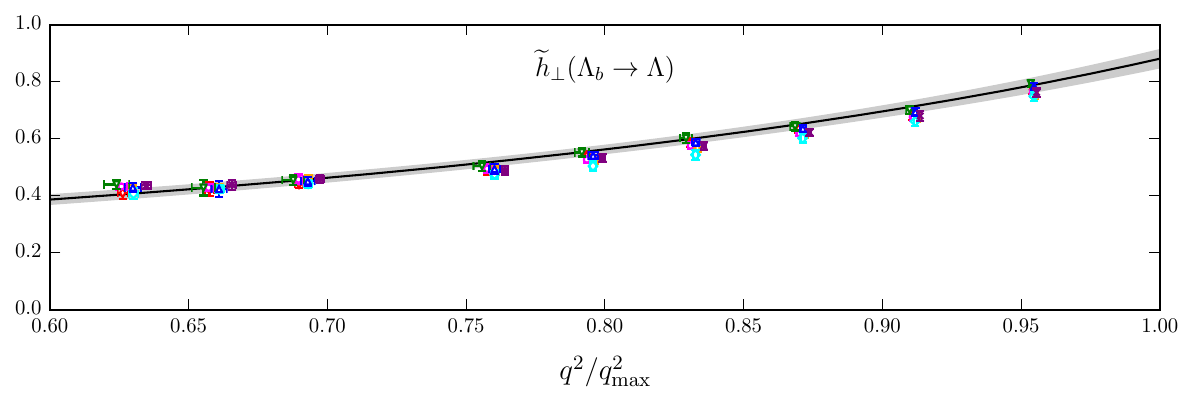}
 
 \caption{\label{fig:tensorFFnominal}Tensor form factors in the high-$q^2$ region: lattice results and nominal fit function in the physical limit.
 The bands indicate the statistical uncertainty.}
 
\end{figure}

We refer to the fit using Eq.~(\ref{eq:FFccfit}) as the ``nominal'' fit.
In the physical limit $a=0$, $m_\pi=m_{\pi,{\rm phys}}$, $m_{\eta_s}=m_{\eta_s,{\rm phys}}$, these functions reduce to the simple form
\begin{equation}
 f(q^2) = \frac{1}{1-q^2/(m_{\rm pole}^f)^2} \big[ a_0^f + a_1^f\:z(q^2) \big]. \label{eq:nominalfitphys}
\end{equation}
The values and uncertainties of the parameters $a_0^f$ and $a_1^f$ from the nominal fit are given in Table \ref{tab:nominal}; their correlation
matrix is given in Tables \ref{tab:nominalcorr1} and \ref{tab:nominalcorr2}. Plots of the lattice data along with the nominal fit functions evaluated in the physical limit
are shown in Figs.~\ref{fig:vectorFFnominal}, \ref{fig:axialvectorFFnominal}, and \ref{fig:tensorFFnominal}.

As in Ref.~\cite{Detmold:2015aaa}, we estimate systematic uncertainties in the extrapolated form factors from the changes in the values and increases in the uncertainties
when redoing the fit with added higher-order terms. Here we use the following functions for the higher-order fit:
\begin{eqnarray}
\nonumber f_{\rm HO}(q^2) &=& \frac{1}{1-q^2/(m_{\rm pole}^f)^2}
\bigg[ a_0^f\bigg(1+c_0^f \frac{m_\pi^2-m_{\pi,{\rm phys}}^2}{\Lambda_\chi^2}+\widetilde{c}_0^f \frac{m_\pi^3-m_{\pi,{\rm phys}}^3}{\Lambda_\chi^3}+c_{s,0}^f \frac{m_{\eta_s}^2-m_{{\eta_s},{\rm phys}}^2}{\Lambda_\chi^2}+\widetilde{c}_{s,0}^f \frac{m_{\eta_s}^3-m_{{\eta_s},{\rm phys}}^3}{\Lambda_\chi^3}\bigg) \\
\nonumber && \hspace{15ex} +\: a_1^f\bigg(1+c_1^f\frac{m_\pi^2-m_{\pi,{\rm phys}}^2}{\Lambda_\chi^2}+c_{s,1}^f\frac{m_{\eta_s}^2-m_{{\eta_s},{\rm phys}}^2}{\Lambda_\chi^2}\bigg)\:z(q^2)  + a_2^f\:[z(q^2)]^2 \bigg] \\
&& \times\: \bigg[1  + b^f\, a^2|\mathbf{p^\prime}|^2 + d^f\, a^2\Lambda_{\rm QCD}^2
                     + \widetilde{b}^f\, a^4 |\mathbf{p^\prime}|^4
                     + \widehat{d}^f\, a^3 \Lambda_{\rm QCD}^3
                     + \widetilde{d}^f a^4 \Lambda_{\rm QCD}^4
                     + j^f   a^4 |\mathbf{p^\prime}|^2\Lambda_{\rm QCD}^2 \bigg]. \hspace{5ex}  \label{eq:FFccfitHO}
\end{eqnarray}
Unlike in Ref.~\cite{Detmold:2015aaa}, here we do not include
terms corresponding to discretization errors proportional to odd powers of $\mathbf{p^\prime}$, as such terms cannot contribute to the ratios used
to extract the form factors because of $O_h$ symmetry (we thank Urs Heller for pointing this out). Because the data themselves do not determine the more complex form (\ref{eq:FFccfitHO}) sufficiently
well, we constrain the higher-order coefficients to be natural-sized using Gaussian priors with the following central values and widths:
\begin{eqnarray}
 a_2^f               &=& 0 \pm 2\, a_1^f\Big|_{\rm nominal}, \label{eq:a2prior} \\
 \widetilde{c}_0^f       &=& 0 \pm 10, \\
 \widetilde{c}_{s,0}^f   &=& 0 \pm 10, \\
 c_1^f               &=& 0 \pm 10, \\
 c_{s,1}^f           &=& 0 \pm 10, \\
 \widetilde{b}^f         &=& 0 \pm \frac{10}{3^4}, \\
 \widehat{d}^f         &=& 0 \pm 10, \\
 \widetilde{d}^f &=& 0 \pm 10, \\
 j^f                 &=& 0 \pm \frac{10}{3^2}.
\end{eqnarray}
Here, Eq.~(\ref{eq:a2prior}) means that we set the widths of $a_2^f$ equal to two times the fit results for $a_1^f$ from the nominal fit.
In the higher-order fit, we impose the constraints (\ref{eq:FFC1}) and (\ref{eq:FFC2}) at $q^2=0$
with widths equal to $|z(0)|^3$. The factors of $1/3^n$ for the prior widths of the coefficients of discretization-error terms proportional to $|\mathbf{p^\prime}|^n$ are motivated by the physical picture that the quarks/gluons in the $\Lambda$ baryon
on average carry only some fraction of the momentum $\mathbf{p^\prime}$, estimated to be of order $1/3$.

\begin{figure}
 \includegraphics[width=0.49\linewidth]{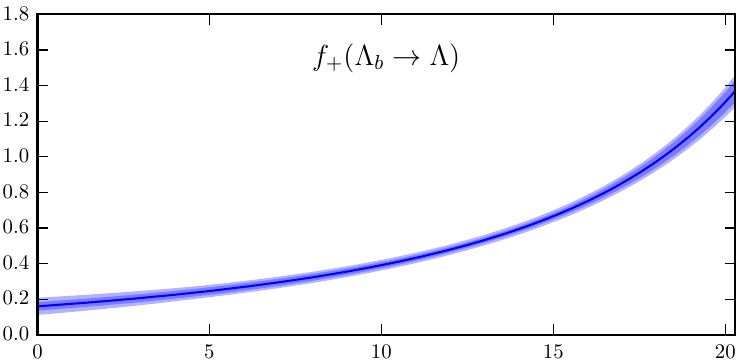} \hfill
 \includegraphics[width=0.49\linewidth]{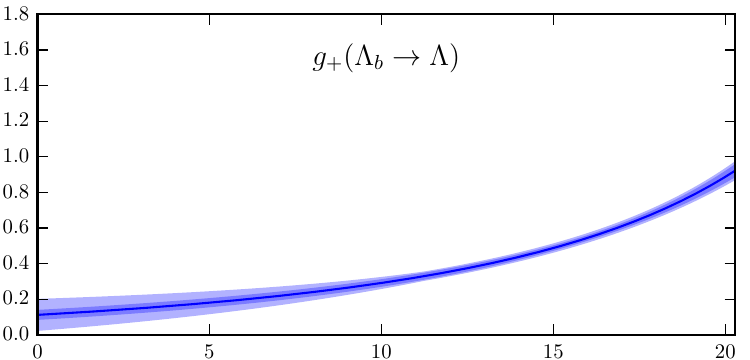} \\
 \includegraphics[width=0.49\linewidth]{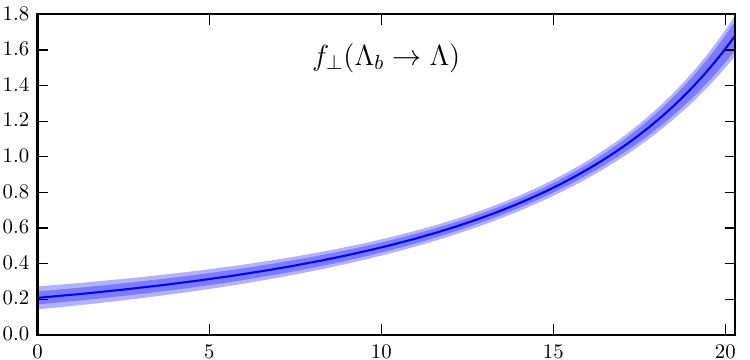} \hfill
 \includegraphics[width=0.49\linewidth]{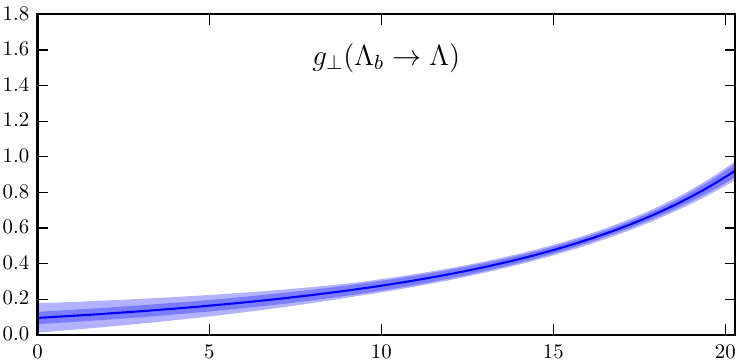} \\
 \includegraphics[width=0.49\linewidth]{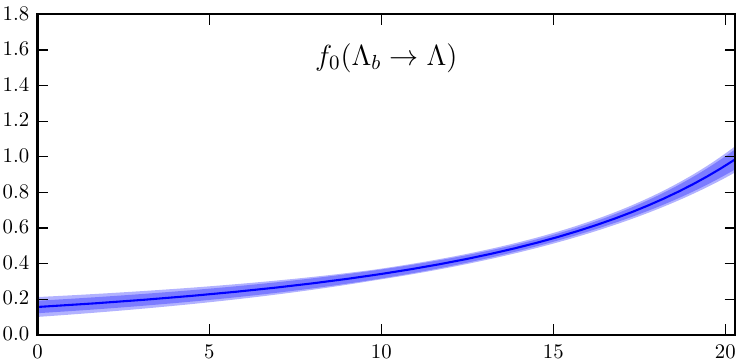} \hfill
 \includegraphics[width=0.49\linewidth]{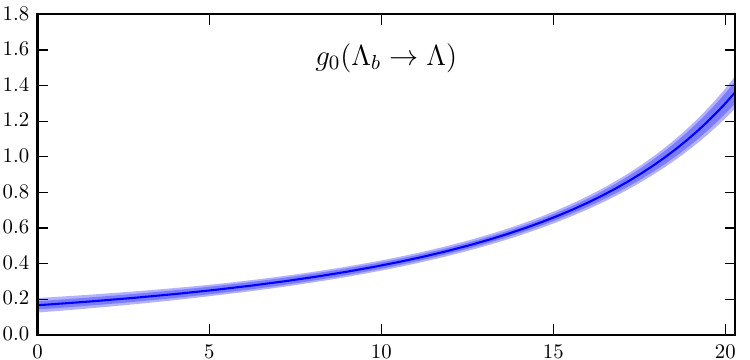} \\
 \includegraphics[width=0.49\linewidth]{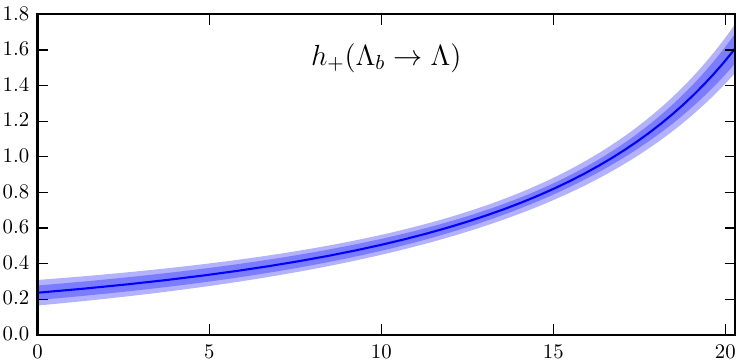} \hfill
 \includegraphics[width=0.49\linewidth]{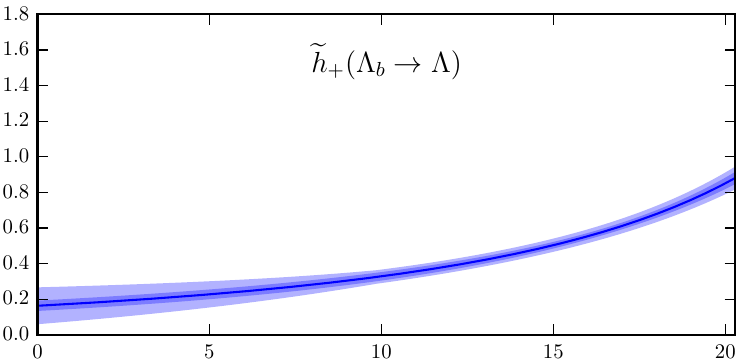} \\
 \includegraphics[width=0.49\linewidth]{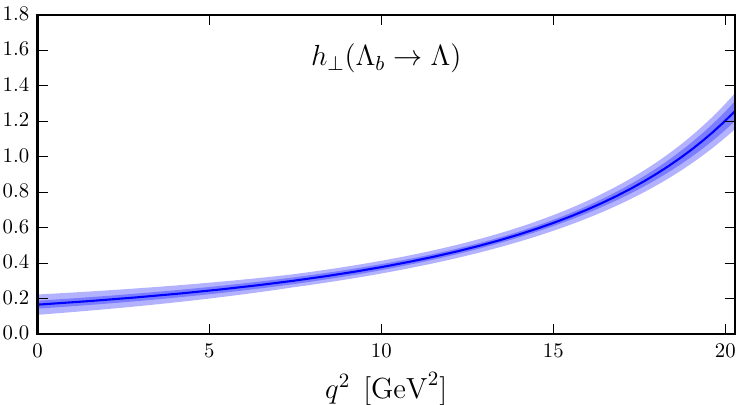} \hfill
 \includegraphics[width=0.49\linewidth]{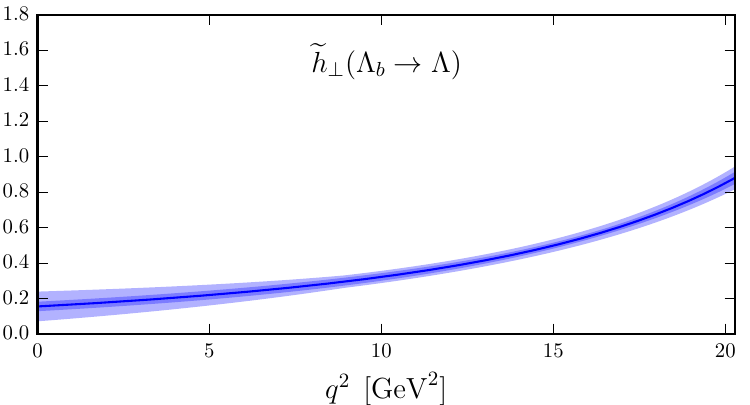}
 \caption{\label{fig:finalFFs}Form factors in the physical limit,
 shown over the entire kinematic range. The inner bands show the statistical uncertainty and the outer
 bands show the total uncertainty.}
\end{figure}

\begin{figure}
 \hspace{1cm}\includegraphics[width=\linewidth]{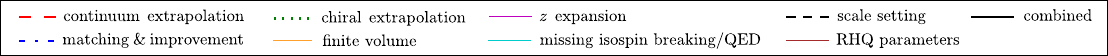}
 
 \vspace{0.5ex}
 
 \includegraphics[width=0.48\linewidth]{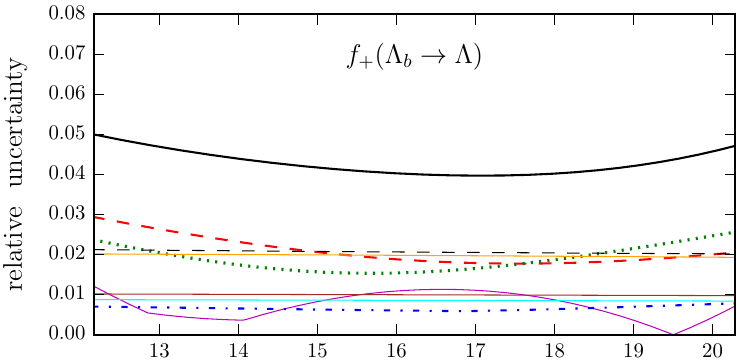} \hfill
 \includegraphics[width=0.48\linewidth]{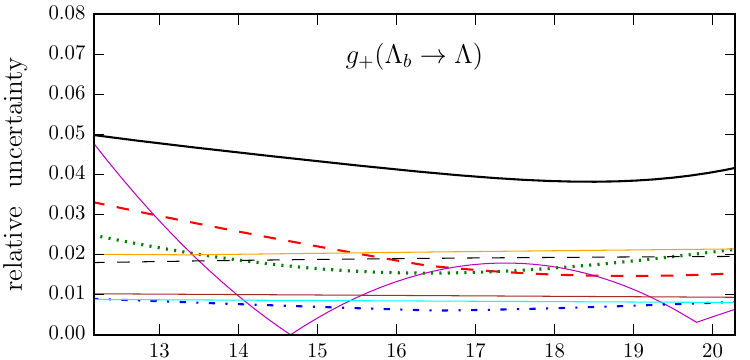} \\
 \includegraphics[width=0.48\linewidth]{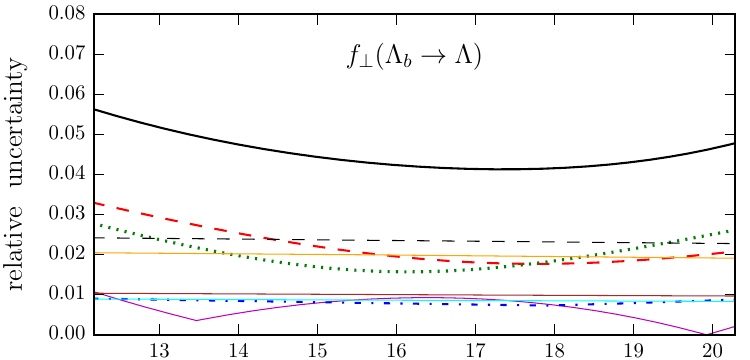} \hfill
 \includegraphics[width=0.48\linewidth]{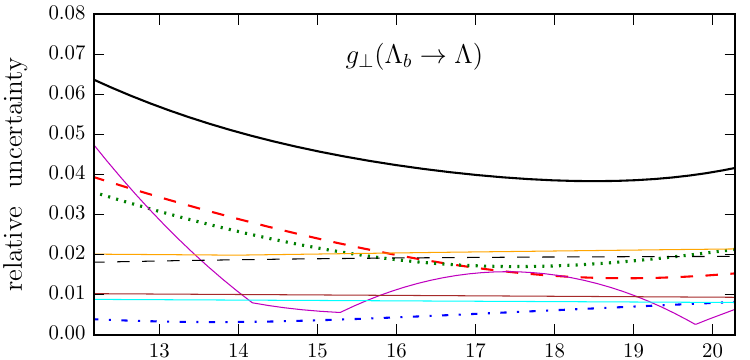} \\
 \includegraphics[width=0.48\linewidth]{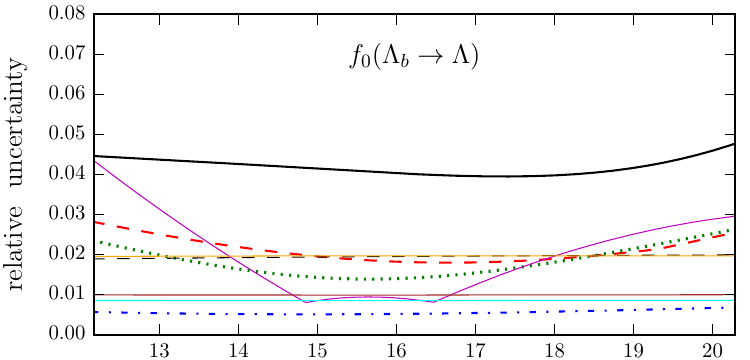} \hfill
 \includegraphics[width=0.48\linewidth]{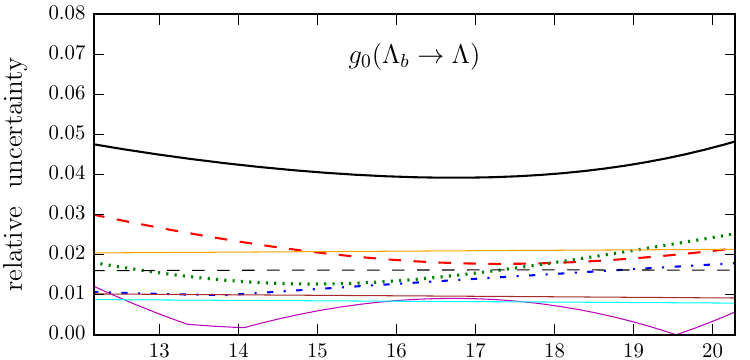} \\
 \includegraphics[width=0.48\linewidth]{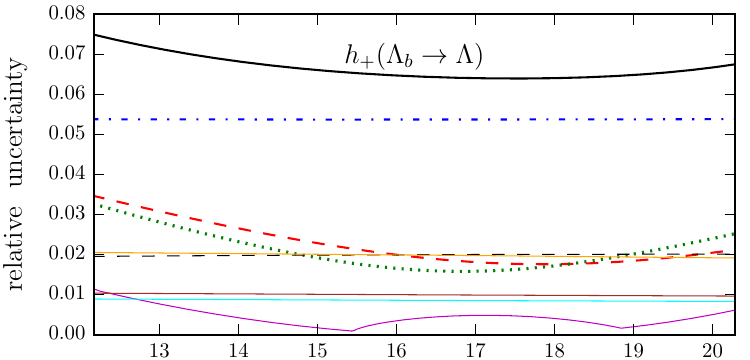} \hfill
 \includegraphics[width=0.48\linewidth]{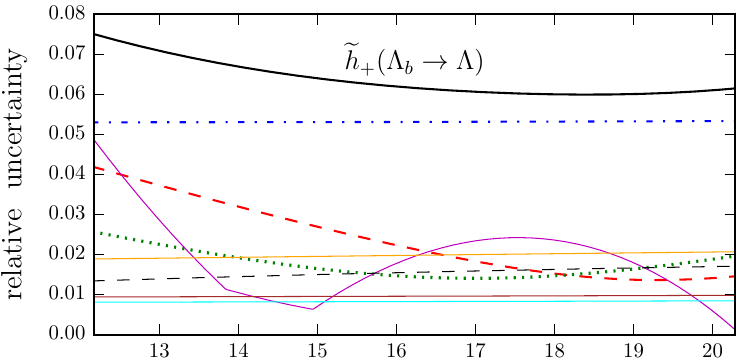} \\
 \includegraphics[width=0.48\linewidth]{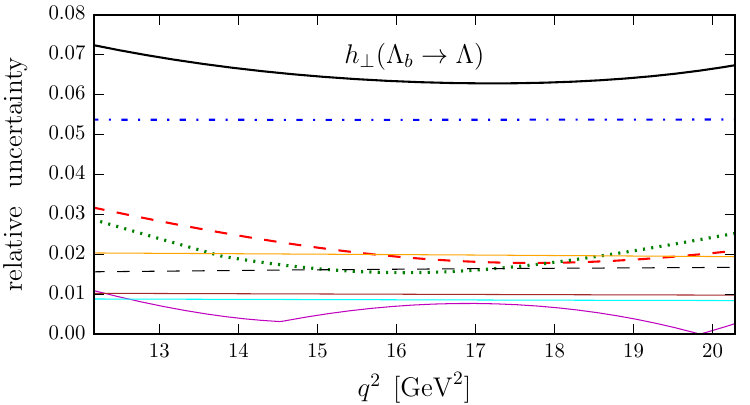} \hfill
 \includegraphics[width=0.48\linewidth]{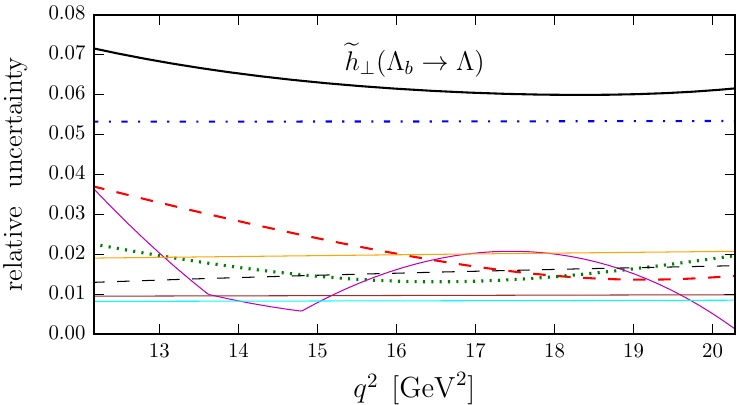}
 
 \caption{\label{fig:finalFFssyst}Systematic uncertainties in the form factors in the high-$q^2$ region.  As explained in the main text and in Ref.~\cite{Detmold:2015aaa},
 the combined uncertainty is not simply the quadratic sum of the individual uncertainties.}
\end{figure}

In the higher-order fit, we simultaneously made the following modifications to account for additional sources of systematic uncertainty:
\begin{itemize}
 \item For the vector and axial vector form factors, in which the residual matching factors in the mostly nonperturbative renormalization procedure and the $\mathcal{O}(a)$-improvement coefficients were computed at one loop,
 we use bootstrap data for the correlator ratios in which these coefficients were drawn from Gaussian random distributions with central values and widths according to Table III of Ref.~\cite{Detmold:2015aaa}.
 \item For the tensor form factors, the renormalization uncertainty is dominated by the use of the tree-level values, $\rho_{T^{\mu\nu}}=1$, for the residual matching factors in the mostly nonperturbative renormalization procedure. We
 estimate the systematic uncertainty in $\rho_{T^{\mu\nu}}$ to be equal to 2 times the maximum value of $|\rho_{V^{\mu}}-1|$, $|\rho_{A^{\mu}}-1|$, which is equal to $0.05316$ \cite{Detmold:2015aaa}. This estimate
 is larger than one-loop results for $|\rho_{T^{\mu\nu}}-1|$ obtained in Ref.~\cite{Bailey:2015dka} for the case of staggered light quarks at comparable lattice spacings. Note that $\rho_{T^{\mu\nu}}$ for the tensor current is scale-dependent, and
 our estimate of the matching uncertainty (and the values of the form factors themselves) should be interpreted as corresponding to $\mu=4.2$ GeV. To incorporate the tensor-current matching uncertainty in the
 fit, we introduced nuisance parameters multiplying the tensor form factors, with Gaussian priors equal to $1\pm 0.05316$.
 \item To propagate the uncertainties in the lattice spacings, lattice pion masses, and lattice $\eta_s$ masses, we promoted these precisely determined quantities to parameters in the fit, with Gaussian priors chosen according to their respective central values and uncertainties.
 \item We estimate the systematic uncertainties in the form factors resulting from the neglected $d-u$ quark-mass difference and QED to be of order $\mathcal{O}((m_d-m_u)/\Lambda_{\rm QCD})\approx 0.5\%$ and $\mathcal{O}(\alpha_{\rm e.m.})\approx 0.7\%$.
 The systematic uncertainty in the $\Lambda_b\to\Lambda$ form factors due to finite-volume effects is expected to be larger than for $\Lambda_b \to \Lambda_c$ (estimated to be 1.5\% in Ref.~\cite{Detmold:2015aaa}) but smaller than for $\Lambda_b \to p$ (estimated to be 3\% in Ref.~\cite{Detmold:2015aaa}),
 so we take this uncertainty to be 2\% here. The systematic uncertainty resulting from the tuning of the relativistic heavy-quark (RHQ) action for the $b$ quark is estimated to be 1\% as in Ref.~\cite{Detmold:2015aaa},
 based on the analysis of $B\to\pi$ form factors using the same $b$-quark action and parameters in Ref.~\cite{Aoki:2012xaa}. To incorporate all of these sources of uncertainties in the higher-order fit, we added
 them to the data correlation matrix used in the fit, treating them as 100\% correlated within each of the following groups of form factors: $(f_+, f_\perp, f_0)$,  $(g_+, g_\perp, g_0)$, $(h_+, h_\perp)$, and $(\widetilde{h}_+, \widetilde{h}_\perp)$.
\end{itemize}
In the physical limit $a=0$, $m_\pi=m_{\pi,{\rm phys}}$, $m_{\eta_s}=m_{\eta_s,{\rm phys}}$, the higher-order fit functions reduce to
\begin{equation}
 f(q^2) = \frac{1}{1-q^2/(m_{\rm pole}^f)^2} \big[ a_0^f + a_1^f\:z(q^2) + a_2^f\:[z(q^2)]^2 \big]. \label{eq:HOfitphys}
\end{equation}
The values and uncertainties of the parameters $a_0^f$, $a_1^f$, and $a_2^f$ from the higher-order fit are given in Table \ref{tab:HO}; their correlation
matrix is given in Tables \ref{tab:HOcorr1} and \ref{tab:HOcorr2}.
As in Ref.~\cite{Detmold:2015aaa}, the recommended procedure for computing the central value, statistical uncertainty, and total systematic uncertainty of any observable depending on the form factors
is the following:
\begin{enumerate}
 \item Compute the observable and its uncertainty using the nominal form factors given by Eq.~(\ref{eq:nominalfitphys}), with the parameter values and correlation matrices from Tables
 \ref{tab:nominal}, \ref{tab:nominalcorr1}, and \ref{tab:nominalcorr2}. Denote the so-obtained central value and uncertainty as
 \begin{equation}
  O, \:\:\sigma_O. \label{eq:O}
 \end{equation}
  \item Compute the same observable and its uncertainty using the higher-order form factors given by Eq.~(\ref{eq:HOfitphys}), with the parameter values and correlation
  matrices from Tables \ref{tab:HO}, \ref{tab:HOcorr1}, and \ref{tab:HOcorr2}. Denote the so-obtained central value and uncertainty as
 \begin{equation}
  O_{\rm HO}, \:\:\sigma_{O,{\rm HO}}.  \label{eq:OHO}
 \end{equation}
  \item The central value, statistical uncertainty, and systematic uncertainty of the observable are then given by
  \begin{equation}
  O \:\pm\: \sigma_{O,{\rm stat}} \:\pm \:\sigma_{O,{\rm syst}}
 \end{equation}
   with
  \begin{eqnarray}
  \sigma_{O,{\rm stat}} &=& \sigma_O, \\
  \sigma_{O,{\rm syst}} &=& {\rm max}\left( |O_{\rm HO}-O|,\:\: H(\sigma_{O,{\rm HO}}-\sigma_O)\sqrt{\sigma_{O,{\rm HO}}^2-\sigma_O^2}  \right), \label{eq:sigmasyst}
 \end{eqnarray}
 where $H$ is the Heaviside step function. To obtain the total uncertainty, the statistical and systematic uncertainties should be added in quadrature,
 \begin{equation}
  \sigma_{O,{\rm tot}} = \sqrt{ \sigma_{O,{\rm stat}}^2 + \sigma_{O,{\rm syst}}^2 }.
 \end{equation}
 More generally, the total \emph{covariance} between two observables $O_1$ and $O_2$ can be computed as
 \begin{equation}
  {\rm cov}_{\rm tot}(O_1,\,O_2) = {\rm cov}_{\rm HO}(O_1,\,O_2)\:\frac{\sigma_{O_1,{\rm tot}}}{\sigma_{O_1,{\rm HO}}}\: \frac{\sigma_{O_2,{\rm tot}}}{\sigma_{O_2,{\rm HO}}}.
 \end{equation}

\end{enumerate}
Plots of the form factors including the total uncertainties are given in Fig.~\ref{fig:finalFFs}. Additionally,
Fig.~\ref{fig:finalFFssyst} shows estimates of the individual sources of the systematic uncertainties in the form factors, obtained
by performing additional fits where each one of the above modifications to the fit functions or data correlation matrix was done individually.\footnote{
The peculiar shape of the $z$-expansion systematic uncertainty is caused by a switching between the two terms $|O_{\rm HO}-O|$ and $H(\sigma_{O,{\rm HO}}-\sigma_O)\sqrt{\sigma_{O,{\rm HO}}^2-\sigma_O^2}$
in Eq.~(\ref{eq:sigmasyst}), which at certain $q^2$-points both happen to be zero.}\textsuperscript{,}\footnote{As an alternative method of estimating the systematic uncertainty associated with discretization effects, 
we have also performed fits using the nominal form, Eq.~(\ref{eq:FFccfit}), but setting $d^f=0$. This more simplistic method
of addressing these effects results in comparable uncertainties.} We stress
that these plots are for illustration only, and the correct procedure for obtaining the total systematic uncertainty in a correlated way is from
the full higher-order fit, in which all modifications were done simultaneously.

\begin{table}
\begin{tabular}{ccccc}
\hline \hline
 Parameter         &  Value  & \hspace{2ex} &  Parameter         & Value  \\
\hline
$a_0^{f_+}$ & $\wm 0.4221\pm 0.0188$ &&  $a_1^{g_0}$ & $-1.0290\pm 0.1614$ \\  
$a_1^{f_+}$ & $-1.1386\pm 0.1683$ &&  $a_1^{g_\perp}$ & $-1.1357\pm 0.1911$ \\  
$a_0^{f_0}$ & $\wm 0.3725\pm 0.0213$ &&  $a_0^{h_+}$ & $\wm 0.4960\pm 0.0258$ \\  
$a_1^{f_0}$ & $-0.9389\pm 0.2250$ &&  $a_1^{h_+}$ & $-1.1275\pm 0.2537$ \\  
$a_0^{f_\perp}$ & $\wm 0.5182\pm 0.0251$ &&  $a_0^{h_\perp}$ & $\wm 0.3876\pm 0.0172$ \\  
$a_1^{f_\perp}$ & $-1.3495\pm 0.2413$ &&  $a_1^{h_\perp}$ & $-0.9623\pm 0.1550$ \\  
$a_0^{g_\perp,g_+}$ & $\wm 0.3563\pm 0.0142$ &&  $a_0^{\widetilde{h}_{\perp},\widetilde{h}_+}$ & $\wm 0.3403\pm 0.0133$ \\  
$a_1^{g_+}$ & $-1.0612\pm 0.1678$ &&  $a_1^{\widetilde{h}_+}$ & $-0.7697\pm 0.1612$ \\  
$a_0^{g_0}$ & $\wm 0.4028\pm 0.0182$ &&  $a_1^{\widetilde{h}_\perp}$ & $-0.8008\pm 0.1537$ \\  
\hline\hline
\end{tabular}
\caption{\label{tab:nominal}Central values and uncertainties of the nominal form factor parameters.
The correlation matrix is provided in Tables \protect\ref{tab:nominalcorr1} and \protect\ref{tab:nominalcorr2}.}
\end{table}

\begin{table}
\begin{tabular}{ccccc}
\hline \hline
 Parameter         &  Value  & \hspace{2ex} &  Parameter         &  Value    \\
\hline
$a_0^{f_+}$ & $\wm 0.4229\pm 0.0274$ &&  $a_2^{g_0}$ & $\wm 1.1490\pm 1.0327$ \\ 
$a_1^{f_+}$ & $-1.3728\pm 0.3068$ &&  $a_1^{g_\perp}$ & $-1.3607\pm 0.2949$ \\ 
$a_2^{f_+}$ & $\wm 1.7972\pm 1.1506$ &&  $a_2^{g_\perp}$ & $\wm 2.4621\pm 1.3711$ \\ 
$a_0^{f_0}$ & $\wm 0.3604\pm 0.0277$ &&  $a_0^{h_+}$ & $\wm 0.4753\pm 0.0423$ \\ 
$a_1^{f_0}$ & $-0.9248\pm 0.3453$ &&  $a_1^{h_+}$ & $-0.8840\pm 0.3997$ \\ 
$a_2^{f_0}$ & $\wm 0.9861\pm 1.1988$ &&  $a_2^{h_+}$ & $-0.8190\pm 1.6760$ \\ 
$a_0^{f_\perp}$ & $\wm 0.5148\pm 0.0353$ &&  $a_0^{h_\perp}$ & $\wm 0.3745\pm 0.0313$ \\ 
$a_1^{f_\perp}$ & $-1.4781\pm 0.4030$ &&  $a_1^{h_\perp}$ & $-0.9439\pm 0.2766$ \\ 
$a_2^{f_\perp}$ & $\wm 1.2496\pm 1.6396$ &&  $a_2^{h_\perp}$ & $\wm 1.1606\pm 1.0757$ \\ 
$a_0^{g_\perp,g_+}$ & $\wm 0.3522\pm 0.0205$ &&  $a_0^{\widetilde{h}_{\perp},\widetilde{h}_+}$ & $\wm 0.3256\pm 0.0248$ \\ 
$a_1^{g_+}$ & $-1.2968\pm 0.2732$ &&  $a_1^{\widetilde{h}_+}$ & $-0.9603\pm 0.2303$ \\ 
$a_2^{g_+}$ & $\wm 2.7106\pm 1.0665$ &&  $a_2^{\widetilde{h}_+}$ & $\wm 2.9780\pm 1.0041$ \\ 
$a_0^{g_0}$ & $\wm 0.4059\pm 0.0267$ &&  $a_1^{\widetilde{h}_\perp}$ & $-0.9634\pm 0.2268$ \\ 
$a_1^{g_0}$ & $-1.1622\pm 0.2929$ &&  $a_2^{\widetilde{h}_\perp}$ & $\wm 2.4782\pm 0.9549$ \\ 
\hline\hline
\end{tabular}
\caption{\label{tab:HO}Central values and uncertainties of the higher-order form factor parameters.
The correlation matrix is provided in Tables \protect\ref{tab:HOcorr1} and \protect\ref{tab:HOcorr2}.}
\end{table}

\FloatBarrier
\section{\label{sec:observables}Calculation of $\Lambda_b \to \Lambda(\to p^+ \pi^-) \, \mu^+ \mu^-$ observables in the Standard Model}
\FloatBarrier

The four-fold differential rate of the decay $\Lambda_b \to \Lambda(\to p^+ \pi^-) \, \ell^+ \ell^-$ with unpolarized $\Lambda_b$ can be written as \cite{Gutsche:2013pp, Boer:2014kda}
\begin{eqnarray}
\nonumber \frac{\mathrm{d}^4 \Gamma}{\mathrm{d} q^2\: \mathrm{d} \cos\theta_\ell\: \mathrm{d} \cos\theta_\Lambda\: \mathrm{d} \phi}
         &=&          \frac{3}{8\pi} \bigg[  \hspace{2.8ex}  \big( K_{1ss} \sin^2\theta_\ell +\, K_{1cc} \cos^2\theta_\ell + K_{1c} \cos\theta_\ell\big)       \\
\nonumber && \phantom{\frac{3}{8\pi} \bigg[} + \big( K_{2ss} \sin^2\theta_\ell +\, K_{2cc} \cos^2\theta_\ell + K_{2c} \cos\theta_\ell\big) \cos\theta_\Lambda  \\
\nonumber && \phantom{\frac{3}{8\pi} \bigg[} + \big( K_{3sc}\sin\theta_\ell \cos\theta_\ell + K_{3s} \sin\theta_\ell\big) \sin\theta_\Lambda \sin\phi          \\
          && \phantom{\frac{3}{8\pi} \bigg[} + \big( K_{4sc}\sin\theta_\ell \cos\theta_\ell + K_{4s} \sin\theta_\ell\big) \sin\theta_\Lambda \cos\phi \bigg], \label{eq:dGamma4f}
\end{eqnarray}
where the angles $\theta_l$ and $\theta_\Lambda$ describe the polar directions of the negatively charged lepton and the proton, respectively, $\phi$ is the azimuthal angle
between the $\ell^+\ell^-$ and $p^+ \pi^-$ decay planes, and the coefficients $K_i$ are functions of $q^2$. The integral of Eq.~(\ref{eq:dGamma4f}) over the angles gives the
$q^2$-differential decay rate,
\begin{equation}
 \frac{\mathrm{d}\Gamma}{\mathrm{d}q^2}=2 K_{1ss} + K_{1cc}.
\end{equation}
We also define the normalized angular observables
\begin{equation}
 \hat{K}_i = \frac{K_i}{\mathrm{d}\Gamma/\mathrm{d}q^2},
\end{equation}
of which the combinations
\begin{eqnarray}
F_L&=&2\hat{K}_{1ss}-\hat{K}_{1cc}, \\
A_{\rm FB}^\ell&=&\frac32\hat{K}_{1c},  \\
A_{\rm FB}^\Lambda&=&\hat{K}_{2ss} + \frac12\hat{K}_{2cc},  \\
A_{\rm FB}^{\ell\Lambda}&=&\frac34\hat{K}_{2c}
\end{eqnarray}
correspond to the fraction of longitudinally polarized dileptons, the lepton-side forward-backward asymmetry, the hadron-side forward-backward asymmetry, and a combined
lepton-hadron forward-backward asymmetry \cite{Boer:2014kda}.

In the approximation where all nonlocal hadronic matrix elements
are absorbed via the ``effective Wilson coefficients'' $C_7^{\rm eff}(q^2)$ and $C_9^{\rm eff}(q^2)$, the hadronic contributions to the functions $K_i$ are
given by the $\Lambda_b \to \Lambda$ form factors computed here and the $\Lambda \to p^+\pi^-$ decay asymmetry parameter,
which is known from experiment to be \cite{Agashe:2014kda}
\begin{equation}
\alpha_\Lambda = 0.642 \pm 0.013. \label{eq:alphaLambda}
\end{equation}
The expressions for $K_i$ are given in Ref.~\cite{Boer:2014kda} for the case $m_\ell=0$, and can be obtained for $m_\ell\neq0$ from Eqs.~(A1) and (A2) of Ref.~\cite{Gutsche:2013pp}.
In the following we focus on the case $\ell=\mu$, which is the most accessible mode at hadron colliders \cite{Aaltonen:2011qs, Aaij:2013mna, Aaij:2015xza}.
Even though lepton-mass effects are not important for $\ell=\mu$ in most of the kinematic range, we include them here.

Following Ref.~\cite{Du:2015tda}, we set the effective Wilson coefficients to
\begin{eqnarray}
 C_7^{\rm eff}(q^2) & = &  C_7 - \frac{1}{3} \left[ C_3 + \frac{4}{3}\,C_4 + 20\,C_5  + \frac{80}{3}\, C_6 \right]
    - \frac{\alpha_s}{4 \pi} \left[ \left(C_1 - 6\,C_2\right) F_{1,c}^{(7)}(q^2) + C_8\, F_8^{(7)}(q^2) \right], \label{eq:C7eff} \\
\nonumber C_9^{\rm eff}(q^2) & = &  C_9 + \frac{4}{3}\, C_3 + \frac{64}{9}\, C_5 + \frac{64}{27}\, C_6
    + h(0,q^2) \left( -\frac{1}{2}\, C_3 - \frac{2}{3}\, C_4 - 8\, C_5 - \frac{32}{3}\, C_6 \right) \\
\nonumber && + h(m_b,q^2) \left( -\frac{7}{2}\, C_3 - \frac{2}{3}\, C_4 - 38\, C_5 - \frac{32}{3}\, C_6 \right)  
    + h(m_c,q^2) \left( \frac{4}{3}\, C_1 + C_2 + 6\, C_3 + 60\, C_5 \right)  \\
&& - \frac{\alpha_s}{4 \pi} \left[ C_1\, F_{1,c}^{(9)}(q^2) + C_2\, F_{2,c}^{(9)}(q^2) + C_8\, F_8^{(9)}(q^2) \right], \label{eq:C9eff}
\end{eqnarray}
where the functions $h(m_q, q^2)$ and $F_8^{(7,9)}(q^2)$ are defined in Eqs.~(11), (82), (83) of Ref.~\cite{Beneke:2001at}, and the functions $F_{1,c}^{(7,9)}(q^2)$
and $F_{2,c}^{(7,9)}(q^2)$ are evaluated using the Mathematica packages for high $q^2$ and low $q^2$ provided in Ref.~\cite{Greub:2008cy} (the low-$q^2$ versions are based on Ref.~\cite{Asatryan:2001zw}).
The effective Wilson coefficients $C_7^{\rm eff}(q^2)$ and $C_9^{\rm eff}(q^2)$ incorporate the leading contributions from an operator
product expansion (OPE) of the nonlocal product of $O_{1,...,6;8}$ with the quark electromagnetic current \cite{Grinstein:2004vb, Beylich:2011aq}.
Unlike in Ref.~\cite{Grinstein:2004vb}, we have not expanded the functions $h(m_c,q^2)$ and $F_{i,c}^{(7,9)}(q^2)$
in powers of $m_c^2/q^2$. The subleading contributions from the OPE arise from dimension-5 operators \cite{Beylich:2011aq}, whose matrix elements in the high-$q^2$ region are suppressed
by $\Lambda_{\rm had}^2/Q^2\sim 1\%$, where $Q^2\sim\{q^2,m_b^2\}$.
Nonfactorizable spectator-scattering effects that are expected to be relevant at low $q^2$ \cite{Beneke:2001at, Wang:2015ndk} have not yet been derived for the baryonic decay, and
we neglect them here.

We evaluated the Wilson coefficients $C_1$-$C_{10}$ in the $\overline{\rm MS}$ scheme at next-to-next-to-leading-logarithmic order \cite{Bobeth:1999mk, Gambino:2003zm, Gorbahn:2004my, Gorbahn:2005sa} using the \texttt{EOS} software \cite{Beaujean:2012uj, EOS};
their values are listed in Table \ref{tab:Wilson}. The charm and bottom masses appearing in the functions $h(m_q, q^2)$, $F_{1,c}^{(7,9)}(q^2)$, and $F_{2,c}^{(7,9)}(q^2)$ are defined
in the pole scheme \cite{Greub:2008cy, Asatryan:2001zw}; we use
\begin{eqnarray}
 m_c^{\rm pole} &=& 1.5953\:\:{\rm GeV}, \\
 m_b^{\rm pole} &=& 4.7417\:\:{\rm GeV},
\end{eqnarray}
also evaluated with \texttt{EOS}. The values of $m_b^{\overline{\rm MS}}$ (which multiplies the operator $O_7$), $\alpha_s$, and $\alpha_e$
are given in Table \ref{tab:Wilson}. We take the CKM matrix elements from the Summer-2014 Standard-Model fit of the UTFit Collaboration \cite{UTfit},
\begin{equation}
|V_{ts}^* V_{tb}| = 0.04088 \pm 0.00055, \label{eq:VtsVtb}
\end{equation}
and use the $\Lambda_b$ lifetime
\begin{equation}
 \tau_{\Lambda_b} = (1.466 \pm 0.010)\:\:{\rm ps} \label{eq:tauLb}
\end{equation}
from the 2015 update of the Review of Particle Physics \cite{Agashe:2014kda}.

\begin{table}
\begin{tabular}{lcccccc}
\hline\hline
            & & $\mu=2.1\:{\rm GeV}$ & & $\mu=4.2\:{\rm GeV}$ & & $\mu=8.4\:{\rm GeV}$ \\
\hline
$C_1$       && $-0.4965$     && $-0.2877$      && $-0.1488$      \\
$C_2$       && $\wm1.0246$   && $\wm1.0101$    && $\wm1.0036$    \\
$C_3$       && $-0.0143$     && $-0.0060$      && $-0.0027$      \\
$C_4$       && $-0.1500$     && $-0.0860$      && $-0.0543$      \\
$C_5$       && $\wm0.0010$   && $\wm0.0004$    && $\wm0.0002$    \\
$C_6$       && $\wm0.0032$   && $\wm0.0011$    && $\wm0.0004$    \\
$C_7$       && $-0.3782$     && $-0.3361$      && $-0.3036$      \\
$C_8$       && $-0.2133$     && $-0.1821$      && $-0.1629$      \\
$C_9$       && $\wm4.5692$   && $\wm4.2745$    && $\wm3.8698$    \\
$C_{10}$    && $-4.1602$     && $-4.1602$      && $-4.1602$      \\
$m_b^{\overline{\rm MS}}\:[{\rm GeV}]$
            && $\wm4.9236$   && $\wm4.2000$    &&  $\wm3.7504$   \\
$\alpha_s$  && $\wm0.2945$   && $\wm0.2233$    &&  $\wm0.1851$   \\
$\alpha_e$  && $\wm1/134.44$ && $\wm1/133.28$  &&  $\wm1/132.51$ \\
\hline\hline
\end{tabular}
\caption{\label{tab:Wilson}Wilson coefficients, $b$-quark mass, and strong and electromagnetic couplings in the $\overline{\rm MS}$ scheme at the nominal scale $\mu=4.2\:{\rm GeV}$ and at the low and high scales
used to estimate the perturbative uncertainties. The values shown here were computed using the \texttt{EOS} \cite{Beaujean:2012uj, EOS} and \texttt{alphaQED} \cite{Jegerlehner:2011mw, alphaQED} packages.
Even though some of the quantities in this table are strongly scale-dependent, most of this dependence cancels in the physical observables.}
\end{table}

\begin{figure}
 \includegraphics[width=0.49\linewidth]{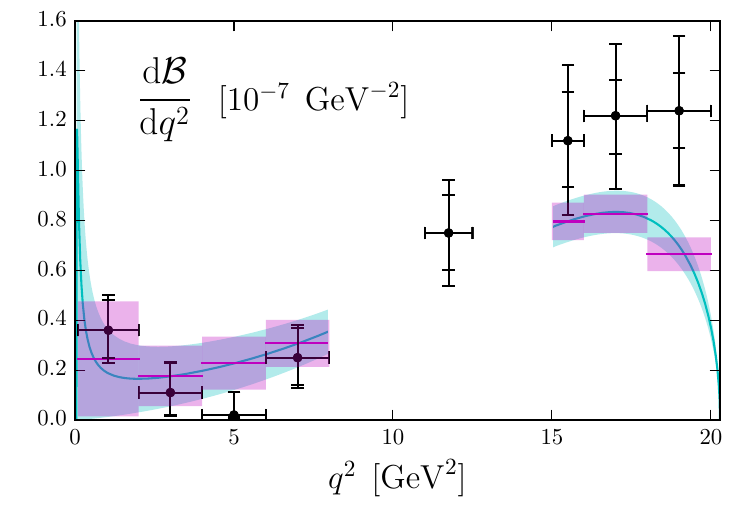}
 
 \caption{\label{fig:dB}$\Lambda_b\to\Lambda\:\mu^+\mu^-$ differential branching fraction calculated in the Standard Model, compared to
 experimental data from LHCb \cite{Aaij:2015xza} (black points; error bars are shown both including and excluding the uncertainty from
 the normalization mode $\Lambda_b \to J/\psi\,\Lambda$ \cite{Agashe:2014kda}).}
\end{figure}

\begin{figure}
 \includegraphics[width=0.49\linewidth]{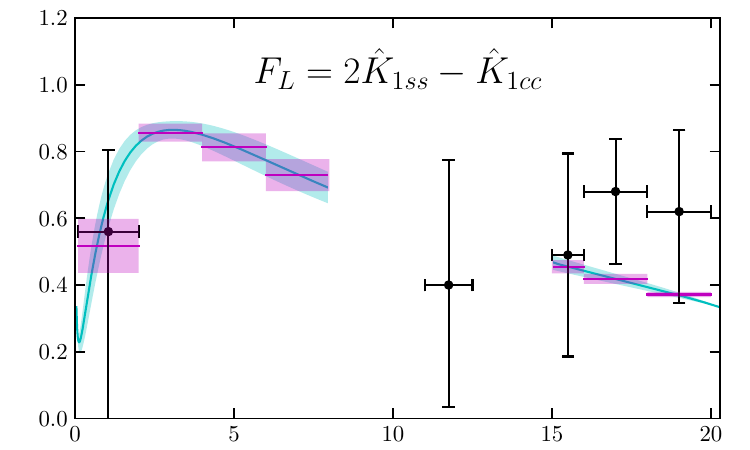} \hfill \includegraphics[width=0.49\linewidth]{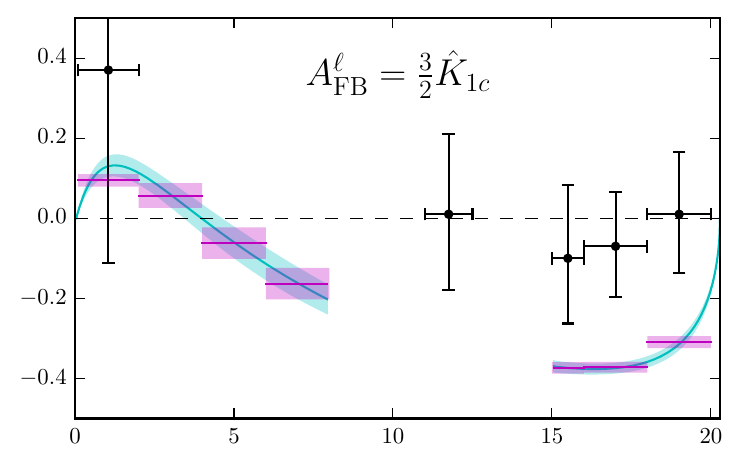}

 \includegraphics[width=0.49\linewidth]{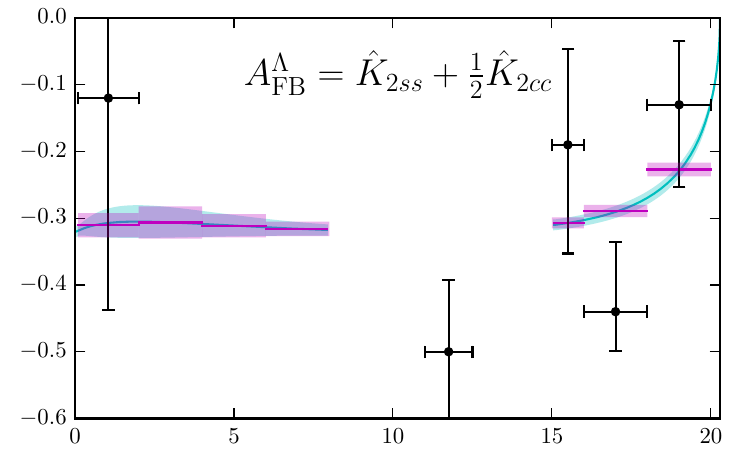} \hfill \includegraphics[width=0.49\linewidth]{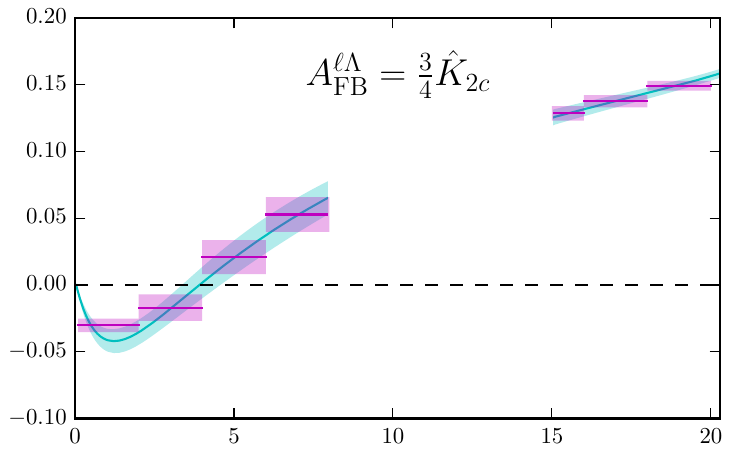}

 \includegraphics[width=0.49\linewidth]{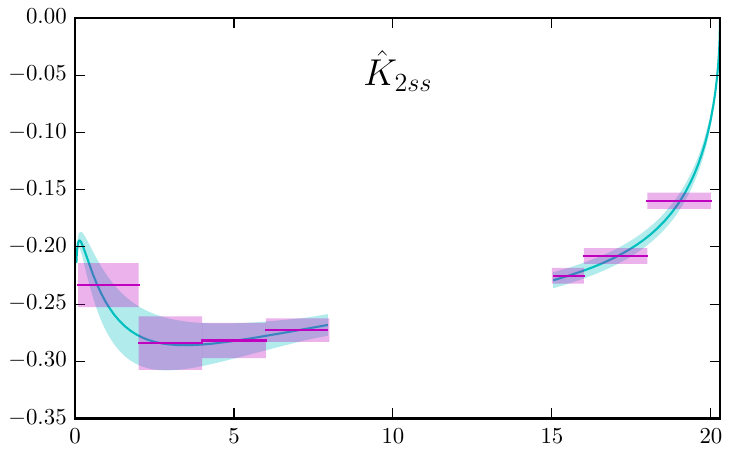} \hfill \includegraphics[width=0.49\linewidth]{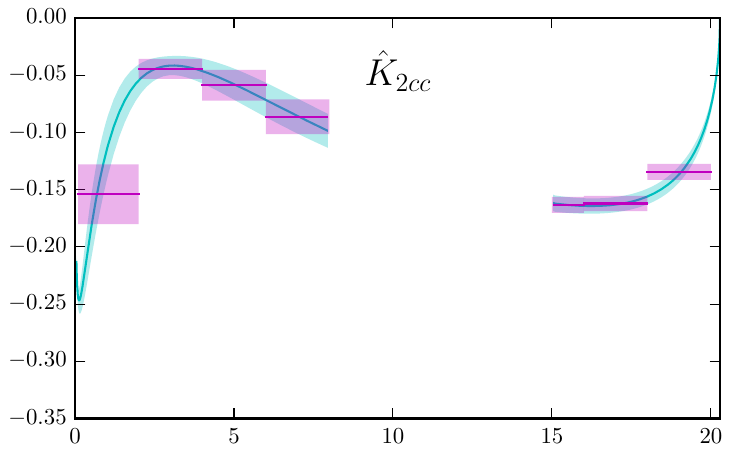}

 \includegraphics[width=0.49\linewidth]{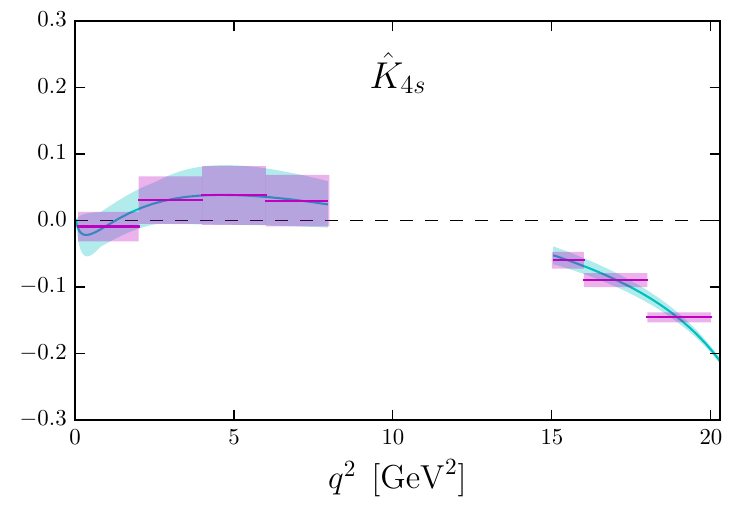} \hfill \includegraphics[width=0.49\linewidth]{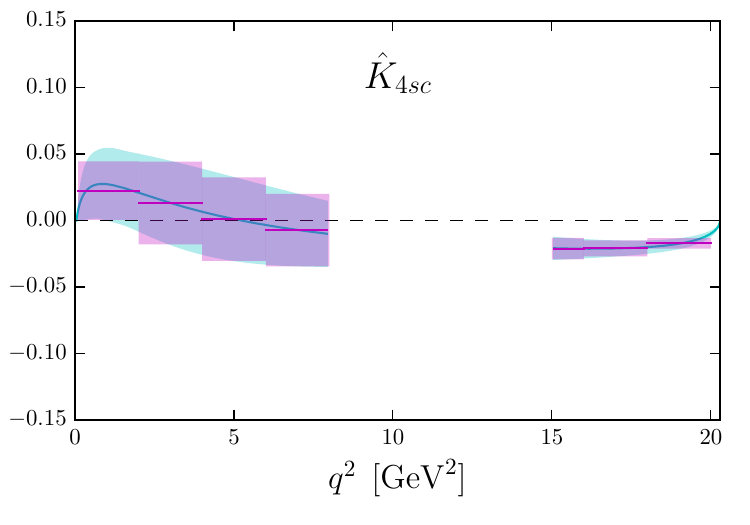}
 
 \caption{\label{fig:angular}$\Lambda_b\to\Lambda(\to p^+ \pi^-)\mu^+\mu^-$ angular observables calculated in the Standard Model (for unpolarized $\Lambda_b$),
 compared to experimental data from LHCb, where available \cite{Aaij:2015xza} (black points). The observables $\hat{K}_{3s}$ and $\hat{K}_{3sc}$ are negligibly small in the Standard Model and are therefore
 not shown here.}
\end{figure}

Our results for the $\Lambda_b\to\Lambda\:\mu^+\mu^-$ differential branching fraction and the $\Lambda_b\to\Lambda(\to p^+ \pi^-)\mu^+\mu^-$ angular observables
are plotted in Figs.~\ref{fig:dB} and Figs.~\ref{fig:angular} as the cyan curves (without binning) and the magenta curves (binned). Where available,
experimental data from LHCb \cite{Aaij:2015xza} are included in the figures. Note that for the angular observables, numerator and denominator are binned separately,
\begin{equation}
 \langle\hat{K}_i\rangle_{[q^2_{\rm min},\: q^2_{\rm max}]} = \frac{\displaystyle\int_{q^2_{\rm min}}^{q^2_{\rm max}} K_i\:\mathrm{d}q^2}{\displaystyle\int_{q^2_{\rm min}}^{q^2_{\rm max}} (\mathrm{d}\Gamma/\mathrm{d}q^2)\:\mathrm{d}q^2}.
\end{equation}
The binned observables are also listed numerically in Table \ref{tab:binnedobs}, including two additional wider bins at low and high $q^2$.
The observables $\hat{K}_{3s}$ and $\hat{K}_{3sc}$ are negligibly small in the Standard Model and are not shown here.

\begin{table}
\begin{tabular}{llllllllll}
\hline \hline
              & $\wm\langle \mathrm{d}\mathcal{B}/\mathrm{d} q^2\rangle$ & $\wm\langle F_L \rangle $ & $\wm\langle A_{\rm FB}^\ell \rangle $ & $\wm\langle A_{\rm FB}^\Lambda \rangle $ &  $\wm\langle A_{\rm FB}^{\ell\Lambda} \rangle $ & $\wm\langle \hat{K}_{2ss} \rangle $ & $\wm\langle \hat{K}_{2cc} \rangle $ & $\wm\langle \hat{K}_{4s} \rangle $ &  $\wm\langle \hat{K}_{4sc} \rangle $  \\
\hline
$[0.1, 2]$    & $\wm0.25(23)$  & $\wm0.517(81)$  & $\wm0.095(15)$  & $-0.310(18)$   & $-0.0302(51)$  & $-0.233(19)$   & $-0.154(26)$  & $-0.009(22)$   & $\wm0.022(22)$ \\ 
$[2, 4]$      & $\wm0.18(12)$  & $\wm0.856(27)$  & $\wm0.057(31)$  & $-0.306(24)$   & $-0.0169(99)$  & $-0.284(23)$   & $-0.0444(87)$ & $\wm0.031(36)$   & $\wm0.013(31)$ \\ 
$[4, 6]$      & $\wm0.23(11)$  & $\wm0.813(42)$  & $-0.062(39)$ & $-0.311(17)$   & $\wm0.021(13)$    & $-0.282(15)$   & $-0.059(13)$  & $\wm0.038(44)$   & $\wm0.001(31)$ \\ 
$[6, 8]$      & $\wm0.307(94)$ & $\wm0.730(48)$  & $-0.163(40)$ & $-0.316(11)$   & $\wm0.053(13)$    & $-0.273(10)$   & $-0.086(15)$  & $\wm0.030(39)$   & $-0.007(27)$ \\ 
$[1.1, 6]$    & $\wm0.20(12)$  & $\wm0.820(32)$  & $\wm0.012(31)$  & $-0.309(21)$   & $-0.0027(99)$  & $-0.280(20)$   & $-0.056(10)$  & $\wm0.030(35)$   & $\wm0.009(30)$ \\ 
$[15, 16]$    & $\wm0.796(75)$ & $\wm0.455(20)$  & $-0.374(14)$ & $-0.3069(83)$  & $\wm0.1286(55)$   & $-0.2253(69)$  & $-0.1633(69)$ & $-0.060(13)$   & $-0.0211(80)$ \\ 
$[16, 18]$    & $\wm0.827(76)$ & $\wm0.418(15)$  & $-0.372(13)$ & $-0.2891(90)$  & $\wm0.1377(46)$   & $-0.2080(69)$  & $-0.1621(66)$ & $-0.090(10)$   & $-0.0209(60)$ \\ 
$[18, 20]$    & $\wm0.665(68)$ & $\wm0.3714(79)$ & $-0.309(15)$ & $-0.227(10)$   & $\wm0.1492(37)$   & $-0.1598(71)$  & $-0.1344(70)$ & $-0.1457(74)$   & $-0.0172(40)$ \\ 
$[15, 20]$    & $\wm0.756(70)$ & $\wm0.410(13)$  & $-0.350(13)$ & $-0.2710(92)$  & $\wm0.1398(43)$   & $-0.1947(68)$  & $-0.1526(65)$ & $-0.1031(97)$   & $-0.0196(55)$ \\ 
\hline \hline
\end{tabular}
\caption{\label{tab:binnedobs}Standard-Model predictions for the binned $\Lambda_b \to \Lambda\, \mu^+ \mu^-$ differential branching fraction (in units of $10^{-7}\:\:{\rm GeV}^{-2}$)
and for the binned $\Lambda_b\to\Lambda(\to p^+ \pi^-)\mu^+\mu^-$ angular observables (with unpolarized $\Lambda_b$). The first column specifies the bin ranges $[q^2_{\rm min},\: q^2_{\rm max}]$ in units of ${\rm GeV}^2$.}
\end{table}

The uncertainties given for the Standard-Model predictions are the total uncertainties, which include the statistical and systematic uncertainties from the form factors
(propagated to the observables using the procedure explained in Sec.~\ref{sec:ccextrap}), the perturbative uncertainties, an estimate of quark-hadron duality violations
(discussed further below), and the parametric uncertainties from Eqs.~(\ref{eq:alphaLambda}), (\ref{eq:VtsVtb}), and (\ref{eq:tauLb}). For all observables
considered here (but not for $\hat{K}_{3s}$ and $\hat{K}_{3sc}$), the uncertainties associated with the subleading contributions from the OPE (at high $q^2$) are negligible compared to the other uncertainties.
The central values of the observables were computed at the renormalization scale $\mu=4.2\:{\rm GeV}$; to estimate
the perturbative uncertainties, we varied the renormalization scale from $\mu=2.1\:{\rm GeV}$ to $\mu=8.4\:{\rm GeV}$. When doing this scale variation, we also
included the renormalization-group running of the tensor form factors from the nominal scale $\mu_0=4.2\:{\rm GeV}$ to the scale $\mu$, by multiplying
these form factors with
\begin{equation}
 \left(\frac{\alpha_s(\mu)}{\alpha_s(\mu_0)} \right)^{-\gamma_T^{(0)}/(2\beta_0)}
\end{equation}
(as in Ref.~\cite{Du:2015tda}), where $\gamma_T^{(0)}=2\, C_F = 8/3$ is the anomalous dimension of the tensor current \cite{Broadhurst:1994se}, and $\beta_0=(11\, N_c - 2\, N_f)/3=23/3$ is the leading-order QCD beta function \cite{Gross:1973id} for 5 active flavors.
Even though we did not perform a one-loop calculation of the residual lattice-to-continuum matching factors for the tensor currents, our estimates of the renormalization uncertainties in the tensor form factors
as discussed in Sec.~\ref{sec:ccextrap} are specific for $\mu=4.2\:{\rm GeV}$, and doing the RG running avoids a double-counting of these uncertainties. Note that
the contributions of the tensor form factors to the observables are proportional to $1/q^2$ (because of the photon propagator connecting $O_7$ to the lepton current), and are
suppressed relative to those from the vector and axial vector form factors at high $q^2$. At low $q^2$, the other uncertainties (statistical uncertainties, $z$-expansion uncertainties, etc.) in the tensor form factors dominate
over the uncertainties from the matching factors.

The functions $C_7^{\rm eff}(q^2)$ and $C_{9}^{\rm eff}(q^2)$ have been computed in perturbation theory and do not correctly describe the local $q^2$-dependence resulting
from charmonium resonances \cite{Lyon:2014hpa}. The $q^2$-region near $q^2=m_{J/\psi}^2$ and $q^2=m_{\psi^\prime}^2$ resonances is excluded for this reason. In the high-$q^2$ region, which
is affected by multiple broad charmonium resonances \cite{Lyon:2014hpa}, it is has been argued using quark-hadron duality
that these functions correctly describe the charm effects to some extent for observables binned over a wide enough range \cite{Beylich:2011aq}. The duality violations
in the $B \to K \ell^+\ell^-$ decay rate integrated over the high-$q^2$ region have been estimated in Ref.~\cite{Beylich:2011aq} to be of order 2\%. We instead
include an overall 5\% uncertainty on both $C_7^{\rm eff}(q^2)$ and $C_{9}^{\rm eff}(q^2)$ to account for these effects, which increases the uncertainty in the
$\Lambda_b \to \Lambda \ell^+\ell^-$ decay rate at high $q^2$ by approximately 5\%. We use the same 5\% uncertainty on $C_7^{\rm eff}(q^2)$ and $C_{9}^{\rm eff}(q^2)$ at low $q^2$,
even though there is no good theoretical basis for it. We are unable to estimate the uncertainties resulting from the neglected spectator-scattering effects at low $q^2$, but it is likely
that these uncertainties are smaller than the large form factor uncertainties in that region, at least for the differential branching fraction.

In our calculation of the $\Lambda_b \to \Lambda(\to p^+\,\pi^-) \mu^+ \mu^-$ angular observables, we have treated the $\Lambda_b$'s as unpolarized. The polarization vector of $\Lambda_b$ baryons
produced through the strong interaction is expected to be perpendicular to the production plane spanned by the $\Lambda_b$ momentum and the beam direction, as a consequence of parity conservation \cite{Lednicky:1985zx}.
Taking into account this polarization direction, the $\Lambda_b \to \Lambda(\to p^+\,\pi^-) \mu^+ \mu^-$ angular distribution then becomes dependent on five angles \cite{Korner:2014bca}, as illustrated
 in Fig.~1 of Ref.~\cite{Aaij:2013oxa} for the similar decay $\Lambda_b \to \Lambda(\to p^+\,\pi^-) J/\psi (\to \mu^+ \mu^-)$. For a particle detector with isotropic acceptance,
the polarization effects will average out if the additional angles are not reconstructed. This is not the case for the asymmetric LHCb detector \cite{Alves:2008zz},
but the polarization parameter for $\Lambda_b$ baryons produced at the LHC was measured to be small and consistent with zero, $P_b=0.06\pm0.07\pm0.02$ (at $\sqrt{s}=7\:{\rm TeV}$) \cite{Aaij:2013oxa}.
The resulting effects are therefore expected to be small. Because the polarization effects depend on the details of the detector and on the experimental analysis, their
study is beyond the scope of the present work.

\FloatBarrier
\section{\label{sec:discussion}Discussion}
\FloatBarrier

The baryonic decay $\Lambda_b \to \Lambda(\to p^+\,\pi^-) \mu^+ \mu^-$ can provide valuable new information on the physics of the $b \to s \mu^+ \mu^-$ transition,
and hence on possible physics beyond the Standard Model. Here we have presented a high-precision lattice QCD calculation of the relevant $\Lambda_b\to\Lambda$ form factors. Using a relativistic heavy-quark action,
we have computed the ten helicity form factors describing the $\Lambda_b\to\Lambda$ matrix elements of the $b\to s$ vector, axial vector, and tensor currents directly at the physical
$b$-quark mass. Compared to Ref.~\cite{Detmold:2012vy}, where we treated the $b$ quark in the static limit, this work significantly reduces the theoretical uncertainties
in the form factors and hence in phenomenological applications. As in Ref.~\cite{Detmold:2015aaa}, we provide the form factors in terms of two sets of $z$-expansion
parameters. The ``nominal'' parameters and their correlation matrix are given in Tables \ref{tab:nominal}, \ref{tab:nominalcorr1}, \ref{tab:nominalcorr2}, and are used
to compute the central values and statistical uncertainties. The ``higher-order'' parameters and their correlation matrix are given in Tables \ref{tab:HO}, \ref{tab:HOcorr1}, \ref{tab:HOcorr2},
and are used in combination with the nominal parameters to compute the systematic uncertainties in a fully correlated way. Data files containing all form factor parameters
and their covariance matrices are also available as supplemental material \cite{supplementalmaterial}. Plots of the form factors with error bands
indicating the total uncertainties are shown in Fig.~\ref{fig:finalFFs}. Our results for all ten form factors are consistent with those of the recent
QCD light-cone sum-rule calculation \cite{Wang:2015ndk}, and our uncertainties are much smaller in most of the kinematic range.

Standard-Model predictions for the $\Lambda_b\to\Lambda(\to p^+ \pi^-)\mu^+\mu^-$ differential branching fraction and
angular observables using our form factor results are presented in Figs.~\ref{fig:dB}, \ref{fig:angular} and Table \ref{tab:binnedobs}.
In Table \ref{tab:SMvsLHCb1520} below, we show a further comparison with LHCb results \cite{Aaij:2015xza} in the bin $[15\:{\rm GeV}^2, 20\:{\rm GeV}^2]$, where
both our predictions and the experimental data are most precise.

\begin{table}[!h]
 \begin{tabular}{ccccc}
 \hline\hline
                                                                & \hspace{1ex}  & This work            & \hspace{1ex}   & LHCb              \\
 \hline
 $\langle \mathrm{d}\mathcal{B}/\mathrm{d} q^2\rangle_{[15,\,20]}$             && $\wm0.756 \pm 0.070$    && $\wm1.20\pm0.27$  \\
 $\wm\langle F_L \rangle_{[15,\,20]} $                                         && $\wm0.410 \pm 0.013$    && $\wm0.61^{+\,0.11}_{-\,0.14}$   \\
 $\wm\langle A_{\rm FB}^\ell \rangle_{[15,\,20]} $                             && $-0.350 \pm 0.013$      && $-0.05\pm0.09$    \\
 $\wm\langle A_{\rm FB}^\Lambda \rangle_{[15,\,20]} $                          && $-0.2710 \pm 0.0092$    && $-0.28\pm0.08$      \\
 \hline\hline
 \end{tabular}
 \caption{\label{tab:SMvsLHCb1520}Comparison of our results for the $\Lambda_b\to\Lambda(\to p^+ \pi^-)\mu^+\mu^-$ observables with LHCb data \cite{Aaij:2015xza} in the bin $[15\:{\rm GeV}^2, 20\:{\rm GeV}^2]$.}
\end{table}

In this bin, the magnitude of the lepton-side forward-backward asymmetry measured by LHCb is smaller than the theoretical value by $3.3\sigma$. The measured
differential branching fraction exceeds the theoretical value by $1.6\sigma$. While the latter deviation is not yet statistically significant, it is in the opposite
direction to what has been observed for the decays $B \to K^{(*)} \mu^+\mu^-$ and $B_s \to \phi \mu^+\mu^-$ \cite{Bouchard:2013mia, Horgan:2013pva, Straub:2015ica, Du:2015tda, Aaij:2013iag, Aaij:2013aln, Aaij:2015esa}.
A negative shift in the Wilson coefficient $C_9$ alone, the simplest scenario that significantly improves the agreement between theory and experiment in global fits of mesonic observables \cite{Descotes-Genon:2013wba, Altmannshofer:2014rta, Descotes-Genon:2015uva}, would
further lower the predicted $\Lambda_b\to\Lambda\:\mu^+\mu^-$ differential branching fraction. It is possible that the tensions are caused by the
broad charmonium resonances in this kinematic region \cite{Lyon:2014hpa}, which may contribute to the baryonic decay rate with the opposite sign. This would
correspond to duality violations in the operator product expansion used here to treat the nonlocal matrix elements of $O_1$ and $O_2$ that are much larger than estimated previously \cite{Beylich:2011aq}.
More precise experimental results, including for the branching ratio of the normalization mode $\Lambda_b \to J/\psi\,\Lambda$, would be very useful in clarifying this aspect of the results.

\FloatBarrier
\section*{Acknowledgments}
\FloatBarrier

SM would like to thank Danny van Dyk for discussions. We are grateful to the RBC and UKQCD collaborations for making their gauge field configurations available.
The lattice calculations were carried out using the Chroma software \cite{Edwards:2004sx} on high-performance computing resources provided by XSEDE (supported by National Science Foundation Grant Number OCI-1053575) and NERSC (supported by U.S.~Department of Energy Grant Number DE-AC02-05CH11231).
SM is supported by National Science Foundation Grant Number PHY-1520996, and by the RHIC Physics Fellow Program of the RIKEN BNL Research Center.
WD was partially supported by the U.S.~Department of Energy Early Career Research Award DE-S{C}0010495 and under Grant No.~DE-S{C}0011090.

\FloatBarrier
\section*{Notes added}
\FloatBarrier

\begin{itemize}
 \item In this version, we corrected an error in the lepton-mass contributions to the Standard-Model predictions for $F_L$. We thank Gudrun Hiller for bringing this to our attention.
 \item In 2018, the LHCb Collaboration published an erratum to Ref.~\cite{Aaij:2015xza}, explaining that the measured values reported for $A_{\rm FB}^\ell$ were accidentally $CP$ asymmetries instead of $CP$ averages. This nullifies the apparent tension in $A_{\rm FB}^\ell$ discussed above.
\end{itemize}

\newpage
\appendix

\FloatBarrier
\section{\label{sec:tables}Additional tables}
\FloatBarrier

\begin{table}[!h]
\small
\begin{tabular}{lccccccccc}
\hline\hline
 & $a_0^{f_+}$ & $a_1^{f_+}$ & $a_0^{f_0}$ & $a_1^{f_0}$ & $a_0^{f_\perp}$ & $a_1^{f_\perp}$ & $a_0^{g_\perp,g_+}$ & $a_1^{g_+}$ & $a_0^{g_0}$ \\ 
\hline
$a_0^{f_+}$ & $\wm 1$ & $-0.8899$ & $\wm 0.5529$ & $-0.4746$ & $\wm 0.7636$ & $-0.5961$ & $\wm 0.6314$ & $-0.4590$ & $\wm 0.7879$ \\ 
$a_1^{f_+}$ & $-0.8899$ & $\wm 1$ & $-0.5184$ & $\wm 0.5200$ & $-0.6740$ & $\wm 0.6363$ & $-0.5654$ & $\wm 0.4889$ & $-0.6672$ \\ 
$a_0^{f_0}$ & $\wm 0.5529$ & $-0.5184$ & $\wm 1$ & $-0.8684$ & $\wm 0.4563$ & $-0.3557$ & $\wm 0.6190$ & $-0.4499$ & $\wm 0.5184$ \\ 
$a_1^{f_0}$ & $-0.4746$ & $\wm 0.5200$ & $-0.8684$ & $\wm 1$ & $-0.3870$ & $\wm 0.3385$ & $-0.5356$ & $\wm 0.5178$ & $-0.4440$ \\ 
$a_0^{f_\perp}$ & $\wm 0.7636$ & $-0.6740$ & $\wm 0.4563$ & $-0.3870$ & $\wm 1$ & $-0.8522$ & $\wm 0.5521$ & $-0.4026$ & $\wm 0.7538$ \\ 
$a_1^{f_\perp}$ & $-0.5961$ & $\wm 0.6363$ & $-0.3557$ & $\wm 0.3385$ & $-0.8522$ & $\wm 1$ & $-0.4408$ & $\wm 0.3880$ & $-0.5619$ \\ 
$a_0^{g_\perp,g_+}$ & $\wm 0.6314$ & $-0.5654$ & $\wm 0.6190$ & $-0.5356$ & $\wm 0.5521$ & $-0.4408$ & $\wm 1$ & $-0.8228$ & $\wm 0.6553$ \\ 
$a_1^{g_+}$ & $-0.4590$ & $\wm 0.4889$ & $-0.4499$ & $\wm 0.5178$ & $-0.4026$ & $\wm 0.3880$ & $-0.8228$ & $\wm 1$ & $-0.4987$ \\ 
$a_0^{g_0}$ & $\wm 0.7879$ & $-0.6672$ & $\wm 0.5184$ & $-0.4440$ & $\wm 0.7538$ & $-0.5619$ & $\wm 0.6553$ & $-0.4987$ & $\wm 1$ \\ 
$a_1^{g_0}$ & $-0.6552$ & $\wm 0.6780$ & $-0.4417$ & $\wm 0.4364$ & $-0.6481$ & $\wm 0.6409$ & $-0.5819$ & $\wm 0.5587$ & $-0.8668$ \\ 
$a_1^{g_\perp}$ & $-0.4118$ & $\wm 0.4324$ & $-0.3960$ & $\wm 0.4799$ & $-0.3575$ & $\wm 0.3304$ & $-0.7533$ & $\wm 0.8361$ & $-0.4530$ \\ 
$a_0^{h_+}$ & $\wm 0.7013$ & $-0.6149$ & $\wm 0.3954$ & $-0.3347$ & $\wm 0.7084$ & $-0.5691$ & $\wm 0.4943$ & $-0.3617$ & $\wm 0.6995$ \\ 
$a_1^{h_+}$ & $-0.4922$ & $\wm 0.5349$ & $-0.2781$ & $\wm 0.2624$ & $-0.5217$ & $\wm 0.5722$ & $-0.3669$ & $\wm 0.3343$ & $-0.4721$ \\ 
$a_0^{h_\perp}$ & $\wm 0.7958$ & $-0.6896$ & $\wm 0.4976$ & $-0.4157$ & $\wm 0.7737$ & $-0.6057$ & $\wm 0.5981$ & $-0.4331$ & $\wm 0.7868$ \\ 
$a_1^{h_\perp}$ & $-0.6561$ & $\wm 0.6714$ & $-0.4222$ & $\wm 0.4030$ & $-0.6574$ & $\wm 0.6439$ & $-0.5073$ & $\wm 0.4432$ & $-0.6300$ \\ 
$a_0^{\widetilde{h}_{\perp},\widetilde{h}_+}$ & $\wm 0.6359$ & $-0.5614$ & $\wm 0.6563$ & $-0.5538$ & $\wm 0.5483$ & $-0.4313$ & $\wm 0.7480$ & $-0.5742$ & $\wm 0.6416$ \\ 
$a_1^{\widetilde{h}_+}$ & $-0.3785$ & $\wm 0.4062$ & $-0.3576$ & $\wm 0.4487$ & $-0.3298$ & $\wm 0.3217$ & $-0.4788$ & $\wm 0.5313$ & $-0.3976$ \\ 
$a_1^{\widetilde{h}_\perp}$ & $-0.3969$ & $\wm 0.4324$ & $-0.3952$ & $\wm 0.4721$ & $-0.3445$ & $\wm 0.3482$ & $-0.5159$ & $\wm 0.5621$ & $-0.4100$ \\ 
\hline\hline
\end{tabular}
\caption{\label{tab:nominalcorr1}Correlation matrix of the nominal form factor parameters, part 1.}
\end{table}

\begin{table}[!h]
\small
\begin{tabular}{lccccccccc}
\hline\hline
 & $a_1^{g_0}$ & $a_1^{g_\perp}$ & $a_0^{h_+}$ & $a_1^{h_+}$ & $a_0^{h_\perp}$ & $a_1^{h_\perp}$ & $a_0^{\widetilde{h}_{\perp},\widetilde{h}_+}$ & $a_1^{\widetilde{h}_+}$ & $a_1^{\widetilde{h}_\perp}$ \\ 
\hline
$a_0^{f_+}$ & $-0.6552$ & $-0.4118$ & $\wm 0.7013$ & $-0.4922$ & $\wm 0.7958$ & $-0.6561$ & $\wm 0.6359$ & $-0.3785$ & $-0.3969$ \\ 
$a_1^{f_+}$ & $\wm 0.6780$ & $\wm 0.4324$ & $-0.6149$ & $\wm 0.5349$ & $-0.6896$ & $\wm 0.6714$ & $-0.5614$ & $\wm 0.4062$ & $\wm 0.4324$ \\ 
$a_0^{f_0}$ & $-0.4417$ & $-0.3960$ & $\wm 0.3954$ & $-0.2781$ & $\wm 0.4976$ & $-0.4222$ & $\wm 0.6563$ & $-0.3576$ & $-0.3952$ \\ 
$a_1^{f_0}$ & $\wm 0.4364$ & $\wm 0.4799$ & $-0.3347$ & $\wm 0.2624$ & $-0.4157$ & $\wm 0.4030$ & $-0.5538$ & $\wm 0.4487$ & $\wm 0.4721$ \\ 
$a_0^{f_\perp}$ & $-0.6481$ & $-0.3575$ & $\wm 0.7084$ & $-0.5217$ & $\wm 0.7737$ & $-0.6574$ & $\wm 0.5483$ & $-0.3298$ & $-0.3445$ \\ 
$a_1^{f_\perp}$ & $\wm 0.6409$ & $\wm 0.3304$ & $-0.5691$ & $\wm 0.5722$ & $-0.6057$ & $\wm 0.6439$ & $-0.4313$ & $\wm 0.3217$ & $\wm 0.3482$ \\ 
$a_0^{g_\perp,g_+}$ & $-0.5819$ & $-0.7533$ & $\wm 0.4943$ & $-0.3669$ & $\wm 0.5981$ & $-0.5073$ & $\wm 0.7480$ & $-0.4788$ & $-0.5159$ \\ 
$a_1^{g_+}$ & $\wm 0.5587$ & $\wm 0.8361$ & $-0.3617$ & $\wm 0.3343$ & $-0.4331$ & $\wm 0.4432$ & $-0.5742$ & $\wm 0.5313$ & $\wm 0.5621$ \\ 
$a_0^{g_0}$ & $-0.8668$ & $-0.4530$ & $\wm 0.6995$ & $-0.4721$ & $\wm 0.7868$ & $-0.6300$ & $\wm 0.6416$ & $-0.3976$ & $-0.4100$ \\ 
$a_1^{g_0}$ & $\wm 1$ & $\wm 0.4923$ & $-0.5981$ & $\wm 0.5503$ & $-0.6558$ & $\wm 0.6600$ & $-0.5503$ & $\wm 0.4241$ & $\wm 0.4542$ \\ 
$a_1^{g_\perp}$ & $\wm 0.4923$ & $\wm 1$ & $-0.3233$ & $\wm 0.2886$ & $-0.3896$ & $\wm 0.3907$ & $-0.5186$ & $\wm 0.5070$ & $\wm 0.5339$ \\ 
$a_0^{h_+}$ & $-0.5981$ & $-0.3233$ & $\wm 1$ & $-0.8048$ & $\wm 0.7363$ & $-0.6286$ & $\wm 0.4842$ & $-0.2961$ & $-0.3080$ \\ 
$a_1^{h_+}$ & $\wm 0.5503$ & $\wm 0.2886$ & $-0.8048$ & $\wm 1$ & $-0.5298$ & $\wm 0.5920$ & $-0.3582$ & $\wm 0.2782$ & $\wm 0.3046$ \\ 
$a_0^{h_\perp}$ & $-0.6558$ & $-0.3896$ & $\wm 0.7363$ & $-0.5298$ & $\wm 1$ & $-0.8738$ & $\wm 0.5995$ & $-0.3660$ & $-0.3791$ \\ 
$a_1^{h_\perp}$ & $\wm 0.6600$ & $\wm 0.3907$ & $-0.6286$ & $\wm 0.5920$ & $-0.8738$ & $\wm 1$ & $-0.5044$ & $\wm 0.3793$ & $\wm 0.4023$ \\ 
$a_0^{\widetilde{h}_{\perp},\widetilde{h}_+}$ & $-0.5503$ & $-0.5186$ & $\wm 0.4842$ & $-0.3582$ & $\wm 0.5995$ & $-0.5044$ & $\wm 1$ & $-0.7185$ & $-0.7634$ \\ 
$a_1^{\widetilde{h}_+}$ & $\wm 0.4241$ & $\wm 0.5070$ & $-0.2961$ & $\wm 0.2782$ & $-0.3660$ & $\wm 0.3793$ & $-0.7185$ & $\wm 1$ & $\wm 0.9448$ \\ 
$a_1^{\widetilde{h}_\perp}$ & $\wm 0.4542$ & $\wm 0.5339$ & $-0.3080$ & $\wm 0.3046$ & $-0.3791$ & $\wm 0.4023$ & $-0.7634$ & $\wm 0.9448$ & $\wm 1$ \\
\hline\hline
\end{tabular}
\caption{\label{tab:nominalcorr2}Correlation matrix of the nominal form factor parameters, part 2.}
\end{table}

\begin{turnpage}

\begin{table}
\small
\begin{tabular}{lcccccccccccccccccccccccccccc}
\hline\hline
 & $a_0^{f_+}$ & $a_1^{f_+}$ & $a_2^{f_+}$ & $a_0^{f_0}$ & $a_1^{f_0}$ & $a_2^{f_0}$ & $a_0^{f_\perp}$ & $a_1^{f_\perp}$ & $a_2^{f_\perp}$ & $a_0^{g_\perp,g_+}$ & $a_1^{g_+}$ & $a_2^{g_+}$ & $a_0^{g_0}$ & $a_1^{g_0}$ \\ 
\hline
$a_0^{f_+}$ & $\wm 1$ & $-0.7339$ & $\wm 0.2311$ & $\wm 0.5619$ & $-0.3171$ & $-0.0104$ & $\wm 0.6938$ & $-0.3846$ & $\wm 0.0292$ & $\wm 0.4710$ & $-0.2594$ & $-0.0054$ & $\wm 0.5532$ & $-0.3705$ \\ 
$a_1^{f_+}$ & $-0.7339$ & $\wm 1$ & $-0.6732$ & $-0.3259$ & $\wm 0.2406$ & $-0.0108$ & $-0.4352$ & $\wm 0.3415$ & $-0.0397$ & $-0.3060$ & $\wm 0.2112$ & $\wm 0.0169$ & $-0.3805$ & $\wm 0.3368$ \\ 
$a_2^{f_+}$ & $\wm 0.2311$ & $-0.6732$ & $\wm 1$ & $-0.0709$ & $\wm 0.0673$ & $\wm 0.2478$ & $\wm 0.0784$ & $-0.1001$ & $\wm 0.0886$ & $-0.0058$ & $\wm 0.0161$ & $\wm 0.0399$ & $\wm 0.0399$ & $-0.0797$ \\ 
$a_0^{f_0}$ & $\wm 0.5619$ & $-0.3259$ & $-0.0709$ & $\wm 1$ & $-0.7563$ & $\wm 0.3096$ & $\wm 0.4994$ & $-0.2483$ & $-0.0045$ & $\wm 0.4748$ & $-0.2686$ & $\wm 0.0179$ & $\wm 0.4091$ & $-0.2573$ \\ 
$a_1^{f_0}$ & $-0.3171$ & $\wm 0.2406$ & $\wm 0.0673$ & $-0.7563$ & $\wm 1$ & $-0.7158$ & $-0.2727$ & $\wm 0.1794$ & $\wm 0.0104$ & $-0.3182$ & $\wm 0.2729$ & $-0.0497$ & $-0.2673$ & $\wm 0.2101$ \\ 
$a_2^{f_0}$ & $-0.0104$ & $-0.0108$ & $\wm 0.2478$ & $\wm 0.3096$ & $-0.7158$ & $\wm 1$ & $\wm 0.0135$ & $\wm 0.0004$ & $\wm 0.0417$ & $\wm 0.0446$ & $-0.0818$ & $\wm 0.1067$ & $\wm 0.0086$ & $-0.0129$ \\ 
$a_0^{f_\perp}$ & $\wm 0.6938$ & $-0.4352$ & $\wm 0.0784$ & $\wm 0.4994$ & $-0.2727$ & $\wm 0.0135$ & $\wm 1$ & $-0.7106$ & $\wm 0.1636$ & $\wm 0.4375$ & $-0.2323$ & $-0.0217$ & $\wm 0.5475$ & $-0.3836$ \\ 
$a_1^{f_\perp}$ & $-0.3846$ & $\wm 0.3415$ & $-0.1001$ & $-0.2483$ & $\wm 0.1794$ & $\wm 0.0004$ & $-0.7106$ & $\wm 1$ & $-0.5951$ & $-0.2540$ & $\wm 0.1760$ & $\wm 0.0317$ & $-0.3372$ & $\wm 0.3297$ \\ 
$a_2^{f_\perp}$ & $\wm 0.0292$ & $-0.0397$ & $\wm 0.0886$ & $-0.0045$ & $\wm 0.0104$ & $\wm 0.0417$ & $\wm 0.1636$ & $-0.5951$ & $\wm 1$ & $-0.0062$ & $\wm 0.0012$ & $\wm 0.0709$ & $\wm 0.0097$ & $-0.0288$ \\ 
$a_0^{g_\perp,g_+}$ & $\wm 0.4710$ & $-0.3060$ & $-0.0058$ & $\wm 0.4748$ & $-0.3182$ & $\wm 0.0446$ & $\wm 0.4375$ & $-0.2540$ & $-0.0062$ & $\wm 1$ & $-0.6700$ & $\wm 0.2588$ & $\wm 0.6125$ & $-0.3281$ \\ 
$a_1^{g_+}$ & $-0.2594$ & $\wm 0.2112$ & $\wm 0.0161$ & $-0.2686$ & $\wm 0.2729$ & $-0.0818$ & $-0.2323$ & $\wm 0.1760$ & $\wm 0.0012$ & $-0.6700$ & $\wm 1$ & $-0.7221$ & $-0.3059$ & $\wm 0.2228$ \\ 
$a_2^{g_+}$ & $-0.0054$ & $\wm 0.0169$ & $\wm 0.0399$ & $\wm 0.0179$ & $-0.0497$ & $\wm 0.1067$ & $-0.0217$ & $\wm 0.0317$ & $\wm 0.0709$ & $\wm 0.2588$ & $-0.7221$ & $\wm 1$ & $-0.0097$ & $\wm 0.0306$ \\ 
$a_0^{g_0}$ & $\wm 0.5532$ & $-0.3805$ & $\wm 0.0399$ & $\wm 0.4091$ & $-0.2673$ & $\wm 0.0086$ & $\wm 0.5475$ & $-0.3372$ & $\wm 0.0097$ & $\wm 0.6125$ & $-0.3059$ & $-0.0097$ & $\wm 1$ & $-0.7498$ \\ 
$a_1^{g_0}$ & $-0.3705$ & $\wm 0.3368$ & $-0.0797$ & $-0.2573$ & $\wm 0.2101$ & $-0.0129$ & $-0.3836$ & $\wm 0.3297$ & $-0.0288$ & $-0.3281$ & $\wm 0.2228$ & $\wm 0.0306$ & $-0.7498$ & $\wm 1$ \\ 
$a_2^{g_0}$ & $\wm 0.0577$ & $-0.0808$ & $\wm 0.1347$ & $\wm 0.0055$ & $\wm 0.0125$ & $\wm 0.0479$ & $\wm 0.0769$ & $-0.0978$ & $\wm 0.1067$ & $-0.0163$ & $\wm 0.0246$ & $\wm 0.2410$ & $\wm 0.3002$ & $-0.6973$ \\ 
$a_1^{g_\perp}$ & $-0.2617$ & $\wm 0.2120$ & $\wm 0.0179$ & $-0.2593$ & $\wm 0.2696$ & $-0.0805$ & $-0.2358$ & $\wm 0.1774$ & $\wm 0.0082$ & $-0.6393$ & $\wm 0.6447$ & $-0.3001$ & $-0.3186$ & $\wm 0.2276$ \\ 
$a_2^{g_\perp}$ & $\wm 0.0270$ & $-0.0266$ & $\wm 0.0411$ & $\wm 0.0295$ & $-0.0421$ & $\wm 0.0621$ & $\wm 0.0234$ & $-0.0206$ & $\wm 0.0237$ & $\wm 0.1673$ & $-0.1942$ & $\wm 0.2013$ & $\wm 0.0413$ & $-0.0259$ \\ 
$a_0^{h_+}$ & $\wm 0.4320$ & $-0.2957$ & $\wm 0.0442$ & $\wm 0.2896$ & $-0.1765$ & $-0.0057$ & $\wm 0.4519$ & $-0.2842$ & $\wm 0.0133$ & $\wm 0.3361$ & $-0.1753$ & $-0.0184$ & $\wm 0.4208$ & $-0.2903$ \\ 
$a_1^{h_+}$ & $-0.2916$ & $\wm 0.2563$ & $-0.0692$ & $-0.1879$ & $\wm 0.1305$ & $\wm 0.0104$ & $-0.3212$ & $\wm 0.2789$ & $-0.0261$ & $-0.2291$ & $\wm 0.1513$ & $\wm 0.0285$ & $-0.2745$ & $\wm 0.2529$ \\ 
$a_2^{h_+}$ & $-0.0357$ & $\wm 0.0142$ & $\wm 0.0352$ & $-0.0462$ & $\wm 0.0393$ & $\wm 0.0042$ & $-0.0374$ & $\wm 0.0073$ & $\wm 0.0704$ & $-0.0407$ & $\wm 0.0179$ & $\wm 0.0451$ & $-0.0417$ & $\wm 0.0204$ \\ 
$a_0^{h_\perp}$ & $\wm 0.4485$ & $-0.3123$ & $\wm 0.0454$ & $\wm 0.3170$ & $-0.2001$ & $\wm 0.0013$ & $\wm 0.4560$ & $-0.2870$ & $\wm 0.0148$ & $\wm 0.3552$ & $-0.1921$ & $-0.0113$ & $\wm 0.4346$ & $-0.2975$ \\ 
$a_1^{h_\perp}$ & $-0.3831$ & $\wm 0.3277$ & $-0.0854$ & $-0.2633$ & $\wm 0.1966$ & $-0.0142$ & $-0.4047$ & $\wm 0.3250$ & $-0.0403$ & $-0.2890$ & $\wm 0.1881$ & $\wm 0.0125$ & $-0.3616$ & $\wm 0.3153$ \\ 
$a_2^{h_\perp}$ & $\wm 0.0149$ & $-0.0429$ & $\wm 0.0852$ & $-0.0129$ & $\wm 0.0074$ & $\wm 0.0366$ & $\wm 0.0197$ & $-0.0473$ & $\wm 0.0898$ & $-0.0164$ & $\wm 0.0058$ & $\wm 0.0545$ & $\wm 0.0112$ & $-0.0390$ \\ 
$a_0^{\widetilde{h}_{\perp},\widetilde{h}_+}$ & $\wm 0.3554$ & $-0.2258$ & $-0.0057$ & $\wm 0.3704$ & $-0.2463$ & $\wm 0.0401$ & $\wm 0.3277$ & $-0.1850$ & $-0.0030$ & $\wm 0.4066$ & $-0.2438$ & $\wm 0.0268$ & $\wm 0.3407$ & $-0.2088$ \\ 
$a_1^{\widetilde{h}_+}$ & $-0.2489$ & $\wm 0.1846$ & $\wm 0.0063$ & $-0.2451$ & $\wm 0.2691$ & $-0.1224$ & $-0.2250$ & $\wm 0.1514$ & $\wm 0.0006$ & $-0.3005$ & $\wm 0.2828$ & $-0.0927$ & $-0.2442$ & $\wm 0.1771$ \\ 
$a_2^{\widetilde{h}_+}$ & $-0.0086$ & $\wm 0.0188$ & $\wm 0.0369$ & $-0.0109$ & $\wm 0.0012$ & $\wm 0.0578$ & $-0.0160$ & $\wm 0.0276$ & $\wm 0.0148$ & $-0.0210$ & $\wm 0.0074$ & $\wm 0.0700$ & $-0.0081$ & $\wm 0.0150$ \\ 
$a_1^{\widetilde{h}_\perp}$ & $-0.2546$ & $\wm 0.1962$ & $\wm 0.0040$ & $-0.2643$ & $\wm 0.2785$ & $-0.1175$ & $-0.2284$ & $\wm 0.1638$ & $\wm 0.0003$ & $-0.3173$ & $\wm 0.2952$ & $-0.0894$ & $-0.2437$ & $\wm 0.1851$ \\ 
$a_2^{\widetilde{h}_\perp}$ & $-0.0114$ & $\wm 0.0174$ & $\wm 0.0358$ & $-0.0081$ & $-0.0125$ & $\wm 0.0694$ & $-0.0201$ & $\wm 0.0250$ & $\wm 0.0212$ & $-0.0198$ & $-0.0027$ & $\wm 0.0853$ & $-0.0167$ & $\wm 0.0204$ \\ 
\hline\hline
\end{tabular}
\caption{\label{tab:HOcorr1}Correlation matrix of the higher-order form factor parameters, part 1.}
\end{table}

\end{turnpage}

\begin{turnpage}

\begin{table}
\small
\begin{tabular}{lcccccccccccccccccccccccccccc}
\hline\hline
 & $a_2^{g_0}$ & $a_1^{g_\perp}$ & $a_2^{g_\perp}$ & $a_0^{h_+}$ & $a_1^{h_+}$ & $a_2^{h_+}$ & $a_0^{h_\perp}$ & $a_1^{h_\perp}$ & $a_2^{h_\perp}$ & $a_0^{\widetilde{h}_{\perp},\widetilde{h}_+}$ & $a_1^{\widetilde{h}_+}$ & $a_2^{\widetilde{h}_+}$ & $a_1^{\widetilde{h}_\perp}$ & $a_2^{\widetilde{h}_\perp}$ \\ 
\hline
$a_0^{f_+}$ & $\wm 0.0577$ & $-0.2617$ & $\wm 0.0270$ & $\wm 0.4320$ & $-0.2916$ & $-0.0357$ & $\wm 0.4485$ & $-0.3831$ & $\wm 0.0149$ & $\wm 0.3554$ & $-0.2489$ & $-0.0086$ & $-0.2546$ & $-0.0114$ \\ 
$a_1^{f_+}$ & $-0.0808$ & $\wm 0.2120$ & $-0.0266$ & $-0.2957$ & $\wm 0.2563$ & $\wm 0.0142$ & $-0.3123$ & $\wm 0.3277$ & $-0.0429$ & $-0.2258$ & $\wm 0.1846$ & $\wm 0.0188$ & $\wm 0.1962$ & $\wm 0.0174$ \\ 
$a_2^{f_+}$ & $\wm 0.1347$ & $\wm 0.0179$ & $\wm 0.0411$ & $\wm 0.0442$ & $-0.0692$ & $\wm 0.0352$ & $\wm 0.0454$ & $-0.0854$ & $\wm 0.0852$ & $-0.0057$ & $\wm 0.0063$ & $\wm 0.0369$ & $\wm 0.0040$ & $\wm 0.0358$ \\ 
$a_0^{f_0}$ & $\wm 0.0055$ & $-0.2593$ & $\wm 0.0295$ & $\wm 0.2896$ & $-0.1879$ & $-0.0462$ & $\wm 0.3170$ & $-0.2633$ & $-0.0129$ & $\wm 0.3704$ & $-0.2451$ & $-0.0109$ & $-0.2643$ & $-0.0081$ \\ 
$a_1^{f_0}$ & $\wm 0.0125$ & $\wm 0.2696$ & $-0.0421$ & $-0.1765$ & $\wm 0.1305$ & $\wm 0.0393$ & $-0.2001$ & $\wm 0.1966$ & $\wm 0.0074$ & $-0.2463$ & $\wm 0.2691$ & $\wm 0.0012$ & $\wm 0.2785$ & $-0.0125$ \\ 
$a_2^{f_0}$ & $\wm 0.0479$ & $-0.0805$ & $\wm 0.0621$ & $-0.0057$ & $\wm 0.0104$ & $\wm 0.0042$ & $\wm 0.0013$ & $-0.0142$ & $\wm 0.0366$ & $\wm 0.0401$ & $-0.1224$ & $\wm 0.0578$ & $-0.1175$ & $\wm 0.0694$ \\ 
$a_0^{f_\perp}$ & $\wm 0.0769$ & $-0.2358$ & $\wm 0.0234$ & $\wm 0.4519$ & $-0.3212$ & $-0.0374$ & $\wm 0.4560$ & $-0.4047$ & $\wm 0.0197$ & $\wm 0.3277$ & $-0.2250$ & $-0.0160$ & $-0.2284$ & $-0.0201$ \\ 
$a_1^{f_\perp}$ & $-0.0978$ & $\wm 0.1774$ & $-0.0206$ & $-0.2842$ & $\wm 0.2789$ & $\wm 0.0073$ & $-0.2870$ & $\wm 0.3250$ & $-0.0473$ & $-0.1850$ & $\wm 0.1514$ & $\wm 0.0276$ & $\wm 0.1638$ & $\wm 0.0250$ \\ 
$a_2^{f_\perp}$ & $\wm 0.1067$ & $\wm 0.0082$ & $\wm 0.0237$ & $\wm 0.0133$ & $-0.0261$ & $\wm 0.0704$ & $\wm 0.0148$ & $-0.0403$ & $\wm 0.0898$ & $-0.0030$ & $\wm 0.0006$ & $\wm 0.0148$ & $\wm 0.0003$ & $\wm 0.0212$ \\ 
$a_0^{g_\perp,g_+}$ & $-0.0163$ & $-0.6393$ & $\wm 0.1673$ & $\wm 0.3361$ & $-0.2291$ & $-0.0407$ & $\wm 0.3552$ & $-0.2890$ & $-0.0164$ & $\wm 0.4066$ & $-0.3005$ & $-0.0210$ & $-0.3173$ & $-0.0198$ \\ 
$a_1^{g_+}$ & $\wm 0.0246$ & $\wm 0.6447$ & $-0.1942$ & $-0.1753$ & $\wm 0.1513$ & $\wm 0.0179$ & $-0.1921$ & $\wm 0.1881$ & $\wm 0.0058$ & $-0.2438$ & $\wm 0.2828$ & $\wm 0.0074$ & $\wm 0.2952$ & $-0.0027$ \\ 
$a_2^{g_+}$ & $\wm 0.2410$ & $-0.3001$ & $\wm 0.2013$ & $-0.0184$ & $\wm 0.0285$ & $\wm 0.0451$ & $-0.0113$ & $\wm 0.0125$ & $\wm 0.0545$ & $\wm 0.0268$ & $-0.0927$ & $\wm 0.0700$ & $-0.0894$ & $\wm 0.0853$ \\ 
$a_0^{g_0}$ & $\wm 0.3002$ & $-0.3186$ & $\wm 0.0413$ & $\wm 0.4208$ & $-0.2745$ & $-0.0417$ & $\wm 0.4346$ & $-0.3616$ & $\wm 0.0112$ & $\wm 0.3407$ & $-0.2442$ & $-0.0081$ & $-0.2437$ & $-0.0167$ \\ 
$a_1^{g_0}$ & $-0.6973$ & $\wm 0.2276$ & $-0.0259$ & $-0.2903$ & $\wm 0.2529$ & $\wm 0.0204$ & $-0.2975$ & $\wm 0.3153$ & $-0.0390$ & $-0.2088$ & $\wm 0.1771$ & $\wm 0.0150$ & $\wm 0.1851$ & $\wm 0.0204$ \\ 
$a_2^{g_0}$ & $\wm 1$ & $\wm 0.0596$ & $\wm 0.0566$ & $\wm 0.0557$ & $-0.0566$ & $\wm 0.0488$ & $\wm 0.0536$ & $-0.0887$ & $\wm 0.1011$ & $-0.0066$ & $\wm 0.0138$ & $\wm 0.0560$ & $\wm 0.0153$ & $\wm 0.0591$ \\ 
$a_1^{g_\perp}$ & $\wm 0.0596$ & $\wm 1$ & $-0.6002$ & $-0.1793$ & $\wm 0.1546$ & $\wm 0.0228$ & $-0.1944$ & $\wm 0.1894$ & $\wm 0.0106$ & $-0.2341$ & $\wm 0.2793$ & $\wm 0.0028$ & $\wm 0.2914$ & $-0.0052$ \\ 
$a_2^{g_\perp}$ & $\wm 0.0566$ & $-0.6002$ & $\wm 1$ & $\wm 0.0130$ & $-0.0169$ & $\wm 0.0193$ & $\wm 0.0158$ & $-0.0238$ & $\wm 0.0271$ & $\wm 0.0153$ & $-0.0596$ & $\wm 0.0756$ & $-0.0585$ & $\wm 0.0801$ \\ 
$a_0^{h_+}$ & $\wm 0.0557$ & $-0.1793$ & $\wm 0.0130$ & $\wm 1$ & $-0.6176$ & $\wm 0.0511$ & $\wm 0.7980$ & $-0.4369$ & $\wm 0.0429$ & $\wm 0.6438$ & $-0.2746$ & $\wm 0.0256$ & $-0.2824$ & $\wm 0.0178$ \\ 
$a_1^{h_+}$ & $-0.0566$ & $\wm 0.1546$ & $-0.0169$ & $-0.6176$ & $\wm 1$ & $-0.5243$ & $-0.3409$ & $\wm 0.3166$ & $-0.0457$ & $-0.2502$ & $\wm 0.1550$ & $\wm 0.0209$ & $\wm 0.1710$ & $\wm 0.0192$ \\ 
$a_2^{h_+}$ & $\wm 0.0488$ & $\wm 0.0228$ & $\wm 0.0193$ & $\wm 0.0511$ & $-0.5243$ & $\wm 1$ & $-0.0408$ & $\wm 0.0050$ & $\wm 0.0879$ & $-0.0406$ & $\wm 0.0261$ & $\wm 0.0128$ & $\wm 0.0249$ & $\wm 0.0180$ \\ 
$a_0^{h_\perp}$ & $\wm 0.0536$ & $-0.1944$ & $\wm 0.0158$ & $\wm 0.7980$ & $-0.3409$ & $-0.0408$ & $\wm 1$ & $-0.6707$ & $\wm 0.1528$ & $\wm 0.6864$ & $-0.2997$ & $\wm 0.0313$ & $-0.3079$ & $\wm 0.0255$ \\ 
$a_1^{h_\perp}$ & $-0.0887$ & $\wm 0.1894$ & $-0.0238$ & $-0.4369$ & $\wm 0.3166$ & $\wm 0.0050$ & $-0.6707$ & $\wm 1$ & $-0.5892$ & $-0.3283$ & $\wm 0.2053$ & $\wm 0.0143$ & $\wm 0.2167$ & $\wm 0.0099$ \\ 
$a_2^{h_\perp}$ & $\wm 0.1011$ & $\wm 0.0106$ & $\wm 0.0271$ & $\wm 0.0429$ & $-0.0457$ & $\wm 0.0879$ & $\wm 0.1528$ & $-0.5892$ & $\wm 1$ & $\wm 0.0151$ & $\wm 0.0017$ & $\wm 0.0199$ & $\wm 0.0001$ & $\wm 0.0246$ \\ 
$a_0^{\widetilde{h}_{\perp},\widetilde{h}_+}$ & $-0.0066$ & $-0.2341$ & $\wm 0.0153$ & $\wm 0.6438$ & $-0.2502$ & $-0.0406$ & $\wm 0.6864$ & $-0.3283$ & $\wm 0.0151$ & $\wm 1$ & $-0.5627$ & $\wm 0.1115$ & $-0.5850$ & $\wm 0.1175$ \\ 
$a_1^{\widetilde{h}_+}$ & $\wm 0.0138$ & $\wm 0.2793$ & $-0.0596$ & $-0.2746$ & $\wm 0.1550$ & $\wm 0.0261$ & $-0.2997$ & $\wm 0.2053$ & $\wm 0.0017$ & $-0.5627$ & $\wm 1$ & $-0.4627$ & $\wm 0.7130$ & $-0.2025$ \\ 
$a_2^{\widetilde{h}_+}$ & $\wm 0.0560$ & $\wm 0.0028$ & $\wm 0.0756$ & $\wm 0.0256$ & $\wm 0.0209$ & $\wm 0.0128$ & $\wm 0.0313$ & $\wm 0.0143$ & $\wm 0.0199$ & $\wm 0.1115$ & $-0.4627$ & $\wm 1$ & $-0.1630$ & $\wm 0.4328$ \\ 
$a_1^{\widetilde{h}_\perp}$ & $\wm 0.0153$ & $\wm 0.2914$ & $-0.0585$ & $-0.2824$ & $\wm 0.1710$ & $\wm 0.0249$ & $-0.3079$ & $\wm 0.2167$ & $\wm 0.0001$ & $-0.5850$ & $\wm 0.7130$ & $-0.1630$ & $\wm 1$ & $-0.5189$ \\ 
$a_2^{\widetilde{h}_\perp}$ & $\wm 0.0591$ & $-0.0052$ & $\wm 0.0801$ & $\wm 0.0178$ & $\wm 0.0192$ & $\wm 0.0180$ & $\wm 0.0255$ & $\wm 0.0099$ & $\wm 0.0246$ & $\wm 0.1175$ & $-0.2025$ & $\wm 0.4328$ & $-0.5189$ & $\wm 1$ \\ \hline\hline
\end{tabular}
\caption{\label{tab:HOcorr2}Correlation matrix of the higher-order form factor parameters, part 2.}
\end{table}

\end{turnpage}

\begin{table}
\footnotesize
\begin{tabular}{ccccllllllllllllll}
\hline\hline
         & & $|\mathbf{p'}|^2/(2\pi/L)^2$ && \hspace{2.5ex} \texttt{C14} & \hspace{2ex} & \hspace{2.5ex} \texttt{C24} & \hspace{2ex} & \hspace{2.5ex} \texttt{C54} & \hspace{2ex} & \hspace{2.5ex} \texttt{C53} & \hspace{2ex} & \hspace{2.5ex} \texttt{F23} & \hspace{2ex} & \hspace{2.5ex} \texttt{F43} & \hspace{2ex} & \hspace{2.5ex} \texttt{F63} \\
\hline
$f_+$                  && 1  &&  1.189(25) &&  1.174(21) &&  1.168(22) &&  1.145(20) &&  1.218(27) &&  1.198(21) &&  1.172(19)  \\ 
                       && 2  &&  1.024(35) &&  1.013(23) &&  1.014(24) &&  1.006(17) &&  1.052(25) &&  1.044(23) &&  1.032(22)  \\ 
                       && 3  &&  0.911(44) &&  0.901(28) &&  0.902(28) &&  0.885(34) &&  0.935(43) &&  0.930(34) &&  0.920(28)  \\ 
                       && 4  &&  0.820(40) &&  0.816(22) &&  0.814(19) &&  0.782(20) &&  0.868(32) &&  0.852(22) &&  0.833(14)  \\ 
                       && 5  &&  0.737(20) &&  0.734(13) &&  0.735(14) &&  0.701(11) &&  0.773(16) &&  0.766(13) &&  0.7529(98)  \\ 
                       && 6  &&  0.672(19) &&  0.672(13) &&  0.673(13) &&  0.644(10) &&  0.705(14) &&  0.702(12) &&  0.6960(94)  \\ 
                       && 8  &&  0.572(19) &&  0.586(13) &&  0.588(14) &&  0.546(11) &&  0.608(15) &&  0.604(13) &&  0.6064(98)  \\ 
                       && 9  &&  0.537(20) &&  0.540(14) &&  0.543(14) &&  0.501(12) &&  0.567(15) &&  0.564(13) &&  0.567(10)  \\ 
                       && 10 &&  0.503(20) &&  0.527(15) &&  0.522(15) &&  0.472(14) &&  0.539(15) &&  0.536(13) &&  0.544(10)  \\ 
\hline
$f_\perp$              && 1  &&  1.479(32) &&  1.466(28) &&  1.470(31) &&  1.424(27) &&  1.512(38) &&  1.488(37) &&  1.461(32)  \\ 
                       && 2  &&  1.297(41) &&  1.281(30) &&  1.297(30) &&  1.275(25) &&  1.325(35) &&  1.321(33) &&  1.312(25)  \\ 
                       && 3  &&  1.157(67) &&  1.148(42) &&  1.160(40) &&  1.134(56) &&  1.186(63) &&  1.183(51) &&  1.175(37)  \\ 
                       && 4  &&  1.030(58) &&  1.024(36) &&  1.034(31) &&  0.994(29) &&  1.093(49) &&  1.079(36) &&  1.059(25)  \\ 
                       && 5  &&  0.917(25) &&  0.915(20) &&  0.927(20) &&  0.888(18) &&  0.963(22) &&  0.962(20) &&  0.954(16)  \\ 
                       && 6  &&  0.839(25) &&  0.838(19) &&  0.848(20) &&  0.816(18) &&  0.883(21) &&  0.886(20) &&  0.879(16)  \\ 
                       && 8  &&  0.712(25) &&  0.726(20) &&  0.738(20) &&  0.682(21) &&  0.761(22) &&  0.762(20) &&  0.766(17)  \\ 
                       && 9  &&  0.666(25) &&  0.670(20) &&  0.681(20) &&  0.620(20) &&  0.711(22) &&  0.712(20) &&  0.714(17)  \\ 
                       && 10 &&  0.630(26) &&  0.654(22) &&  0.654(22) &&  0.590(24) &&  0.682(24) &&  0.679(22) &&  0.683(18)  \\ 
\hline
$f_0$                  && 1  &&  0.863(33) &&  0.845(36) &&  0.837(42) &&  0.830(42) &&  0.895(28) &&  0.881(29) &&  0.857(36)  \\ 
                       && 2  &&  0.768(34) &&  0.765(24) &&  0.756(30) &&  0.741(35) &&  0.794(27) &&  0.785(27) &&  0.770(32)  \\ 
                       && 3  &&  0.709(26) &&  0.699(19) &&  0.694(23) &&  0.672(27) &&  0.723(27) &&  0.716(26) &&  0.704(26)  \\ 
                       && 4  &&  0.656(23) &&  0.650(16) &&  0.640(17) &&  0.604(23) &&  0.676(23) &&  0.663(21) &&  0.654(18)  \\ 
                       && 5  &&  0.605(21) &&  0.605(13) &&  0.599(13) &&  0.563(20) &&  0.608(20) &&  0.602(18) &&  0.599(16)  \\ 
                       && 6  &&  0.560(21) &&  0.564(13) &&  0.558(13) &&  0.528(20) &&  0.565(20) &&  0.558(19) &&  0.561(16)  \\ 
                       && 8  &&  0.488(22) &&  0.505(13) &&  0.503(14) &&  0.471(20) &&  0.499(21) &&  0.496(20) &&  0.508(17)  \\ 
                       && 9  &&  0.468(22) &&  0.475(14) &&  0.474(14) &&  0.450(20) &&  0.469(22) &&  0.465(21) &&  0.485(17)  \\ 
                       && 10 &&  0.444(24) &&  0.473(14) &&  0.467(14) &&  0.423(25) &&  0.455(23) &&  0.454(21) &&  0.473(18)  \\ 
\hline
$g_+$                  && 1  &&  0.802(19) &&  0.790(12) &&  0.785(12) &&  0.777(18) &&  0.824(19) &&  0.815(18) &&  0.788(17)  \\ 
                       && 2  &&  0.715(20) &&  0.706(12) &&  0.701(15) &&  0.685(19) &&  0.736(19) &&  0.727(19) &&  0.705(17)  \\ 
                       && 3  &&  0.653(20) &&  0.641(14) &&  0.637(16) &&  0.616(20) &&  0.667(19) &&  0.658(19) &&  0.642(17)  \\ 
                       && 4  &&  0.597(19) &&  0.5992(94) &&  0.5907(91) &&  0.560(17) &&  0.601(19) &&  0.592(18) &&  0.582(17)  \\ 
                       && 5  &&  0.552(19) &&  0.5538(86) &&  0.5491(82) &&  0.514(17) &&  0.557(20) &&  0.549(21) &&  0.538(18)  \\ 
                       && 6  &&  0.510(20) &&  0.514(10) &&  0.5084(86) &&  0.478(17) &&  0.516(25) &&  0.508(25) &&  0.499(20)  \\ 
                       && 8  &&  0.445(27) &&  0.461(10) &&  0.4602(87) &&  0.428(22) &&  0.456(23) &&  0.453(22) &&  0.455(19)  \\ 
                       && 9  &&  0.425(23) &&  0.431(12) &&  0.4322(92) &&  0.407(17) &&  0.429(26) &&  0.425(25) &&  0.433(19)  \\ 
                       && 10 &&  0.409(28) &&  0.438(10) &&  0.4317(92) &&  0.393(25) &&  0.422(23) &&  0.419(22) &&  0.428(17)  \\ 
\hline
$g_\perp$              && 1  &&  0.801(23) &&  0.790(14) &&  0.785(14) &&  0.777(21) &&  0.824(23) &&  0.814(23) &&  0.789(21)  \\ 
                       && 2  &&  0.714(24) &&  0.705(14) &&  0.699(17) &&  0.683(21) &&  0.733(24) &&  0.725(23) &&  0.705(21)  \\ 
                       && 3  &&  0.652(24) &&  0.640(16) &&  0.635(18) &&  0.614(22) &&  0.663(24) &&  0.654(23) &&  0.641(22)  \\ 
                       && 4  &&  0.594(23) &&  0.597(13) &&  0.587(12) &&  0.556(21) &&  0.594(24) &&  0.586(23) &&  0.578(21)  \\ 
                       && 5  &&  0.548(23) &&  0.552(12) &&  0.546(11) &&  0.509(21) &&  0.550(25) &&  0.542(25) &&  0.535(23)  \\ 
                       && 6  &&  0.506(23) &&  0.513(14) &&  0.507(12) &&  0.476(21) &&  0.507(31) &&  0.499(30) &&  0.496(26)  \\ 
                       && 8  &&  0.443(35) &&  0.462(15) &&  0.461(13) &&  0.427(28) &&  0.448(32) &&  0.444(30) &&  0.453(26)  \\ 
                       && 9  &&  0.423(32) &&  0.433(19) &&  0.434(14) &&  0.412(24) &&  0.415(39) &&  0.412(37) &&  0.429(29)  \\ 
                       && 10 &&  0.405(40) &&  0.440(15) &&  0.433(13) &&  0.393(34) &&  0.417(34) &&  0.414(32) &&  0.429(25)  \\ 
\hline
$g_0$                  && 1  &&  1.200(25) &&  1.184(24) &&  1.172(19) &&  1.178(24) &&  1.208(32) &&  1.184(32) &&  1.146(24)  \\ 
                       && 2  &&  1.022(34) &&  1.009(26) &&  1.014(22) &&  1.020(20) &&  1.050(29) &&  1.038(30) &&  1.008(22)  \\ 
                       && 3  &&  0.897(17) &&  0.885(13) &&  0.892(13) &&  0.880(14) &&  0.941(17) &&  0.931(14) &&  0.900(11)  \\ 
                       && 4  &&  0.808(14) &&  0.802(12) &&  0.801(11) &&  0.795(19) &&  0.843(17) &&  0.834(12) &&  0.8090(91)  \\ 
                       && 5  &&  0.725(13) &&  0.722(11) &&  0.726(10) &&  0.710(12) &&  0.768(12) &&  0.760(11) &&  0.7370(82)  \\ 
                       && 6  &&  0.656(13) &&  0.652(11) &&  0.655(10) &&  0.634(11) &&  0.694(12) &&  0.690(11) &&  0.6705(88)  \\ 
                       && 8  &&  0.553(13) &&  0.563(11) &&  0.567(10) &&  0.526(12) &&  0.597(12) &&  0.594(11) &&  0.5800(92)  \\ 
                       && 9  &&  0.520(13) &&  0.517(11) &&  0.523(12) &&  0.479(11) &&  0.556(12) &&  0.554(12) &&  0.5451(95)  \\ 
                       && 10 &&  0.492(14) &&  0.510(14) &&  0.506(16) &&  0.465(15) &&  0.527(14) &&  0.523(13) &&  0.518(10)  \\ 
\hline\hline
\end{tabular}
\normalsize
\caption{\label{tab:VAhelicitylat}Lattice results for the vector and axial vector current helicity form factors.}
\end{table}

\begin{table}
\footnotesize
\begin{tabular}{ccccllllllllllllll}
\hline\hline
         & & $|\mathbf{p'}|^2/(2\pi/L)^2$ && \hspace{2.5ex} \texttt{C14} & \hspace{2ex} & \hspace{2.5ex} \texttt{C24} & \hspace{2ex} & \hspace{2.5ex} \texttt{C54} & \hspace{2ex} & \hspace{2.5ex} \texttt{C53} & \hspace{2ex} & \hspace{2.5ex} \texttt{F23} & \hspace{2ex} & \hspace{2.5ex} \texttt{F43} & \hspace{2ex} & \hspace{2.5ex} \texttt{F63} \\
\hline
$h_+$                  && 1  &&  1.402(37) &&  1.391(31) &&  1.393(34) &&  1.358(32) &&  1.437(42) &&  1.403(46) &&  1.378(35)  \\ 
                       && 2  &&  1.247(36) &&  1.230(28) &&  1.242(26) &&  1.210(28) &&  1.270(36) &&  1.268(31) &&  1.252(23)  \\ 
                       && 3  &&  1.106(65) &&  1.096(50) &&  1.106(45) &&  1.076(49) &&  1.134(59) &&  1.137(51) &&  1.125(33)  \\ 
                       && 4  &&  0.999(46) &&  0.992(33) &&  0.995(30) &&  0.980(28) &&  1.055(40) &&  1.037(35) &&  1.014(25)  \\ 
                       && 5  &&  0.905(46) &&  0.901(31) &&  0.909(29) &&  0.884(30) &&  0.953(42) &&  0.947(36) &&  0.935(26)  \\ 
                       && 6  &&  0.821(42) &&  0.821(32) &&  0.826(30) &&  0.795(39) &&  0.866(42) &&  0.865(35) &&  0.854(25)  \\ 
                       && 8  &&  0.678(32) &&  0.685(23) &&  0.695(24) &&  0.668(28) &&  0.733(28) &&  0.731(27) &&  0.728(20)  \\ 
                       && 9  &&  0.628(32) &&  0.631(23) &&  0.640(23) &&  0.604(25) &&  0.687(27) &&  0.686(27) &&  0.678(20)  \\ 
                       && 10 &&  0.608(33) &&  0.618(25) &&  0.616(25) &&  0.601(31) &&  0.669(30) &&  0.663(29) &&  0.659(21)  \\ 
\hline
$h_\perp$              && 1  &&  1.074(23) &&  1.054(22) &&  1.052(22) &&  1.043(21) &&  1.101(27) &&  1.076(24) &&  1.055(18)  \\ 
                       && 2  &&  0.938(21) &&  0.919(16) &&  0.922(17) &&  0.913(17) &&  0.960(21) &&  0.952(17) &&  0.939(13)  \\ 
                       && 3  &&  0.829(40) &&  0.817(28) &&  0.819(27) &&  0.808(32) &&  0.850(36) &&  0.847(27) &&  0.837(19)  \\ 
                       && 4  &&  0.750(21) &&  0.740(14) &&  0.739(12) &&  0.729(14) &&  0.786(17) &&  0.772(13) &&  0.757(10)  \\ 
                       && 5  &&  0.678(19) &&  0.674(12) &&  0.676(11) &&  0.662(14) &&  0.712(16) &&  0.706(13) &&  0.693(11)  \\ 
                       && 6  &&  0.618(18) &&  0.616(11) &&  0.616(11) &&  0.600(15) &&  0.651(14) &&  0.647(12) &&  0.638(10)  \\ 
                       && 8  &&  0.522(18) &&  0.528(12) &&  0.531(11) &&  0.517(14) &&  0.559(14) &&  0.555(12) &&  0.5514(91)  \\ 
                       && 9  &&  0.489(18) &&  0.487(12) &&  0.490(11) &&  0.474(14) &&  0.524(14) &&  0.520(12) &&  0.5170(91)  \\ 
                       && 10 &&  0.464(19) &&  0.476(14) &&  0.471(12) &&  0.462(17) &&  0.502(19) &&  0.497(13) &&  0.496(10)  \\ 
\hline
$\widetilde{h}_+$      && 1  &&  0.772(16) &&  0.759(13) &&  0.756(15) &&  0.749(17) &&  0.795(14) &&  0.785(12) &&  0.762(14)  \\ 
                       && 2  &&  0.685(20) &&  0.675(13) &&  0.672(16) &&  0.661(19) &&  0.702(15) &&  0.694(13) &&  0.680(18)  \\ 
                       && 3  &&  0.634(15) &&  0.620(14) &&  0.617(15) &&  0.602(17) &&  0.644(15) &&  0.637(14) &&  0.623(15)  \\ 
                       && 4  &&  0.583(17) &&  0.575(14) &&  0.567(15) &&  0.540(20) &&  0.604(15) &&  0.588(14) &&  0.573(16)  \\ 
                       && 5  &&  0.540(16) &&  0.529(14) &&  0.525(15) &&  0.503(19) &&  0.553(16) &&  0.542(15) &&  0.533(16)  \\ 
                       && 6  &&  0.495(18) &&  0.500(17) &&  0.494(18) &&  0.477(20) &&  0.509(19) &&  0.497(17) &&  0.493(17)  \\ 
                       && 8  &&  0.451(18) &&  0.467(20) &&  0.464(20) &&  0.451(17) &&  0.462(21) &&  0.458(20) &&  0.466(17)  \\ 
                       && 9  &&  0.436(26) &&  0.443(14) &&  0.442(14) &&  0.437(16) &&  0.441(28) &&  0.434(28) &&  0.449(19)  \\ 
                       && 10 &&  0.421(21) &&  0.446(11) &&  0.440(12) &&  0.418(21) &&  0.453(19) &&  0.439(17) &&  0.446(15)  \\ 
\hline
$\widetilde{h}_\perp$  && 1  &&  0.772(16) &&  0.758(12) &&  0.755(14) &&  0.748(17) &&  0.794(13) &&  0.784(12) &&  0.762(14)  \\ 
                       && 2  &&  0.685(20) &&  0.675(12) &&  0.672(15) &&  0.660(18) &&  0.701(14) &&  0.694(13) &&  0.680(17)  \\ 
                       && 3  &&  0.633(15) &&  0.619(11) &&  0.616(12) &&  0.602(15) &&  0.643(14) &&  0.635(13) &&  0.622(13)  \\ 
                       && 4  &&  0.585(16) &&  0.575(12) &&  0.567(13) &&  0.543(17) &&  0.601(17) &&  0.587(13) &&  0.575(14)  \\ 
                       && 5  &&  0.539(15) &&  0.527(13) &&  0.524(13) &&  0.503(16) &&  0.551(14) &&  0.541(14) &&  0.531(15)  \\ 
                       && 6  &&  0.490(17) &&  0.494(16) &&  0.489(17) &&  0.474(17) &&  0.504(17) &&  0.493(16) &&  0.490(16)  \\ 
                       && 8  &&  0.442(17) &&  0.457(17) &&  0.455(17) &&  0.442(14) &&  0.454(17) &&  0.450(16) &&  0.459(14)  \\ 
                       && 9  &&  0.422(24) &&  0.428(13) &&  0.428(13) &&  0.423(15) &&  0.428(28) &&  0.423(28) &&  0.438(18)  \\ 
                       && 10 &&  0.410(21) &&  0.4320(89) &&  0.4271(99) &&  0.406(19) &&  0.439(16) &&  0.428(15) &&  0.436(13)  \\ 
\hline\hline
\end{tabular}
\normalsize
\caption{\label{tab:Thelicitylat}Lattice results for the tensor current helicity form factors.}
\end{table}

\begin{table}
\footnotesize
\begin{tabular}{ccccllllllllllllll}
\hline\hline
         & & $|\mathbf{p'}|^2/(2\pi/L)^2$ && \hspace{2.5ex} \texttt{C14} & \hspace{0.5ex} & \hspace{2.5ex} \texttt{C24} & \hspace{0.5ex} & \hspace{2.5ex} \texttt{C54} & \hspace{0.5ex} & \hspace{2.5ex} \texttt{C53} & \hspace{0.5ex} & \hspace{2.5ex} \texttt{F23} & \hspace{0.5ex} & \hspace{2.5ex} \texttt{F43} & \hspace{0.5ex} & \hspace{2.5ex} \texttt{F63} \\
\hline
$f_1^V$                && 1  && $\wm0.995(24)$ && $\wm0.983(20)$ && $\wm0.976(19)$ && $\wm0.958(21)$ && $\wm1.017(26)$ && $\wm1.005(21)$ && $\wm0.987(17)$  \\ 
                       && 2  && $\wm0.855(32)$ && $\wm0.850(22)$ && $\wm0.847(22)$ && $\wm0.839(17)$ && $\wm0.879(23)$ && $\wm0.874(20)$ && $\wm0.865(22)$  \\ 
                       && 3  && $\wm0.769(32)$ && $\wm0.762(21)$ && $\wm0.761(23)$ && $\wm0.740(24)$ && $\wm0.787(33)$ && $\wm0.785(26)$ && $\wm0.778(24)$  \\ 
                       && 4  && $\wm0.707(31)$ && $\wm0.707(16)$ && $\wm0.701(14)$ && $\wm0.668(18)$ && $\wm0.746(25)$ && $\wm0.732(16)$ && $\wm0.7157(99)$  \\ 
                       && 5  && $\wm0.646(17)$ && $\wm0.645(11)$ && $\wm0.643(11)$ && $\wm0.6073(99)$ && $\wm0.676(15)$ && $\wm0.668(12)$ && $\wm0.6555(84)$  \\ 
                       && 6  && $\wm0.594(16)$ && $\wm0.596(10)$ && $\wm0.594(11)$ && $\wm0.5634(88)$ && $\wm0.620(13)$ && $\wm0.615(10)$ && $\wm0.6129(78)$  \\ 
                       && 8  && $\wm0.515(17)$ && $\wm0.530(11)$ && $\wm0.529(11)$ && $\wm0.490(10)$ && $\wm0.544(14)$ && $\wm0.540(11)$ && $\wm0.5427(84)$  \\ 
                       && 9  && $\wm0.488(17)$ && $\wm0.492(11)$ && $\wm0.492(12)$ && $\wm0.456(11)$ && $\wm0.512(14)$ && $\wm0.507(11)$ && $\wm0.5118(87)$  \\ 
                       && 10 && $\wm0.458(18)$ && $\wm0.483(13)$ && $\wm0.477(13)$ && $\wm0.429(17)$ && $\wm0.487(14)$ && $\wm0.485(12)$ && $\wm0.4949(87)$  \\ 
\hline
$f_2^V$                && 1  && $\wm0.399(19)$ && $\wm0.397(17)$ && $\wm0.404(17)$ && $\wm0.384(24)$ && $\wm0.410(28)$ && $\wm0.398(30)$ && $\wm0.388(22)$  \\ 
                       && 2  && $\wm0.365(18)$ && $\wm0.354(16)$ && $\wm0.369(14)$ && $\wm0.359(19)$ && $\wm0.369(24)$ && $\wm0.368(20)$ && $\wm0.366(14)$  \\ 
                       && 3  && $\wm0.319(31)$ && $\wm0.317(21)$ && $\wm0.326(17)$ && $\wm0.324(31)$ && $\wm0.330(30)$ && $\wm0.328(25)$ && $\wm0.326(15)$  \\ 
                       && 4  && $\wm0.266(24)$ && $\wm0.261(20)$ && $\wm0.272(16)$ && $\wm0.269(17)$ && $\wm0.287(26)$ && $\wm0.286(21)$ && $\wm0.282(16)$  \\ 
                       && 5  && $\wm0.223(12)$ && $\wm0.222(10)$ && $\wm0.2328(93)$ && $\wm0.231(14)$ && $\wm0.237(15)$ && $\wm0.242(15)$ && $\wm0.244(10)$  \\ 
                       && 6  && $\wm0.202(12)$ && $\wm0.199(10)$ && $\wm0.2081(91)$ && $\wm0.209(14)$ && $\wm0.217(15)$ && $\wm0.223(15)$ && $\wm0.218(10)$  \\ 
                       && 8  && $\wm0.163(12)$ && $\wm0.162(10)$ && $\wm0.1714(94)$ && $\wm0.158(16)$ && $\wm0.179(16)$ && $\wm0.183(16)$ && $\wm0.183(11)$  \\ 
                       && 9  && $\wm0.147(13)$ && $\wm0.147(10)$ && $\wm0.1547(93)$ && $\wm0.135(15)$ && $\wm0.164(16)$ && $\wm0.169(16)$ && $\wm0.166(11)$  \\ 
                       && 10 && $\wm0.142(14)$ && $\wm0.140(12)$ && $\wm0.145(11)$ && $\wm0.132(21)$ && $\wm0.161(21)$ && $\wm0.160(18)$ && $\wm0.155(13)$  \\ 
\hline
$f_3^V$                && 1  && $-0.176(40)$ && $-0.185(39)$ && $-0.187(43)$ && $-0.170(65)$ && $-0.162(37)$ && $-0.165(38)$ && $-0.174(43)$  \\ 
                       && 2  && $-0.120(25)$ && $-0.119(23)$ && $-0.128(22)$ && $-0.137(44)$ && $-0.118(28)$ && $-0.124(24)$ && $-0.133(22)$  \\ 
                       && 3  && $-0.089(24)$ && $-0.092(23)$ && $-0.100(22)$ && $-0.100(30)$ && $-0.093(27)$ && $-0.101(24)$ && $-0.108(20)$  \\ 
                       && 4  && $-0.078(34)$ && $-0.087(37)$ && $-0.094(33)$ && $-0.098(36)$ && $-0.107(52)$ && $-0.105(36)$ && $-0.095(21)$  \\ 
                       && 5  && $-0.067(21)$ && $-0.065(20)$ && $-0.070(18)$ && $-0.070(29)$ && $-0.108(27)$ && $-0.105(22)$ && $-0.091(18)$  \\ 
                       && 6  && $-0.057(22)$ && $-0.053(21)$ && $-0.061(19)$ && $-0.059(30)$ && $-0.093(27)$ && $-0.096(23)$ && $-0.087(20)$  \\ 
                       && 8  && $-0.050(29)$ && $-0.046(25)$ && $-0.048(20)$ && $-0.036(32)$ && $-0.084(30)$ && $-0.081(26)$ && $-0.063(22)$  \\ 
                       && 9  && $-0.037(34)$ && $-0.032(26)$ && $-0.035(21)$ && $-0.011(31)$ && $-0.082(33)$ && $-0.080(29)$ && $-0.051(23)$  \\ 
                       && 10 && $-0.028(26)$ && $-0.020(25)$ && $-0.021(21)$ && $-0.013(34)$ && $-0.065(34)$ && $-0.062(30)$ && $-0.044(25)$  \\ 
\hline
$f_1^A$                && 1  && $\wm0.817(13)$ && $\wm0.803(13)$ && $\wm0.804(12)$ && $\wm0.781(16)$ && $\wm0.837(17)$ && $\wm0.832(14)$ && $\wm0.777(12)$  \\ 
                       && 2  && $\wm0.726(11)$ && $\wm0.717(12)$ && $\wm0.716(10)$ && $\wm0.709(15)$ && $\wm0.757(18)$ && $\wm0.750(17)$ && $\wm0.707(12)$  \\ 
                       && 3  && $\wm0.664(11)$ && $\wm0.648(10)$ && $\wm0.6497(96)$ && $\wm0.631(12)$ && $\wm0.692(15)$ && $\wm0.686(13)$ && $\wm0.650(13)$  \\ 
                       && 4  && $\wm0.6139(89)$ && $\wm0.6119(85)$ && $\wm0.6106(84)$ && $\wm0.5823(92)$ && $\wm0.630(18)$ && $\wm0.626(13)$ && $\wm0.5981(61)$  \\ 
                       && 5  && $\wm0.5660(80)$ && $\wm0.5609(96)$ && $\wm0.5600(80)$ && $\wm0.5349(87)$ && $\wm0.5846(87)$ && $\wm0.5789(72)$ && $\wm0.5521(49)$  \\ 
                       && 6  && $\wm0.520(11)$ && $\wm0.5160(77)$ && $\wm0.5140(77)$ && $\wm0.4858(85)$ && $\wm0.541(10)$ && $\wm0.5355(98)$ && $\wm0.5111(73)$  \\ 
                       && 8  && $\wm0.450(12)$ && $\wm0.4589(78)$ && $\wm0.4583(76)$ && $\wm0.430(11)$ && $\wm0.475(11)$ && $\wm0.4710(99)$ && $\wm0.4588(72)$  \\ 
                       && 9  && $\wm0.4298(97)$ && $\wm0.4279(77)$ && $\wm0.4290(87)$ && $\wm0.3976(83)$ && $\wm0.455(10)$ && $\wm0.4515(96)$ && $\wm0.4404(67)$  \\ 
                       && 10 && $\wm0.416(12)$ && $\wm0.435(10)$ && $\wm0.429(13)$ && $\wm0.392(14)$ && $\wm0.431(12)$ && $\wm0.428(11)$ && $\wm0.4256(72)$  \\ 
\hline
$f_2^A$                && 1  && $\wm0.021(25)$ && $\wm0.017(23)$ && $\wm0.024(22)$ && $\wm0.005(27)$ && $\wm0.017(31)$ && $\wm0.023(30)$ && $-0.015(29)$  \\ 
                       && 2  && $\wm0.015(24)$ && $\wm0.015(24)$ && $\wm0.022(27)$ && $\wm0.034(36)$ && $\wm0.030(31)$ && $\wm0.032(32)$ && $\wm0.002(33)$  \\ 
                       && 3  && $\wm0.015(24)$ && $\wm0.011(27)$ && $\wm0.019(29)$ && $\wm0.022(32)$ && $\wm0.038(35)$ && $\wm0.041(36)$ && $\wm0.013(38)$  \\ 
                       && 4  && $\wm0.026(24)$ && $\wm0.020(22)$ && $\wm0.031(21)$ && $\wm0.034(26)$ && $\wm0.045(30)$ && $\wm0.051(29)$ && $\wm0.025(29)$  \\ 
                       && 5  && $\wm0.023(23)$ && $\wm0.011(21)$ && $\wm0.018(20)$ && $\wm0.033(25)$ && $\wm0.044(28)$ && $\wm0.048(29)$ && $\wm0.022(30)$  \\ 
                       && 6  && $\wm0.017(24)$ && $\wm0.004(23)$ && $\wm0.009(22)$ && $\wm0.013(26)$ && $\wm0.043(32)$ && $\wm0.046(32)$ && $\wm0.020(31)$  \\ 
                       && 8  && $\wm0.009(32)$ && $-0.005(25)$ && $-0.003(22)$ && $\wm0.004(28)$ && $\wm0.034(39)$ && $\wm0.034(38)$ && $\wm0.007(32)$  \\ 
                       && 9  && $\wm0.009(32)$ && $-0.007(29)$ && $-0.006(25)$ && $-0.018(29)$ && $\wm0.050(47)$ && $\wm0.050(45)$ && $\wm0.015(39)$  \\ 
                       && 10 && $\wm0.015(40)$ && $-0.007(26)$ && $-0.005(24)$ && $-0.002(35)$ && $\wm0.018(39)$ && $\wm0.017(38)$ && $-0.004(32)$  \\ 
\hline
$f_3^A$                && 1  && $-0.787(44)$ && $-0.793(46)$ && $-0.778(32)$ && $-0.816(48)$ && $-0.753(61)$ && $-0.727(61)$ && $-0.774(38)$  \\ 
                       && 2  && $-0.639(71)$ && $-0.637(69)$ && $-0.657(57)$ && $-0.669(53)$ && $-0.624(81)$ && $-0.623(87)$ && $-0.662(61)$  \\ 
                       && 3  && $-0.527(33)$ && $-0.541(27)$ && $-0.560(19)$ && $-0.560(28)$ && $-0.554(33)$ && $-0.554(30)$ && $-0.572(20)$  \\ 
                       && 4  && $-0.458(27)$ && $-0.454(25)$ && $-0.461(19)$ && $-0.501(45)$ && $-0.497(32)$ && $-0.491(30)$ && $-0.506(21)$  \\ 
                       && 5  && $-0.394(23)$ && $-0.403(22)$ && $-0.420(17)$ && $-0.431(28)$ && $-0.448(30)$ && $-0.447(29)$ && $-0.464(20)$  \\ 
                       && 6  && $-0.351(23)$ && $-0.357(22)$ && $-0.374(16)$ && $-0.384(26)$ && $-0.391(28)$ && $-0.400(28)$ && $-0.418(20)$  \\ 
                       && 8  && $-0.295(24)$ && $-0.299(23)$ && $-0.317(18)$ && $-0.271(27)$ && $-0.344(32)$ && $-0.349(31)$ && $-0.348(21)$  \\ 
                       && 9  && $-0.271(25)$ && $-0.270(23)$ && $-0.288(18)$ && $-0.240(27)$ && $-0.301(31)$ && $-0.306(31)$ && $-0.315(22)$  \\ 
                       && 10 && $-0.237(27)$ && $-0.238(26)$ && $-0.245(21)$ && $-0.229(40)$ && $-0.298(38)$ && $-0.298(36)$ && $-0.293(27)$  \\ 
\hline\hline
\end{tabular}
\normalsize
\caption{\label{tab:VAWeinberglat}Lattice results for the vector and axial vector current Weinberg form factors.}
\end{table}

\begin{table}
\footnotesize
\begin{tabular}{ccccllllllllllllll}
\hline\hline
         & & $|\mathbf{p'}|^2/(2\pi/L)^2$ && \hspace{2.5ex} \texttt{C14} & \hspace{0.5ex} & \hspace{2.5ex} \texttt{C24} & \hspace{0.5ex} & \hspace{2.5ex} \texttt{C54} & \hspace{0.5ex} & \hspace{2.5ex} \texttt{C53} & \hspace{0.5ex} & \hspace{2.5ex} \texttt{F23} & \hspace{0.5ex} & \hspace{2.5ex} \texttt{F43} & \hspace{0.5ex} & \hspace{2.5ex} \texttt{F63} \\
\hline
$f_1^{TV}$             && 1  && $\wm0.451(24)$ && $\wm0.458(22)$ && $\wm0.456(21)$ && $\wm0.432(23)$ && $\wm0.468(33)$ && $\wm0.447(36)$ && $\wm0.434(29)$  \\ 
                       && 2  && $\wm0.412(23)$ && $\wm0.411(21)$ && $\wm0.415(19)$ && $\wm0.398(20)$ && $\wm0.419(29)$ && $\wm0.420(26)$ && $\wm0.410(17)$  \\ 
                       && 3  && $\wm0.360(34)$ && $\wm0.359(29)$ && $\wm0.365(23)$ && $\wm0.348(24)$ && $\wm0.372(34)$ && $\wm0.375(33)$ && $\wm0.367(20)$  \\ 
                       && 4  && $\wm0.316(33)$ && $\wm0.315(26)$ && $\wm0.316(23)$ && $\wm0.318(22)$ && $\wm0.344(35)$ && $\wm0.334(31)$ && $\wm0.320(20)$  \\ 
                       && 5  && $\wm0.280(34)$ && $\wm0.278(26)$ && $\wm0.282(24)$ && $\wm0.274(24)$ && $\wm0.301(36)$ && $\wm0.298(30)$ && $\wm0.294(19)$  \\ 
                       && 6  && $\wm0.245(30)$ && $\wm0.245(25)$ && $\wm0.248(23)$ && $\wm0.236(29)$ && $\wm0.262(35)$ && $\wm0.263(29)$ && $\wm0.258(18)$  \\ 
                       && 8  && $\wm0.180(20)$ && $\wm0.181(17)$ && $\wm0.188(16)$ && $\wm0.176(20)$ && $\wm0.203(22)$ && $\wm0.204(23)$ && $\wm0.202(16)$  \\ 
                       && 9  && $\wm0.158(20)$ && $\wm0.163(17)$ && $\wm0.167(16)$ && $\wm0.148(17)$ && $\wm0.187(22)$ && $\wm0.189(22)$ && $\wm0.182(15)$  \\ 
                       && 10 && $\wm0.161(21)$ && $\wm0.158(18)$ && $\wm0.159(17)$ && $\wm0.155(22)$ && $\wm0.187(24)$ && $\wm0.186(24)$ && $\wm0.181(17)$  \\ 
\hline
$f_2^{TV}$             && 1  && $\wm0.855(20)$ && $\wm0.833(19)$ && $\wm0.836(17)$ && $\wm0.833(20)$ && $\wm0.871(26)$ && $\wm0.860(19)$ && $\wm0.849(15)$  \\ 
                       && 2  && $\wm0.747(16)$ && $\wm0.731(14)$ && $\wm0.734(14)$ && $\wm0.728(16)$ && $\wm0.763(19)$ && $\wm0.758(14)$ && $\wm0.752(12)$  \\ 
                       && 3  && $\wm0.669(26)$ && $\wm0.659(17)$ && $\wm0.661(19)$ && $\wm0.653(24)$ && $\wm0.684(25)$ && $\wm0.681(16)$ && $\wm0.677(13)$  \\ 
                       && 4  && $\wm0.616(11)$ && $\wm0.6083(80)$ && $\wm0.6088(71)$ && $\wm0.5942(94)$ && $\wm0.638(13)$ && $\wm0.6312(94)$ && $\wm0.6233(73)$  \\ 
                       && 5  && $\wm0.565(11)$ && $\wm0.5634(74)$ && $\wm0.5641(68)$ && $\wm0.5508(93)$ && $\wm0.589(11)$ && $\wm0.5854(86)$ && $\wm0.5759(52)$  \\ 
                       && 6  && $\wm0.523(10)$ && $\wm0.5221(67)$ && $\wm0.5227(60)$ && $\wm0.5090(77)$ && $\wm0.5493(82)$ && $\wm0.5460(55)$ && $\wm0.5391(44)$  \\ 
                       && 8  && $\wm0.459(11)$ && $\wm0.4647(77)$ && $\wm0.4662(67)$ && $\wm0.454(10)$ && $\wm0.4870(92)$ && $\wm0.4834(66)$ && $\wm0.4811(57)$  \\ 
                       && 9  && $\wm0.437(11)$ && $\wm0.4327(78)$ && $\wm0.4350(68)$ && $\wm0.424(10)$ && $\wm0.4608(93)$ && $\wm0.4563(67)$ && $\wm0.4566(57)$  \\ 
                       && 10 && $\wm0.413(18)$ && $\wm0.426(12)$ && $\wm0.4219(89)$ && $\wm0.413(15)$ && $\wm0.442(22)$ && $\wm0.437(14)$ && $\wm0.4387(89)$  \\ 
\hline
$f_1^{TA}$             && 1  && $\wm0.010(20)$ && $\wm0.018(19)$ && $\wm0.021(17)$ && $\wm0.011(25)$ && $\wm0.043(33)$ && $\wm0.027(33)$ && $\wm0.009(19)$  \\ 
                       && 2  && $\wm0.000(17)$ && $\wm0.005(18)$ && $\wm0.005(18)$ && $\wm0.003(21)$ && $\wm0.014(22)$ && $\wm0.012(20)$ && $\wm0.008(17)$  \\ 
                       && 3  && $\wm0.004(24)$ && $\wm0.010(30)$ && $\wm0.008(29)$ && $\wm0.003(28)$ && $\wm0.013(25)$ && $\wm0.014(21)$ && $\wm0.006(30)$  \\ 
                       && 4  && $-0.017(27)$ && $-0.002(24)$ && $-0.003(21)$ && $-0.020(26)$ && $\wm0.021(37)$ && $\wm0.006(37)$ && $-0.011(22)$  \\ 
                       && 5  && $\wm0.008(17)$ && $\wm0.016(17)$ && $\wm0.012(16)$ && $\wm0.000(23)$ && $\wm0.012(22)$ && $\wm0.011(20)$ && $\wm0.011(17)$  \\ 
                       && 6  && $\wm0.024(18)$ && $\wm0.031(17)$ && $\wm0.024(16)$ && $\wm0.015(23)$ && $\wm0.027(26)$ && $\wm0.022(21)$ && $\wm0.017(17)$  \\ 
                       && 8  && $\wm0.037(17)$ && $\wm0.044(15)$ && $\wm0.037(16)$ && $\wm0.038(19)$ && $\wm0.033(26)$ && $\wm0.030(24)$ && $\wm0.031(21)$  \\ 
                       && 9  && $\wm0.053(13)$ && $\wm0.057(13)$ && $\wm0.051(13)$ && $\wm0.054(15)$ && $\wm0.047(20)$ && $\wm0.042(18)$ && $\wm0.043(14)$  \\ 
                       && 10 && $\wm0.040(15)$ && $\wm0.049(14)$ && $\wm0.046(14)$ && $\wm0.039(20)$ && $\wm0.047(23)$ && $\wm0.036(20)$ && $\wm0.035(16)$  \\ 
\hline
$f_2^{TA}$             && 1  && $\wm0.764(22)$ && $\wm0.745(15)$ && $\wm0.739(16)$ && $\wm0.740(22)$ && $\wm0.761(28)$ && $\wm0.763(26)$ && $\wm0.755(23)$  \\ 
                       && 2  && $\wm0.685(22)$ && $\wm0.672(10)$ && $\wm0.6681(94)$ && $\wm0.658(13)$ && $\wm0.691(17)$ && $\wm0.685(15)$ && $\wm0.674(10)$  \\ 
                       && 3  && $\wm0.631(17)$ && $\wm0.612(15)$ && $\wm0.611(11)$ && $\wm0.600(12)$ && $\wm0.634(20)$ && $\wm0.626(16)$ && $\wm0.618(12)$  \\ 
                       && 4  && $\wm0.596(17)$ && $\wm0.577(14)$ && $\wm0.5695(97)$ && $\wm0.556(12)$ && $\wm0.588(34)$ && $\wm0.583(28)$ && $\wm0.582(11)$  \\ 
                       && 5  && $\wm0.534(15)$ && $\wm0.517(11)$ && $\wm0.516(11)$ && $\wm0.503(14)$ && $\wm0.543(16)$ && $\wm0.534(15)$ && $\wm0.524(13)$  \\ 
                       && 6  && $\wm0.476(18)$ && $\wm0.476(13)$ && $\wm0.475(14)$ && $\wm0.465(15)$ && $\wm0.487(19)$ && $\wm0.480(19)$ && $\wm0.479(15)$  \\ 
                       && 8  && $\wm0.422(18)$ && $\wm0.433(13)$ && $\wm0.435(11)$ && $\wm0.421(13)$ && $\wm0.437(19)$ && $\wm0.434(18)$ && $\wm0.442(12)$  \\ 
                       && 9  && $\wm0.394(22)$ && $\wm0.398(13)$ && $\wm0.402(12)$ && $\wm0.395(14)$ && $\wm0.404(29)$ && $\wm0.401(29)$ && $\wm0.416(17)$  \\ 
                       && 10 && $\wm0.390(21)$ && $\wm0.4081(95)$ && $\wm0.4045(88)$ && $\wm0.387(18)$ && $\wm0.416(18)$ && $\wm0.410(17)$ && $\wm0.418(11)$  \\ 
\hline\hline
\end{tabular}
\normalsize
\caption{\label{tab:TWeinberglat}Lattice results for the tensor current Weinberg form factors.}
\end{table}

\FloatBarrier

\providecommand{\href}[2]{#2}
\begingroup
\raggedright

\endgroup

\end{document}